\documentclass[fleqn,usenatbib]{mnras}

\usepackage[T1]{fontenc}
\usepackage{ae,aecompl}

\usepackage{longtable,lscape}
\usepackage{multicol}
\usepackage{natbib}
\usepackage{txfonts}
\usepackage[labelformat=default,labelfont=bf,labelsep=period,]{caption,subcaption}
\usepackage{booktabs}
\usepackage{hyperref}
\PassOptionsToPackage{pdfpagelabels=false}{hyperref}

\def\m2s2{\,m$^{2}$\,s$^{-2}$} 

\usepackage{graphicx}	
\usepackage{amsmath}	
\usepackage{amssymb}	
\usepackage{multirow}
\usepackage{afterpage}

\title[Multiplicity statistics of the coolest brown dwarfs]{Constraining the multiplicity statistics of the coolest brown dwarfs:\\
binary fraction continues to decrease with spectral type}

\author[C. Fontanive et al.]{
Cl\'emence Fontanive,$^{1,2}$\thanks{E-mail: \href{mailto:fontan@roe.ac.uk}{fontan@roe.ac.uk}}
Beth Biller,$^{1,2}$
Mariangela Bonavita$^{1,2}$
and Katelyn Allers$^{3}$
\\
$^{1}$Institute for Astronomy, University of Edinburgh, Blackford Hill, Edinburgh EH9 3HJ, UK\\
$^{2}$Centre for Exoplanet Science, University of Edinburgh, Edinburgh EH9 3FD, UK\\
$^{3}$Department of Physics and Astronomy, Bucknell University, Lewisburg, PA 17837, USA 
}

\date{Accepted 2018 June 22. Received 2018 June 19; in original form 2017 December 20}

\pubyear{2018}

\begin{document}
\label{firstpage}
\pagerange{\pageref{firstpage}--\pageref{lastpage}}
\maketitle

\begin{abstract}
Binary statistics of the latest-type T and Y brown dwarfs are sparse and it is unclear whether the trends seen in the multiplicity properties of their more massive counterparts hold for the very coolest brown dwarfs. We present results from a search for substellar and planetary-mass companions to a sample of 12 ultracool T8$-$Y0 field brown dwarfs with the Hubble Space Telescope/Wide Field Camera 3. We find no evidence for resolved binary companions among our sample down to separations of 0.7$-$2.5 AU. Combining our survey with prior searches, we place some of the first statistically robust constraints to date on the multiplicity properties of the coolest, lowest-mass brown dwarfs in the field. Accounting for observational biases and incompleteness, we derive a binary frequency of $f = 5.5^{+5.2}_{-3.3}$\% for T5$-$Y0 brown dwarfs at separations of 1.5$-$1000 AU, for an overall binary fraction of $f_\mathrm{tot} = 8\pm6$\%. Modelling the projected separation as a lognormal distribution, we find a peak in separation at $\rho_0 = 2.9^{+0.8}_{-1.4}$ AU with a logarithmic width of $\sigma = 0.21^{+0.14}_{-0.08}$. We infer a mass ratio distribution peaking strongly towards unity, with a power law index of $\gamma = 6.1^{+4.0}_{-2.7}$, reinforcing the significance of the detection of a tighter and higher mass ratio companion population around lower-mass primaries. These results are consistent with prior studies and support the idea of a decreasing binary frequency with spectral type in the Galactic field.
\end{abstract}



\section{Introduction}
\label{intro}

There is evidence that binary frequency in the Galactic field decreases as a function of spectral type. Over 70\% of massive B and A-type stars are observed in binary or hierarchical systems \citep{Kouwenhoven2007, Peter2012}. This fraction decreases to 50$-$60\% for Solar-type stars \citep{Duquennoy1991, Raghavan2010} and around 30$-$40\% of M-stars are found in multiple systems \citep{Fischer1992, Delfosse2004, Janson2012}. Surveys probing old (1$-$10 Gyr) brown dwarfs from the field \citep{Close2003, Burgasser2006, Gelino2011, Huelamo2015} observed a substantially lower binary rate ($\sim$10$-$20\%) than in the stellar population, extending the trend of a decreasing binary fraction with later spectral type seen in the stellar regime. As stellar binary frequency decreases with decreasing primary mass, the semi-major axis distribution peaks at closer separations and mass ratios shift towards unity. These trends appear to continue across the boundary between stars and substellar objects and to persist throughout the brown dwarf mass regime \citep{Duchene2007,Duchene2013, Kraus2012}. Indeed, brown dwarf binaries are found to be less prevalent than their stellar analogues, are predominantly found on tightly bound orbits, with an observed peak in separation around $\sim$4 AU, and are highly concentrated near equal-mass systems, with over 75\% of systems having mass ratios $q \equiv M_\mathrm{2}/M_\mathrm{1} \geq 0.8$ \citep{Allen2007, Burgasser2007}.
Surveys investigating binary properties of late-M and L dwarfs in the field \citep{Reid2001, Close2002, Close2003} found binary fractions around 15$-$20\%. These searches also revealed that L dwarfs have fewer binary companions detected on separations $>$10 AU than M-type field objects. \citet{Burgasser2003, Burgasser2006} probed T-dwarfs with spectral types spanning from T0$-$T8 and measured binary rates of $\sim$10\%, with all identified systems having separations $<$5 AU and mass ratios $>$0.8. These results confirm the idea of a decreasing binary fraction within the brown dwarf mass regime and suggest a more compact and symmetric substellar binary population at later spectral types \citep{Huelamo2015}.

Binary statistics of the latest-type ($\geqslant$T8), coolest brown dwarfs ($T_\mathrm{eff} < 800$ K) are still poorly constrained, mainly because the majority of late-T and Y ultracool dwarfs were only discovered in recent years \citep{Cushing2011, Kirkpatrick2011, Kirkpatrick2012, Mace2013, Pinfield2014}. Five brown dwarf binaries with primary spectral types of T8 or later have been discovered so far (see \citealp{Gelino2011, Liu2011b, Liu2012, Dupuy2015}). Two of these systems (W1217$+$1626 and W1711$+$3500; \citealp{Liu2012}) have unusually wide separations (8$-$15 AU) and surprisingly low mass ratios ($q\sim0.5$). It is not clear whether these discoveries signal a change in binary properties at the lowest masses or consist of peculiar systems, thus not representative of the true binary population of ultracool dwarfs. Most formation scenarios for brown dwarfs only allow very tight binaries ($<$10 AU separation) to survive to field ages (e.g. ejection scenario, \citealp{Reipurth2001}; turbulent fragmentation, \citealp{Padoan2004}; disc fragmentation and binary disruption, \citealp{Goodwin2007}). The existence of wide field binaries such as those discovered by \citet{Liu2012} is difficult to explain via such mechanisms but such systems may simply be uncommon.

In this paper we present a search for low-mass companions to some of the coolest brown dwarfs in order to place the first constraints to date on the binary properties of the latest-type T and Y dwarfs in the field. Our multiplicity search is also an attempt to confirm whether wide, low mass ratio systems are indeed more common around $\geqslant$T8 dwarfs than around their more massive, earlier-type counterparts. Section~\ref{sample} describes the probed sample and our observations. The search for companions is detailed in Section~\ref{search} and the achieved sensitivity limits are presented in Section~\ref{limits}. In Section~\ref{additional_samples} we introduce additional samples of T5$-$T7.5 and $\geqslant$T8 brown dwarfs from the multiplicity surveys in \citet{Gelino2011} and \citet{Aberasturi2014}. The latter subset is used to extend the size of our observed sample and set more robust statistical constraints on binary fraction for $\geqslant$T8 brown dwarfs, while the former serves as a comparison with earlier spectral types. A thorough statistical analysis of binary properties is detailed in Section~\ref{stats} for the observed and additional samples. We provide an assessment of the multiplicity properties of mid-T to Y field brown dwarfs in Section~\ref{discussion}, where we discuss our interpretation of the obtained results and compare them to earlier-type stellar and substellar objects. Finally, we summarise the main results of our project in Section~\ref{conclusions}.

\begin{small}
\begin{table*}
\caption{Log of HST observations, F127M photometry and F139M limiting magnitude.}
\begin{tabular}{ l c c c c c c }
\hline\hline
Object ID	    & Short & Obs. Date &   \multicolumn{2}{c}{F127M} & \multicolumn{2}{c}{F139M}  \\
\cmidrule(lr){4-5} \cmidrule(lr){6-7}
        &       & (UT)      &  t (s) & Phot. (mag)     &  t (s) & Phot. (mag) \\
\hline
WISE J014656.66$+$423410.0  & W0146$+$4234  & 2013 Jun 24 & 698.465 & $20.01 \pm 0.03$ & 698.465 & $> 26.5$\\ 
WISE J014807.30$-$720259.0  & W0148$-$7202  & 2012 Oct 30 & 698.465 & $18.51 \pm 0.03$ & 698.465 & $> 24.9$\\
WISE J024714.52$+$372523.5  & W0247$+$3725	& 2012 Nov 14 & 698.465 & $17.65 \pm 0.03$ & 698.465 & $> 24.7$\\
WISE J032120.91$-$734758.8  & W0321$-$7347	& 2013 Aug 18 & 698.465 & $18.35 \pm 0.03$ & 698.465 & $> 24.9$\\
WISE J033515.01$+$431045.1  & W0335$+$4310	& 2013 Jan 01 & 698.465 & $19.02 \pm 0.03$ & 698.465 & $> 27.2$\\
WISE J071322.55$-$291751.9  & W0713$-$2917	& 2013 Aug 15 & 698.465 & $19.33 \pm 0.03$ & 698.465 & $> 25.3$\\
WISE J072312.44$+$340313.5  & W0723$+$3403	& 2013 Apr 04 & 698.465 & $17.84 \pm 0.03$ & 698.465 & $> 26.9$\\
WISE J073444.02$-$715744.0  & W0734$-$7157	& 2013 Sep 23 & 698.465 & $19.82 \pm 0.03$ & 698.465 & $> 24.8$\\
WISE J104245.23$-$384238.3  & W1042$-$3842	& 2012 Oct 26 & 698.465 & $18.39 \pm 0.03$ & 698.465 & $> 25.5$\\
WISE J115013.88$+$630240.7  & W1150$+$6302  & 2012 Nov 10 & 698.465 & $17.36 \pm 0.03$ & 698.465 & $> 24.9$\\
WISE J151721.13$+$052929.3  & W1517$+$0529  & 2013 Jul 22 & 698.465 & $18.18 \pm 0.03$ & 698.465 & $> 26.5$\\
WISE J222055.31$-$362817.4  & W2220$-$3628  & 2013 Jul 21 & 698.465 & $19.84 \pm 0.03$ & 698.465 & $> 26.4$\\
\hline \\ [-2ex]
\label{t:observations}
\end{tabular}
\end{table*}
\end{small}
\begin{small}
\begin{table*}
\caption{Observed late-T and Y brown dwarf targets.}
\begin{tabular}{ l c c c c c c c c c c}
\hline\hline
Object ID & RA      & Dec.    & SpT & Distance & Ref. & $J$   & $H$   & Ref. & log($L_{\mathrm{bol}}/L_\odot$) & Mass \\
   & (J2000) & (J2000) & (NIR)  & (pc) & (dist.)  & (mag) & (mag) & (phot.)  &  & (M$_{\mathrm{Jup}}$) \\
\hline
W0146$+$4234$^{\mathrm{a}}$ & 01:46:56.67  & $+$42:34:10.1 & T9.0 & $10.6\pm1.5$ & (1) & $20.69\pm0.07$ & $20.30\pm0.12$ & (4) & $-6.59\pm0.25$ & $17\pm6$ \\ 
W0148$-$7202 & 01:48:07.30	& $-$72:02:59.0 & T9.5 & $11.0\pm0.4$	& (2) & $18.96\pm0.07$  & $19.22\pm0.04$ & (5) & $-6.05\pm0.27$ & $29\pm7$ \\
W0247$+$3725$^{\mathrm{b}}$  & 02:47:14.52  & +37:25:23.5 & T8.0 & $17.5\pm2.1$ & (3)	& $18.44\pm0.17$ & $18.24\pm0.19$ & (6) & $-5.84\pm0.17$ & $34\pm5$ \\
W0321$-$7347 & 03:21:20.91	& $-$73:47:58.8	& T8.0  & $26.0\pm3.1$ & (3)	& $19.13\pm0.11$  & $19.06\pm0.12$  & (6) & $-5.91\pm0.16$ & $32\pm5$ \\
W0335$+$4310 & 03:35:15.01	& +43:10:45.1	& T9.0  & $14.3\pm1.7$ & (1)	& $20.07\pm0.30$  & $19.60\pm0.26$  & (6) & $-6.50\pm0.24$ & $19\pm6$ \\
W0713$-$2917 & 07:13:22.55	& $-$29:17:51.9	& Y0.0  & $9.4\pm1.2$	& (1) & $19.64\pm0.15$  & $>19.30$ & (7) & $-6.19\pm0.31$ & $26\pm7$ \\
W0723$+$3403$^{\mathrm{b}}$  & 07:23:12.44  & +34:03:13.5	& T9.0 & $14.0\pm1.7$ &	(3) & $18.21\pm0.14$ & $>18.47$ & (6) & $-5.63\pm0.23$ & $40\pm8$ \\
W0734$-$7157 & 07:34:44.02	& $-$71:57:44.0	& Y0.0  & $13.6\pm1.2$ &	(2) & $20.41\pm0.27$  & $\cdots$ & (7) & $-6.16\pm0.28$ & $27\pm7$ \\
W1042$-$3842 & 10:42:45.23	& $-$38:42:38.3	& T8.5 & $15.4\pm0.8$ & (2) & $18.98\pm0.09$  & $19.08\pm0.11$ & (6) & $-6.18\pm0.17$ & $26\pm5$ \\
W1150$+$6302$^{\mathrm{b}}$  & 11:50:13.88  & +63:02:40.7 & T8.0 & $10.1\pm1.2$ & (3) & $17.72\pm0.08$ & $>18.01$ & (5) & $-6.03\pm0.16$ & $29\pm5$ \\
W1517$+$0529 & 15:17:21.13	& +05:29:29.3	& T8.0  & $22.2\pm2.7$ & (3)	& $18.54\pm0.05$ & $18.85\pm0.15$ & (6) & $-5.82\pm0.16$ & $35\pm5$ \\
W2220$-$3628 & 22:20:55.31	& $-$36:28:17.4	& Y0.0  & $7.4\pm0.9$ & (1) & $20.38\pm0.17$  & $20.81\pm0.30$ & (6) & $-6.60\pm0.31$ & $17\pm7$ \\
\hline \\ [-2.5ex]
\multicolumn{11}{l}{
  \begin{minipage}{.96\textwidth}
    \textbf{Notes.}
    $^{\mathrm{a}}$ combined photometry for the binary W0146$+$4234AB (see text).\\
    Magnitudes are on the MKO-NIR filter system except for $^{\mathrm{b}}$ on the 2MASS filter system. Bolometric luminosities and masses were derived in this work (see text). Masses were estimated adopting uniform age distributions in the range 2$-$8 Gyr. \\
    \textbf{References.}\\
    Distances:
    (1) \citet{Beichman2014};
    (2) \citet{Tinney2014};
    (3) \citet{Kirkpatrick2012}.\\
    Photometry:
    (4) \citet{Dupuy2015};
    (5) \citet{Kirkpatrick2011};
    (6) \citet{Mace2013};
    (7) \citet{Kirkpatrick2012}.
  \end{minipage}}
\label{t:sample}
\end{tabular}
\end{table*}
\end{small}

\section{Sample and observations}
\label{sample}

\subsection{Sample selection}

Our sample consists of 12 nearby sources ($d < 30$ pc) identified as isolated field objects in prior searches for brown dwarfs \citep{Kirkpatrick2011,Kirkpatrick2012,Mace2013} via the Wide-Field Infrared Survey Explorer (WISE; \citealp{Wright2010}). With reported spectral types of T8 or later and estimated masses $\lesssim$40 M$_\mathrm{Jup}$ (see Section \ref{mass_estimates}), these objects are some of the coolest and lowest-mass known brown dwarfs in the Solar neighbourhood. The observed targets are listed in Table~\ref{t:observations}. The full WISE designations are given in the table in the form WISE Jhhmmss.ss$\pm$ddmmss.s. We abbreviate source names to the short form Whhmm$\pm$ddmm hereafter. All targets were observed with the Wide Field Camera 3 (WFC3) on the Hubble Space Telescope (HST).

\subsection{HST/WFC3 imaging}
\label{imaging}

A common problem encountered in direct imaging searches for brown dwarfs is the high contamination rate observed in most photometric surveys. The broadband colours of brown dwarfs can be very similar to those of reddened stars in the near-infrared (NIR) and a large number of selected candidates turn out to be background interlopers. True substellar objects may however be distinguished from background stars through specific spectral characteristics. In particular, brown dwarfs have a strong water absorption feature observed at 1.35$-$1.45 $\mu$m \citep{McLean2003}. This H$_2$O spectral signature is found in all objects with spectral types M6 or later, with a deeper absorption observed in later-type objects. Spectra of reddened stars lack this water absorption feature and this attribute can therefore be used to identify brown dwarfs and differentiate them from reddened background stars (see \citealp{Allers2010}).

The WFC3/IR F139M filter on HST is sensitive to this water absorption band and, combined with the F127M filter, provides a unique probe into this substellar characteristic. Brown dwarfs are indeed expected to appear fainter in the F139M water band, while reddened stars will not exhibit any absorption. Comparing photometry in the two adjacent HST filters therefore provides a robust detection method for brown dwarfs, especially for late-type T and Y dwarfs that show particularly deep water absorption features. For this reason, targets in this study were observed with the F127M and F139M bands on WFC3, covering the 1.27 $\mu$m peak observed in late-type brown dwarfs and the H$_2$O absorption band found in substellar spectra, respectively.

Observations were taken between October 2012 and September 2013 with the IR channel of the WFC3 instrument on HST (Snapshot Program 12873, PI Biller). With a field of view of 123\arcsec$\times$136\arcsec, the 1024$\times$1024 pixel array of the IR channel has a plate scale of 0\farcs13 pixel$^{-1}$. At the estimated distances of our targets, this resolution allows us to probe companions down to separations in the range 0.96$-$3.38 AU. Images were taken in MULTIACCUM mode with two 349.233 s exposures along a two-point $\sim$0\farcs6 line dither pattern in each filter, providing a total exposure time of 698.466 s in both filters. All observations were performed so that the targets were roughly located at the centre of the field of view of the camera. 
The pipeline processed flat-field images were used as input in the MultiDrizzle software \citep{Fruchter2002} to correct for geometric distortion, perform cosmic ray rejection and combine all dithered images into a single and final master frame.

The original Snapshot proposal contained a total of 33 science targets, with one orbit per target, from which 13 were executed. From the 13 sources observed, one target (WISE J085716.25+560407.6) was missed due to wrong telescope pointing, providing us with a final sample of 12 objects.
\citet{Dupuy2015} discovered that the brown dwarf WISE J014656.66$+$423410.0 is a close near-equal mass binary with a projected separation of 0\farcs0875 (0.93 AU). However, the binary is not resolved in our HST observations due to the large pixel scale of the WFC3/IR camera and is thus treated as an unresolved single source in our multiplicity analysis.
A log of observations is given in Table~\ref{t:observations}.

We used the PhotUtils Python package to perform aperture photometry on the primaries in the F127M images. The PhotUtils \textit{CircularAperture} and \textit{aperture\_photometry} modules were called in Python to extract the photometry, adopting a 0\farcs4 aperture radius. Following the procedure in \citet{Schneider2015} we estimated the background level and its uncertainty by applying the same 0\farcs4 aperture to 1000 random star-free positions (determined via a 3-$\sigma$ clip) and took the mean and standard deviation of these measurements as the background and its uncertainty. Magnitudes were calculated on the Vega system using the photometric zero point provided in the HST/WFC3 webpages\footnote{\url{http://www.stsci.edu/hst/wfc3/phot_zp_lbn}} for the F127M filter (23.4932). The same method was applied to estimate the limiting background magnitude at the 5-$\sigma$ level in the F139M observations (using a zero point of 23.2093) since all of our targets were found to drop out entirely in the F139M observations. The obtained photometry for the science targets is presented in Table~\ref{t:observations}.

\begin{figure*}
    \centering
    \includegraphics[width=0.65\textwidth]{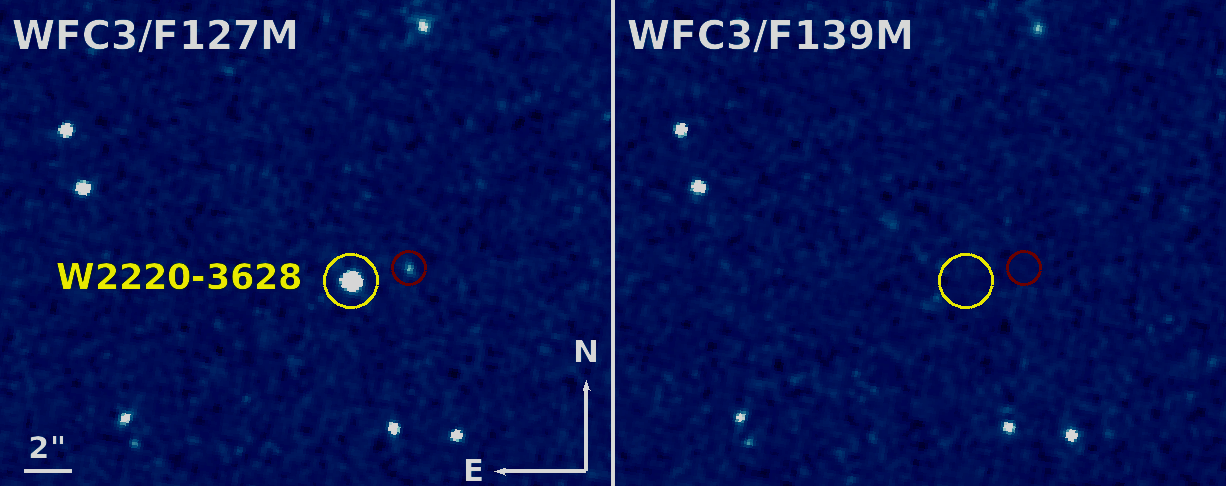}
    \caption{WFC3/IR F127M (left) and F139M (right) images of W2220$-$3628. The science target is in the yellow circle. The identified candidate companion is encircled in red. While most objects in the field of view have roughly similar fluxes in the two bandpasses, the Y dwarf and the selected candidate both exhibit a strong magnitude drop in the water-band F139M filter.}
\label{f:W2220}
\end{figure*}

\subsection{Primary mass estimates}
\label{mass_estimates}

Published information available for all targets was gathered from the literature in order to estimate the masses of our science targets. NIR photometry (Mauna Kea Observatory (MKO) or 2MASS filter system), spectral types and distances are summarised in Table~\ref{t:sample}.  \citet{Filippazzo2015} derived bolometric corrections for brown dwarfs at various ages and found a tighter correlation of spectral type with BC$_J$ rather than BC$_{K_s}$ for old mid to late-T dwarfs. This suggests that the former provides a more reliable correction when estimating luminosities for late-type field objects. We thus used $J$-band photometric data to estimate primary masses. Magnitudes on the MKO-NIR filter system were converted to 2MASS magnitudes using the relations derived in \citet{Stephens2004} based on spectral type.

Absolute magnitudes were computed for all targets adopting the distances in Table~\ref{t:sample}. We used parallax measurements from \citet{Beichman2014} or \citet{Tinney2014} when available and the ``adopted distances'' from table 8 in \citet{Kirkpatrick2012} otherwise. No errors are reported for the distance estimates from \citet{Kirkpatrick2012}, derived from the combination of $H$ and $W2$ spectrophotometric distances. The average relative standard deviation of single band distance estimates around the adopted mean for the full list of objects considered in that work is 11.5\%. We thus chose relative errors of $\pm$12\% on the final distances for these targets. A Monte-Carlo approach was implemented to account for the distance uncertainties. A total of $10^6$ distances were drawn from a Gaussian distribution centred on the distance value and with a standard deviation $\sigma$ set to the error on the distance. Similarly, apparent magnitudes were selected from a Gaussian centred on the measured apparent magnitude values in Table~\ref{t:sample} and with standard deviations set to the errors on the measured apparent magnitudes. An absolute magnitude was obtained for every apparent magnitude and distance generated and the final absolute $J$ magnitude was set to the mean of the output distribution, with an error set to the standard deviation of the output distribution.
Luminosities were then obtained by applying bolometric corrections to the absolute magnitudes. We used the spectral types in Table~\ref{t:sample} to estimate bolometric corrections BC$_J$ and associated errors from the relations in \citet{Filippazzo2015} for field objects, assuming errors in spectral type of $\pm$0.5 subtypes. The relations were extrapolated for spectral types later than T9. The extracted BC$_J$ were used to compute bolometric luminosities $L_\mathrm{bol}$ and their uncertainties, using the same approach to propagate the uncertainties. The final bolometric luminosity for each target is given in Table~\ref{t:sample}.

Precise ages for our targets are not known and are particularly difficult to obtain. We assumed that a similar distribution in age as in the solar neighbourhood \citep{Caloi1999} applied to our sample and adopted typical estimated ages of $5\pm3$ Gyr. We adopted a uniform distribution of ages within this range. For each target, we simulated an input of $10^6$ Gaussian-distributed luminosities, using the value and associated error calculated previously as the mean and standard deviation of the Gaussian, and drew an age value from a uniform distribution between 2 and 8 Gyr for each luminosity. We then interpolated the drawn luminosity and age values into the Lyon/COND evolutionary models for brown dwarfs \citep{Baraffe2003} to infer a corresponding mass. The final mass was taken to be the mean of the output distribution and the associated mass uncertainty was set to the standard deviation of the output distribution. The estimated bolometric luminosities and primary masses for all targets in our sample are shown in Table~\ref{t:sample}. All targets were found to have estimated masses $\lesssim$40 M$_\mathrm{Jup}$ for the adopted ages of 2$-$8 Gyr, making our sample the largest subset of very late-type and exclusively low-mass brown dwarfs studied as part of a multiplicity search.
As the W0146$+$4234 binary system \citep{Dupuy2015} is unresolved in our images and is thus treated as a single source in our analysis, we used the combined photometry of the binary components to estimate the mass of an unresolved object with that apparent magnitude.

\section{Search for candidate companions and selection techniques}
\label{search}

\subsection{The water-band detection method}

Images in the WFC3/IR F127M and F139M filters from our core sample were visually inspected to search for sources other than the science targets exhibiting a significant magnitude drop in the latter bandpass. All targets in our sample were found to drop out entirely in the F139M water-band filter as a result of the deep water absorption feature robustly observed at 1.4 $\mu$m in substellar spectra, which is particularly strong for late spectral types (Figure~\ref{f:W2220}). Assuming a similar or later spectral type for possible companions, potential candidates are expected to drop by the same amount as the primaries and to also be undetected in the F139M band.

\subsubsection{Candidate companion around W2220$-$3628}

\begin{figure}
    \centering
    \includegraphics[width=0.39\textwidth]{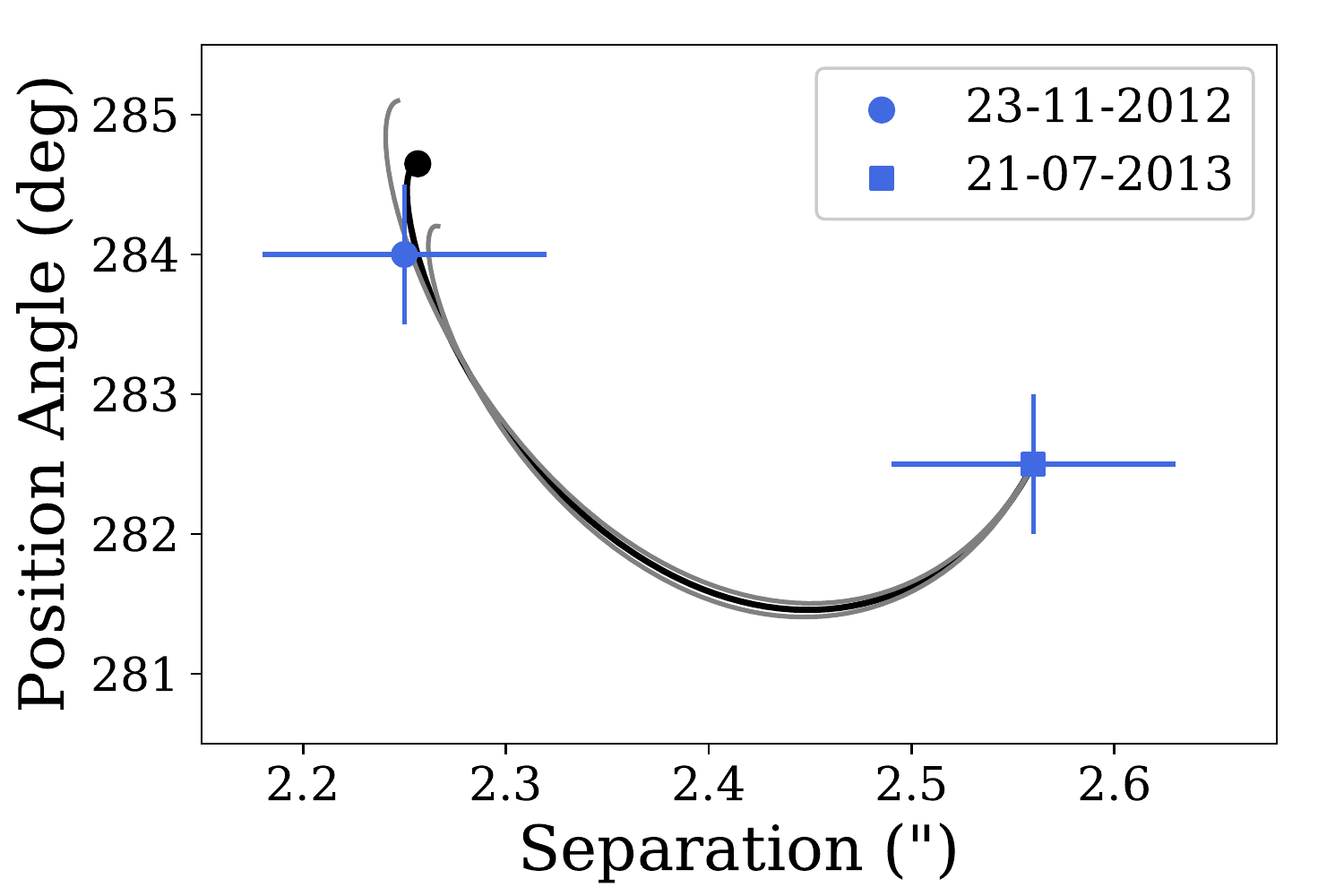}
    \caption{Common proper motion analysis of W2220$-$3628 and the selected candidate. The black solid line represents the motion of a background object relative to the primary, computed using the proper motion and parallax measurements of the science target from \citet{Beichman2014}. The grey lines show the same motion using the 1-$\sigma$ errors on the proper motion and parallax values. The blue symbols mark the measured positions of the candidate relative to the science target in our images (square) and in the past HST epoch (circle). The black circle indicates the expected position of a background source at the date of the first epoch. The relative motion of the candidate between the two epochs is consistent with a background object.}
\label{f:PM_plot}
\end{figure}

Only one candidate companion was identified in our sample, found at 2\farcs56 $\pm$ 0\farcs07 around the Y0 brown dwarf W2220$-$3628. The candidate was detected in each dithered frame in the F127M band but was not retrieved in the F139M images. Figure~\ref{f:W2220} shows the primary and candidate companion in the final F127M and F139M images, highlighting the significant magnitude drop of both objects in the latter bandpass (right panel). To be considered bonafide companions, candidates must have similar red colours to their primary in addition to a robust sign of water absorption at 1.4 $\mu$m. We found no blue source at the position of the candidate in broadband surveys (WISE, Spitzer). A true companion must also possess common proper motion with the primary. Archival HST images of W2220$-$3628 in the WFC3/IR F125W filter (GO Program 12970, PI Cushing) were compared to our images, providing an 8-month baseline between epochs. Figure~\ref{f:PM_plot} shows the position of the candidate relative to W2220$-$3628 in our program (blue square) and in the past HST epoch (blue circle). The black circle shows the expected position of a background object in the first epoch images. With the high proper motion of our science target ($\mu_{\alpha}=+283\pm13$ mas yr$^{-1}$ and $\mu_{\delta}=-97\pm17$ mas yr$^{-1}$; \citealp{Beichman2014}), astrometric measurements in the two epochs proved the candidate to be lacking common proper motion with the primary and to be consistent with a background object.

\begin{small}
\begin{table*}
\caption{HST WFC3/IR photometry and colours of W2220$-$3628 and the background source.}
\begin{tabular}{p{2.3cm} c c c c c c }
\hline\hline
Object & F127M & F139M & F105W & F125W & F127M$-$F139M & F105W$-$F125W \\
 & (mag) & (mag) & (mag) & (mag) & (mag) & (mag) \\
\hline
W2220$-$3628 & $19.84\pm0.03$ & $>$ $26.41\pm0.03$ & $21.64\pm0.03$ & $21.04\pm0.03$ & $<$ $-6.57\pm0.04$ & $0.60\pm0.04$ \\
Background source & $23.29\pm0.03$ & $>$ $26.41\pm0.03$ & $24.18\pm0.03$ & $23.63\pm0.03$ & $<$ $-3.12\pm0.04$ & $0.55\pm0.04$ \\
\hline
\label{t:photometry}
\end{tabular}
\end{table*}
\end{small}

\subsubsection{Nature of the background contaminant}

Possible contaminants with the water-band detection method may be background brown dwarfs or mid-M stars showing water absorption at 1.4 $\mu$m, or faint galaxies undetected in the F139M band with an emission line covered by the F127M filter. While the past HST program used to check for common proper motion with the primary only contained one set of F125W images providing a sufficiently large time baseline to confirm or refute common proper motion for the candidate, additional observations in the F105W and F125W filters were also acquired as part of the same program in June 2013 (one month before observations from our program). We therefore used those images to investigate the photometry and colours of the identified background source. We used the same method as that described in Section~\ref{imaging} to perform aperture photometry on the primary and selected candidate in the F127M, F105W and F125W images, and estimate the limiting background magnitude in the F139M observations. Magnitudes were calculated on the Vega system using the appropriate photometric zero points provided in the HST/WFC3 webpages\footnote{\url{http://www.stsci.edu/hst/wfc3/phot_zp_lbn}} for each of the considered filters. The obtained photometry for the science target and the background source is presented in Table~\ref{t:photometry}. We note that our F105W and F125W photometry for W2220$-$3628 is in good agreement with the values reported in \citet{Schneider2015} for the same HST images ($21.638\pm0.027$ mag and $20.997\pm0.005$ mag, respectively).

\begin{figure}
    \centering
    \includegraphics[width=0.45\textwidth]{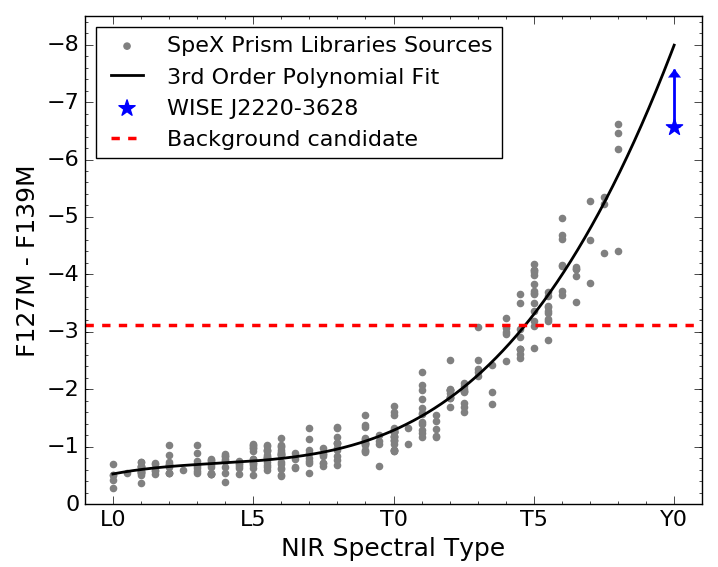}
    \caption{Synthetic F127M$-$F139M colours of L and T dwarfs in the SpeX prism spectral libraries (grey circles) and third order polynomial fit to the data (black line). The minimum colours estimated from the measured photometry of W2220$-$3628 and the background source are shown by the blue symbol and red line, respectively.}
\label{f:water_abs}
\end{figure}

\begin{small}
\begin{table*}
\caption{Summary of the calculation of the number $N_\mathrm{exp}$ of background brown dwarf contaminants expected to be found in our survey for spectral types between T4 and Y0.5. The values of $M_J$ for each spectral type were calculated using the relations in $M_J$ versus spectral type derived by \citet{Dupuy2012}. Distances correspond to the range over which an object of given absolute $J$ magnitude is detectable in the F127M images (upper limit) but not in the F139M observations (lower limit), assuming the average detection limits of our survey. See text for references of space density values.}
\begin{tabular}{ l c c c c c }
\hline\hline
SpT & $M_J$ & $d_\mathrm{min}$ & $d_\mathrm{max}$ & Space Density & $N_\mathrm{exp}$ \\
 & (mag) & (pc) & (pc)& ($\times$ 10$^{-3}$ pc$^{-3}$) & \\
\hline
T4$-$T4.5 & 14.87 & 225 & 600 & $0.47\pm0.27$ & 0.017$-$0.063 \\
T5$-$T5.5 & 14.95 & 139 & 552 & $0.47\pm0.27$ & 0.014$-$0.051 \\
T6$-$T6.5 & 15.50 & 99 & 485  & $0.50\pm0.28$ & 0.010$-$0.037 \\
T7$-$T7.5 & 16.11 & 49 & 381  & $0.73\pm0.42$ & 0.007$-$0.027 \\
T8$-$T8.5 & 17.55 & 18 & 204  & $2.63\pm0.58$ & 0.007$-$0.011 \\
T9$-$T9.5 & 18.41 & 6 & 137   & 1.6           & 0.002 \\
Y0$-$Y0.5 & 20.53 & 1 & 52    & 1.9           & 0.0001 \\
\hline
\label{t:background_prob}
\end{tabular}
\end{table*}
\end{small}


The F105W$-$F125W colour of the background source was found to be comparable to that of the science target, suggesting that it could be a late-type brown dwarf. Photometry in the F127M and F139M bandpasses showed the candidate to be dropping by a minimum of $3.12\pm0.04$ mag between the two filters. In Figure~\ref{f:water_abs} we computed synthetic F127M$-$F139M colours for all L and T dwarfs in the SpeX prism spectral libraries\footnote{\url{http://www.browndwarfs.org/spexprism}} (grey symbols). We excluded targets with no NIR spectral classification as well as spectra flagged as low-quality data. Flux ratios between the F127M and F139M bandpasses were computed for all available sources by taking into account the transmission value of the two filters at each wavelength and integrating the spectra over the relevant spectral regions. A third order polynomial was fit to the data (black line in Figure~\ref{f:water_abs}) yielding:
\begin{equation}
\begin{aligned}
\mathrm{F127M} - &\mathrm{F139M} = 3.719 - \left(9.127 \times 10^{-1} \times \mathrm{SpT}\right)\\
& + \left(6.456 \times 10^{-2} \times \mathrm{SpT}^2\right) - \left(1.572 \times 10^{-3} \times \mathrm{SpT}^3\right),
\end{aligned}
\label{eq:colour_fit}
\end{equation}
where SpT(L0) = 10 and SpT(T8) = 28. The fit was derived for spectral types between L0 and T8 as the SpeX Prism Spectral Libraries do not contain $>$T8 spectra. The derived relation strongly reflects the strengthened H$_2$O absorption band along the L and T substellar sequences. \citet{Burgasser2010} compared SpeX and literature classifications for 189 spectra of 178 L and T sources and found standard deviations between classifications of 1.1 subtypes for L dwarfs and 0.5 subtypes for T dwarfs. We thus assumed uncertainties of 1 and 0.5 subtypes for L and T dwarfs, respectively, when deriving the polynomial fit. The mean scatter in the relation in Eq.~\ref{eq:colour_fit} is 0.3 mag. We note that there are fewer $>$T5 objects relative to earlier spectral types and that these objects show a significantly larger scatter in their synthetic colours. Our measured F127M$-$F139M lower limit for W2220$-$3628 (blue star) appears to be consistent with the extrapolation of the fit at spectral types later than T8. The lower limit for the F127M$-$F139M photometry of the background source is shown by the red line in Figure~\ref{f:water_abs}. The observed drop in the water-band filter suggests a spectral type of $\sim$mid-T or later for this object to be of substellar nature.

To quantify the likelihood that the background source is a brown dwarf, we calculated the probability of finding a background brown dwarf false positive in our survey, for spectral types varying from T4 to Y0.5. We used published brown dwarf space density values to estimate the probability of observing one such background brown dwarf for various spectral type bins. Space densities were taken from \citet{Burningham2013} for the T6 to T8.5 spectral types and from \citet{Kirkpatrick2012} for $\geqslant$T9 brown dwarfs. We used the value for the T3$-$T5.5 space density from \citet{Metchev2008} for T4 to T5.5 objects, assuming a homogeneous distribution of densities across that spectral type range. We used the relation from \citet{Dupuy2012} between spectral type and absolute magnitude to infer expected 2MASS $J$ absolute magnitudes for each spectral type bin. We then applied a filter transform to convert the obtained $J$-band magnitudes to absolute F127M and F139M magnitudes for each spectral type, based on a similar method to the one used to compute synthetic F127M$-$F139M colours. False positives are background sources detected in our F127M data but dropping out in the F139M observations. As a result, we estimated, based on the derived HST absolute magnitudes, the distance ranges in which brown dwarfs of various spectral types are detectable in the former images but not in the latter. We used our average 5-$\sigma$ detection limits in both sets of observations (25.2 mag in F127M and 25.5 mag in F139M, respectively) to infer the maximum distance at which an object of given absolute magnitude can be detected in the F127M images, and compute the minimum distance required for the object to be undetected in the F139M data. Space densities, absolute 2MASS $J$ magnitudes and the estimated minimum and maximum distances are listed in Table~\ref{t:background_prob}.

For each spectral type bin, the expected number of contaminant background brown dwarfs is found by considering the volume of a thick spherical shell located within the distance limits in Table~\ref{t:background_prob}. We then multiplied that volume by the corresponding space density and the fraction of the sky area covered by our program (12 images of 123\arcsec$\times$136\arcsec over 4$\pi$ sr) to obtain an average number of background brown dwarfs expected to be found in our survey for each spectral type bin. The obtained values are listed in the $N_\mathrm{exp}$ column in Table~\ref{t:background_prob}. The expected number of T4$-$Y0.5 background brown dwarfs in our program detected in the F127M images but not retrieved in the F139M data was found to be in the range 0.057$-$0.191, given by the sum of the values in Table~\ref{t:background_prob}. The identified source is thus rather unlikely to be a background mid-T$-$Y brown dwarf. We exclude later spectral types as a later-type Y object would need to be very close to be detected ($<$ 50 pc; see Table~\ref{t:background_prob}) and would likely show high proper motion over the 8-month baseline between epochs, and would thus not be consistent with a background source. As the F139M$-$F127M colour of this object ruled out the possibility of it being an earlier-type star or brown dwarf, we conclude that it is most likely extra-galactic, although there is a small probability of the contaminant being a background mid-T to Y dwarf.
While a large number of extra-galactic reference sources are found in our deep HST observations, only one such object was identified in the total sky area covered by our program. The contamination rate from such sources for the water-band detection method is therefore low and does not present a major concern regarding the reliability of our selection technique for substellar companions.

\subsection{PSF subtraction}
\label{PSF sub}

No well-resolved binary pairs were identified in our F127M images. Point spread function (PSF) subtraction was attempted to search for more closely-separated systems with blended PSFs, at separations $<$0\farcs5. The WFC3/IR PSF is severely undersampled by the 0\farcs13 detector pixel. To mitigate the effect of undersampling, we constructed higher-resolution master frames using the individual F127M dithered frames to recover information lost to undersampling. The pipeline processed flat-field images were used as input in the MultiDrizzle software \citep{Fruchter2002} and recombined into a single output frame with a 0\farcs065 pixel scale, improving the spatial resolution of the final images by a factor of 2.

Tiny Tim models \citep{Krist1995} are generated from pre-launch simulations and still show large discrepancies when applied to on-orbit WFC3/IR data (see \citealp{Biretta2014,Garcia2015}). We therefore generated empirical PSFs from the data and did not attempt synthetic PSF fitting. As the primaries all have roughly similar spectral types (within two subtypes) and are all located near the centre of the detector chip, variations in the PSF due to spatial or spectral variations are expected to be negligible. For each target, we performed PSF subtraction using the PSFs of all other targets in the sample (excluding the known tight binary W0146$+$4234AB) to create an empirical PSF model. We extracted sub-images of 40$\times$40 pixels (2\farcs6$\times$2\farcs6) centred on the primaries. Each sub-image was background-subtracted and normalised to a peak pixel value of 1. To align two individual PSFs, we performed a progressive grid search to identify the best position. One PSF was moved on a coordinate grid of resolution 0.05 pixel and re-sampled onto the original image grid at each position via a cubic interpolation. The optimal position was taken to be the one that minimised the root mean square difference of the two sub-images. All observed PSFs used to create an empirical PSF were aligned via this method and median-combined to generate a final empirical PSF model for each target in our observed sample.

The observed and empirical PSFs were then aligned using the same fitting routine, after scaling the peak of the empirical PSF to that of the target. PSF subtraction was performed at the position that minimised the root mean square difference of the observed PSF and re-binned empirical PSF. Varying amounts of residual flux were found in the resulting images, with relative intensities ranging from 0.01 to 0.25 of the maximum flux of the data. The amount of residuals was generally found to be correlated to the brightness of the primary relative to the rest of the sample. Observed residuals in the final PSF-subtracted images can be due to either the presence of a secondary source or to discrepancies in the shapes of the individual PSFs of our targets. In the case of an unresolved binary, we expect to find similar residuals when using the individual observed PSFs of our targets as separate PSF models. On the other hand, residuals due to large disparities between observed PSFs should vary based on the PSF used to perform PSF subtraction. To check the origin of the observed residuals, we also ran the same PSF-subtraction routine for each target using the PSF of every other object in the sample as a single model PSF. We found that targets of similar magnitude generally provided better fits. We did not find any convincing sign of close-in companions in the PSF-subtracted images around any of the targets. The residuals seen with the median empirical PSF models were not consistently recovered with the single PSF models and were due to disparities between the PSFs of objects with larger magnitude differences. Our PSF subtraction technique did not allow us to recover the tight binary W0146$+$4234AB, which showed large fluctuations in the residual flux based on the model PSF used for the subtraction. This result is not surprising given the 87.5 mas separation of the binary and the large 130 mas pixel scale of the WFC3/IR channel.

\begin{figure}
    \centering
    \includegraphics[width=0.47\textwidth]{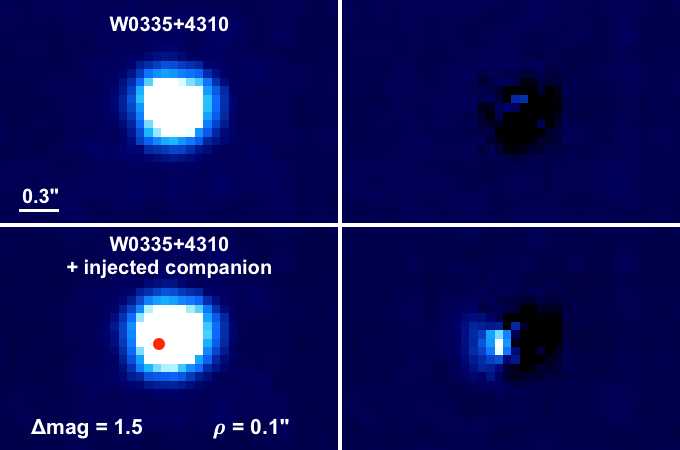}
    \caption{\textbf{Top:} W0335$+$4310 before (left) and after (right) PSF subtraction. PSF subtraction was performed using an empirical PSF constructed with the observed PSFs of all other targets in our sample. \textbf{Bottom:} same as top panels with a fake 0\farcs1 companion with $\Delta$mag = 1.5 injected around the science target. The red dot indicates the position at which the companion was injected. The same PSF subtraction routine was applied to fit the synthetic binary. A clear localised residual flux was found at the position angle of the simulated companion after single PSF subtraction.}
    \label{f:PSFsub}
\end{figure}

\begin{small}
\begin{table*}
\caption{Measured F127M contrasts and minimum detectable mass ratios.}
\begin{tabular}{ l c c c c c c c c c c c c }
\hline\hline
Object ID &  $\Delta$mag & $q$ & $\Delta$mag & $q$ & $\Delta$mag & $q$  & $\Delta$mag & $q$ & $\Delta$mag & $q$ & $\Delta$mag & $q$  \\
        &  (0\farcs2)  & (0\farcs2) & (0\farcs3) & (0\farcs3) & (0\farcs5) & (0\farcs5) & (1\farcs0) & (1\farcs0) & (2\farcs0) & (2\farcs0) & (5\farcs0) & (5\farcs0)\\  
\hline
W0146$+$4234    & 0.58 & 0.68 & 1.80 & 0.53 & 3.93 & 0.35 & 3.85 & 0.36 & 3.86 & 0.36 & 3.95 & 0.35 \\
W0148$-$7202    & 0.84 & 0.65 & 2.60 & 0.37 & 5.05 & 0.22 & 5.36 & 0.21 & 5.52 & 0.20 & 5.61 & 0.20 \\
W0247$+$3725	& 0.63 & 0.85 & 2.11 & 0.58 & 4.81 & 0.29 & 5.79 & 0.24 & 6.13 & 0.22 & 6.22 & 0.22 \\
W0321$-$7347	& 1.30 & 0.81 & 2.69 & 0.57 & 5.25 & 0.27 & 5.78 & 0.24 & 5.85 & 0.24 & 6.01 & 0.23 \\
W0335$+$4310	& 0.79 & 0.83 & 2.46 & 0.57 & 4.64 & 0.35 & 5.08 & 0.33 & 5.12 & 0.33 & 4.89 & 0.34 \\
W0713$-$2917	& 0.73 & 0.55 & 2.09 & 0.36 & 4.11 & 0.24 & 4.59 & 0.23 & 4.38 & 0.23 & 4.42 & 0.23 \\
W0723$+$3403	& 0.73 & 0.75 & 2.16 & 0.51 & 4.90 & 0.23 & 5.68 & 0.20 & 6.01 & 0.18 & 6.08 & 0.18 \\
W0734$-$7157	& 0.87 & 0.56 & 2.26 & 0.35 & 4.06 & 0.24 & 4.33 & 0.23 & 4.28 & 0.23 & 4.24 & 0.23 \\
W1042$-$3842	& 0.54 & 0.78 & 2.32 & 0.48 & 5.04 & 0.31 & 5.18 & 0.30 & 5.38 & 0.29 & 5.37 & 0.29 \\
W1150$+$6302    & 0.62 & 0.80 & 2.35 & 0.51 & 5.03 & 0.29 & 5.88 & 0.25 & 6.35 & 0.23 & 6.43 & 0.23 \\
W1517$+$0529    & 1.11 & 0.74 & 2.88 & 0.49 & 5.05 & 0.28 & 5.45 & 0.26 & 5.84 & 0.24 & 5.83 & 0.24 \\
W2220$-$3628    & 1.46 & 0.51 & 3.02 & 0.30 & 4.50 & 0.23 & 4.69 & 0.22 & 4.48 & 0.23 & 4.64 & 0.23 \\
\hline \\ [-2ex]
\label{t:limits}
\end{tabular}
\end{table*}
\end{small}

To test the ability of our PSF-subtraction technique to recover close-in candidates, we injected fake companions around the primaries and performed the same PSF subtractions. We used scaled-down versions of the PSFs of other primaries in the sample to simulate companions with magnitude differences in the range 0$-$5 mag and separations from 0 to 10 pixels (0\arcsec to 0\farcs65) from the centre of the primary at randomly chosen position angles. For each injected companion we then repeated the same PSF subtractions on the synthetic binaries, using a median and single empirical PSFs, and visually inspected the obtained images. We found that at separations $\geqslant$0\farcs25, observations were background-limited rather than diffraction limited. Injected companions were retrieved with S/N $\geqslant$ 5 in all PSF-subtracted images provided that the magnitude of the fake companion was within the detection limits of the image (see Section~\ref{limits}). At separations in the range 0\farcs10$-$0\farcs25, our PSF-subtraction technique was able to recover companions with $\Delta$mag down to $\sim$1$-$2, with a localised residual flux found at the position angle of the fake companion. Injected companions were consistently retrieved in the image obtained using a median empirical PSF as well as in over two thirds of the images obtained with single PSF models. An example of the typical results achieved is shown in Figure~\ref{f:PSFsub}. The top panels show the observed PSF of W0335$+$4310 before and after PSF subtraction, using a median empirical PSF. In the bottom panel, a fake companion with a magnitude difference of 1.5 was injected at a separation of 0\farcs1. After running the same PSF subtraction routine on the synthetic binary system, the simulated companion was clearly retrieved. Finally, at separations $<$0\farcs1 the amount and position of the residual flux after PSF subtraction were found to vary significantly between the final images for a given target and injected companion. The discrepancies observed were comparable to the disparities obtained after applying PSF subtraction to the original data, with no injected companion, or to the unresolved W0146$+$4234 binary system. These residuals could therefore not be interpreted as an unambiguous sign of binarity. We note that simulated companions similar to the W0146$+$4234 secondary component were never detected. Comparable results were achieved around all primaries in the sample.

We conclude that our PSF subtraction method would have allowed us to detect companions at separations from 0\farcs25 with magnitudes within our detection limits, as well as to uncover closer companions down to 0\farcs1 with $\Delta$mag $\lesssim$ 1$-$2. These results are consistent with the contrast curves derived in Section~\ref{limits}. The lack of obvious signs of companions in our PSF-subtracted images strongly indicates that our sample did not contain any bonafide companion in these separation and magnitude ranges. As PSF-subtraction for closer-in binaries showed significant discrepancies depending on the PSF used as model, our technique could not confidently rule out the presence of $<$0\farcs1 companions in our sample, such as the 0\farcs0875 W0146$+$4234 binary which was not recovered with our PSF-subtraction method.

\begin{figure*}
\addtocounter{figure}{-1}
    \centering
    \begin{subfigure}{0.49\textwidth}
        \includegraphics[width=\textwidth]{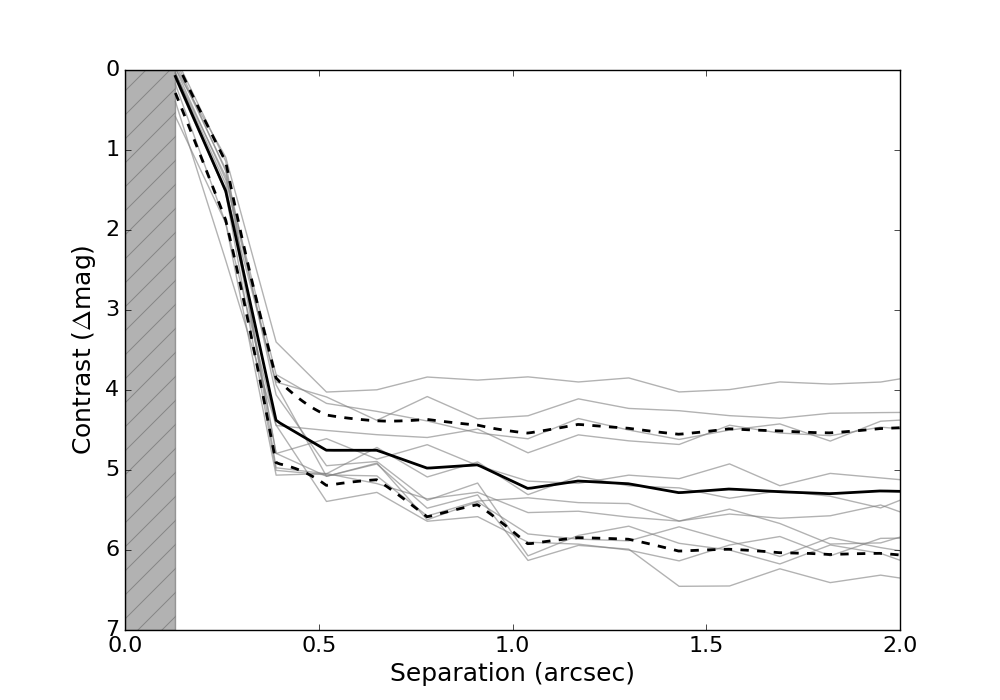}
    \end{subfigure}
    \begin{subfigure}{0.49\textwidth}
        \includegraphics[width=\textwidth]{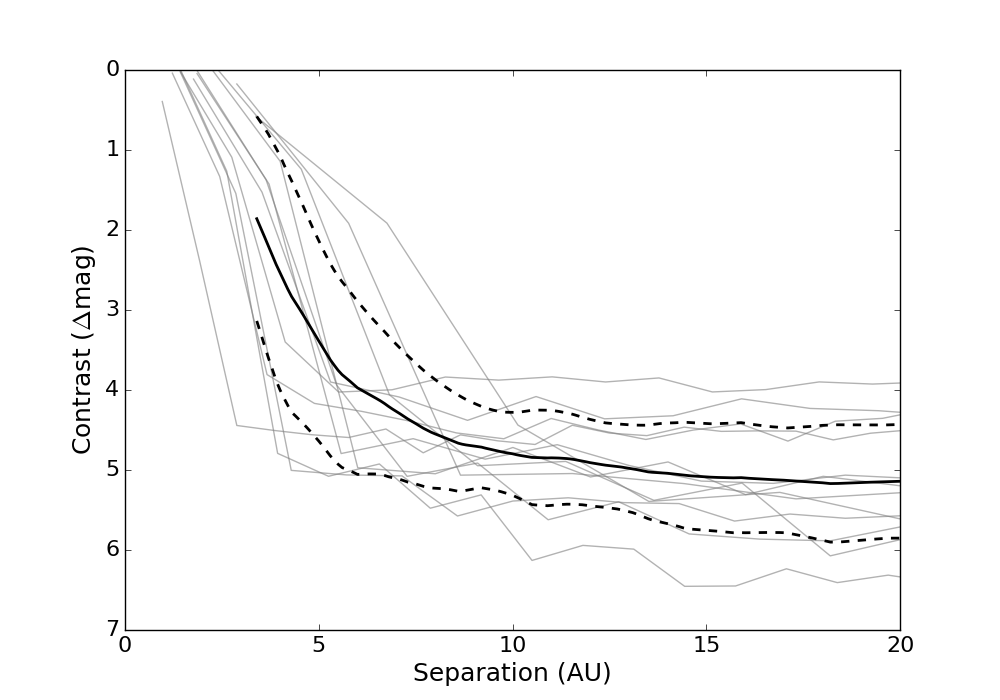}
    \end{subfigure}
    \caption{Magnitude difference limits at the 5-$\sigma$ level around our 12 targets in the WFC3/IR F127M observations (grey lines) as a function of angular (left) and physical (right) projected separation. The solid black lines show the mean detection levels for the sample and the 1-$\sigma$ standard deviations around the mean (dotted lines). The shaded region on the left panel represents the pixel scale of the WFC3/IR camera. On the right panel, the mean was calculated starting at the smallest physical separation resolved for all targets.}
    \label{f:mag_limits_SNAP}
\end{figure*}

\begin{figure*}
    \centering
    \includegraphics[width=0.72\textwidth]{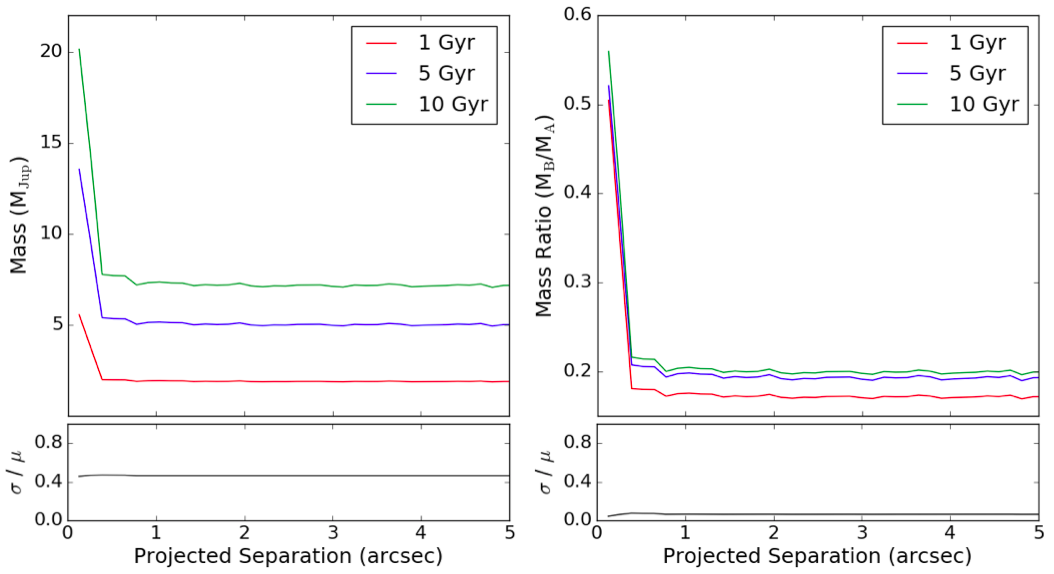}
    \caption{Detection limits reached around our target W0148$-$7202 in terms of minimum detectable masses (left) and mass ratios (right). The magnitude limits for the target (Figure~\ref{f:mag_limits_SNAP}) were converted to masses using the AMES-COND evolutionary models from \citet{Allard2001} at discrete ages of 1, 5 and 10 Gyr. The primary mass was calculated at each age considered when converting the mass limits into mass ratios. The bottom panels show the relative scatter in the masses and mass ratios between adopted ages of 1, 5 and 10 Gyr. The mass limit curves have a mean relative scatter of 0.49. The same curves in mass ratio space have a mean relative scatter of 0.11, therefore significantly reducing the uncertainty introduced by adopting a discrete age of 5 Gyr.}
    \label{f:limits_w_ages}
\end{figure*}

\section{Survey sensitivity limits}
\label{limits}

\subsection{Achieved contrasts}
\label{mag_contrasts}

For each object in the sample, sensitivity limits were computed to establish the full range of detectable companions covered by the survey (Table~\ref{t:limits}). Detection limits were determined from the final F127M images described in Section~\ref{imaging}. The 5-$\sigma$ noise curves were calculated as a function of radius by computing the standard deviation in circular annuli with 1 pixel-widths (0\farcs13), centred on the targets. Limits were calculated from a radius of 1 pixel up to 20 pixels before the closest edge of the image. Noise levels were then converted into magnitude contrasts by dividing the obtained noise levels at each separation by the peak pixel value of the targets and converting the obtained flux ratios into magnitude differences. The achieved magnitude contrasts are presented in Figure~\ref{f:mag_limits_SNAP}. We are complete down to $\Delta$mag $\sim$ 2 at 0\farcs3, and down to $\Delta$mag $\sim$ 4 from angular separations of 0\farcs5 and physical separations of 10 AU.

The results achieved for the injection of simulated companions in Section~\ref{PSF sub} are consistent with our measured contrast curves. Fake companions simulated as scaled-down versions of our primaries with contrasts down to our achieved limits were consistently retrieved with S/N $\geqslant$ 5. We therefore conclude that our measured contrast curves provide reliable estimates for the limits of detectable companions.

\begin{figure*}
\addtocounter{figure}{-1}
    \centering   
    \begin{subfigure}{0.48\textwidth}
        \includegraphics[width=\textwidth]{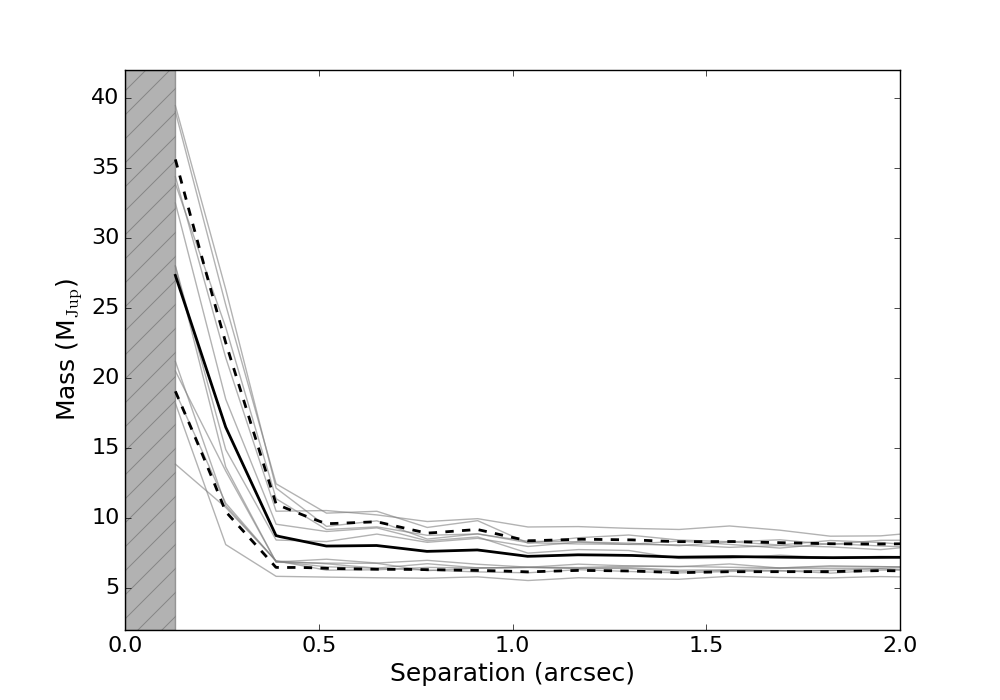}
    \end{subfigure}
    \begin{subfigure}{0.48\textwidth}
        \includegraphics[width=\textwidth]{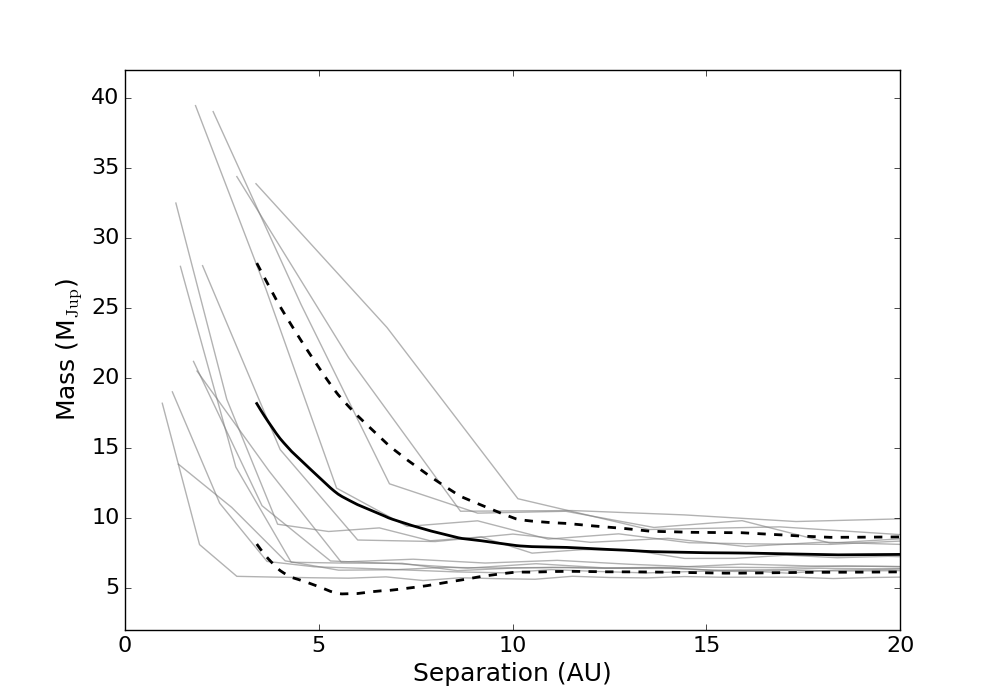}
    \end{subfigure}
    \begin{subfigure}{0.48\textwidth}
        \includegraphics[width=\textwidth]{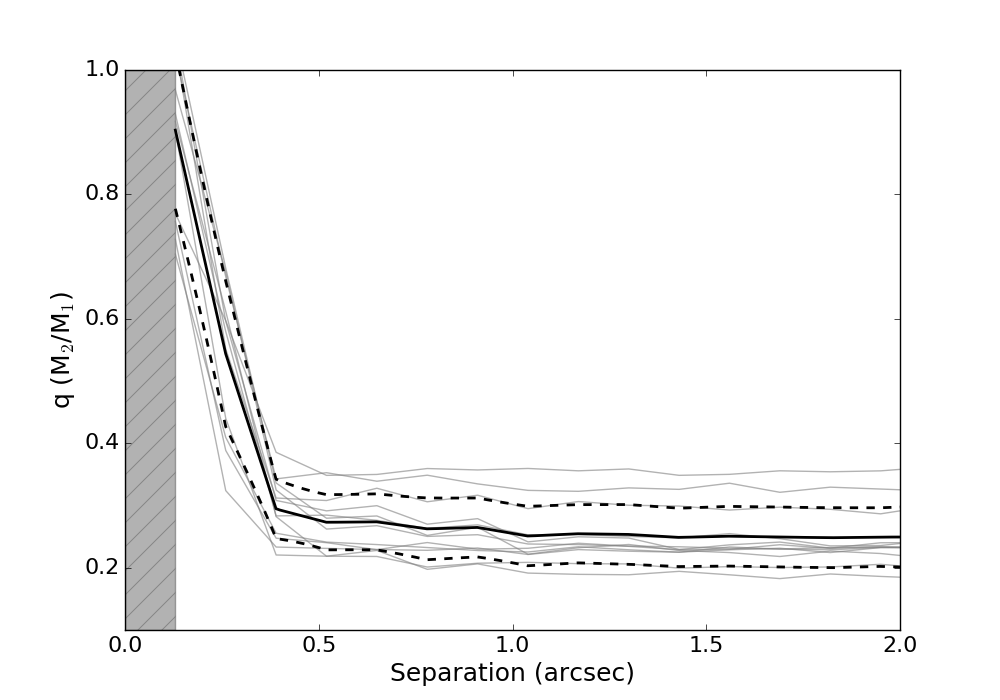}
    \end{subfigure}
    \begin{subfigure}{0.48\textwidth}
        \includegraphics[width=\textwidth]{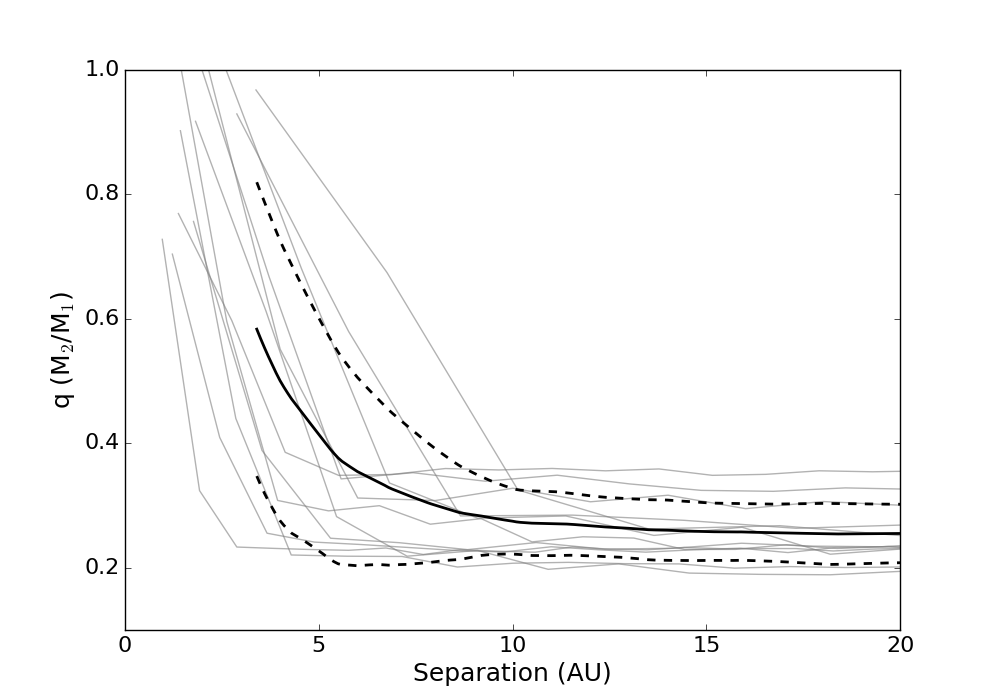}
    \end{subfigure}
    \caption{\textbf{Top:} Minimum masses detectable around our 12 targets at the 5-$\sigma$ level as a function of angular (left) and physical (right) projected separation. Magnitude contrasts were converted to masses using the AMES-COND evolutionary models from \citet{Allard2001} at an adopted age of 5 Gyr. The solid black lines show the mean detection levels for the sample and the 1-$\sigma$ standard deviations around the mean (dotted lines). The shaded region on the left panel represents the pixel scale of the WFC3/IR camera. On the right panel, the mean was calculated starting at the smallest physical separation resolved for all targets. \textbf{Bottom:} Same as top panel for minimum detectable mass ratios. Mass limits were converted to mass ratios using the masses calculated in this work and listed in Table~\ref{t:sample}.}
    \label{f:mass_limits_SNAP}
\end{figure*}

\subsection{Limits on minimum detectable companion masses}
\label{mass_ratios}

The magnitude contrasts $\Delta$mag were converted into apparent magnitudes using the measured F127M photometry of our science targets (Table~\ref{t:observations}). We then converted the apparent magnitude limits into corresponding absolute magnitudes using the parallax and spectrophotometric distances from Table~\ref{t:sample}.

The magnitude$-$mass relationship for brown dwarfs shows a strong age degeneracy. As shown in Figure~\ref{f:limits_w_ages}, the age chosen to convert our detection limits into minimum detectable masses highly affects the obtained results.
For our 12 targets, we found an average scatter in the inferred mass limits of 3.7 M$_\mathrm{Jup}$ when considering discrete ages of 1 Gyr, 5 Gyr and 10 Gyr. This corresponds to a very large mean relative scatter of 0.50 for the low mass limits reached in our survey. Working with mass ratios, on the other hand, significantly reduces the scatter between various adopted ages. Converting the same mass limits into mass ratio curves, we found a mean relative scatter in mass ratio of only 0.12, therefore crucially reducing the scatter seen in the mass domain for the same discrete ages. Figure~\ref{f:limits_w_ages} illustrates this effect for one target from the survey, showing the notably smaller relative scatter obtained in the mass ratio curves (bottom panels). We thus consider mass ratio space rather than companion mass throughout this work and adopted a median age of 5 Gyr to obtain sensitivity limits at the 5-$\sigma$ detection level.

We interpolated the absolute magnitude curves into the AMES-Cond evolutionary models \citet{Allard2001} to infer corresponding mass and mass ratio limits at an adopted age of 5 Gyr for all survey objects. The AMES-Cond luminosity isochrones are similar to those from the Lyon/COND models \citep{Baraffe2003} used to estimate primary masses in Section \ref{mass_estimates}. As the AMES-Cond models provide photometric data specific to the HST/WFC3 filters, not available in the Lyon/COND models, the former are better suited to convert our magnitude limits into masses. The minimum detectable masses and mass ratios around each target are presented in Figure~\ref{f:mass_limits_SNAP}. We are sensitive to systems with secondary masses $>$ 5$-$10 M$_\mathrm{Jup}$ beyond 0\farcs5 assuming ages of 5 Gyr for our targets. In comparison, using ages of 1 Gyr and 10 Gyr yields corresponding limits of $\sim$2$-$5 M$_\mathrm{Jup}$ and $\sim$8$-$15 M$_\mathrm{Jup}$, respectively. In terms of mass ratios, we are complete down to $q\sim0.7$ at 0\farcs3 and $q\sim0.4$ at separations $\geqslant$0\farcs5 assuming a median age of 5 Gyr for our sample. These values vary by less than $\sim$12\% for ages of 1$-$10 Gyr and we consider that they are representative of the true detection limits of our survey regardless of the unknown ages of our targets.

\subsection{Detection probability map}
\label{prob_map}

The obtained sensitivity curves were used to define a detection probability map for our survey. This provides the probability that a companion at a given physical projected separation $\rho$ and mass ratio $q$ would have been detected in our observed program. The 5-$\sigma$ mass ratio limits for each target in the sample (see Figure~\ref{f:mass_limits_SNAP}) were placed using a cubic interpolation onto a grid of separations and mass ratios with a resolution of 0.002 in $q$ and steps of 0.01 in log($\rho$). For every point of the grid, we then identified the number of targets around which a companion of given separation and mass ratio would have been retrieved in our survey. A companion was considered as detectable around a given target if its mass ratio was higher than the detection limit value at the projected separation of the companion. Companions with separations outside the range covered for a given target were counted as undetectable. The number obtained for each cell of the grid was then divided by the total number of objects in our sample, providing a number between 0 and 1 representing the average detection probability in our program for any ($\rho$, $q$) pair at the 5-$\sigma$ detection level.

Figure~\ref{f:detection_map_SNAP} shows the resulting detection probability map for our core sample of 12 objects. Companions inside the 100\% completeness region are detectable around all targets in the survey. We are not sensitive to any companion in the 0\% detection probability region. Using the bolometric luminosity values derived by \citet{Dupuy2015} for the binary components of W0146$+$4234, we inferred masses of 11$\pm$4 M$_\mathrm{Jup}$ and 10$\pm$4 M$_\mathrm{Jup}$ at an age of 5$\pm$3 Gyr for the primary and secondary, respectively, from the Lyon/COND evolutionary models for brown dwarfs \citep{Baraffe2003}. These masses correspond to a mass ratio $q = 0.91\pm0.05$. The unresolved W0146$-$4234AB system is marked by a yellow star in Figure~\ref{f:detection_map_SNAP} and was found to be located outside our sensitivity limits, in the 0\% detection probability region.

\begin{figure}
    \centering
    \includegraphics[trim={0.5cm 1.2cm 1cm 1cm}, width=0.5\textwidth]{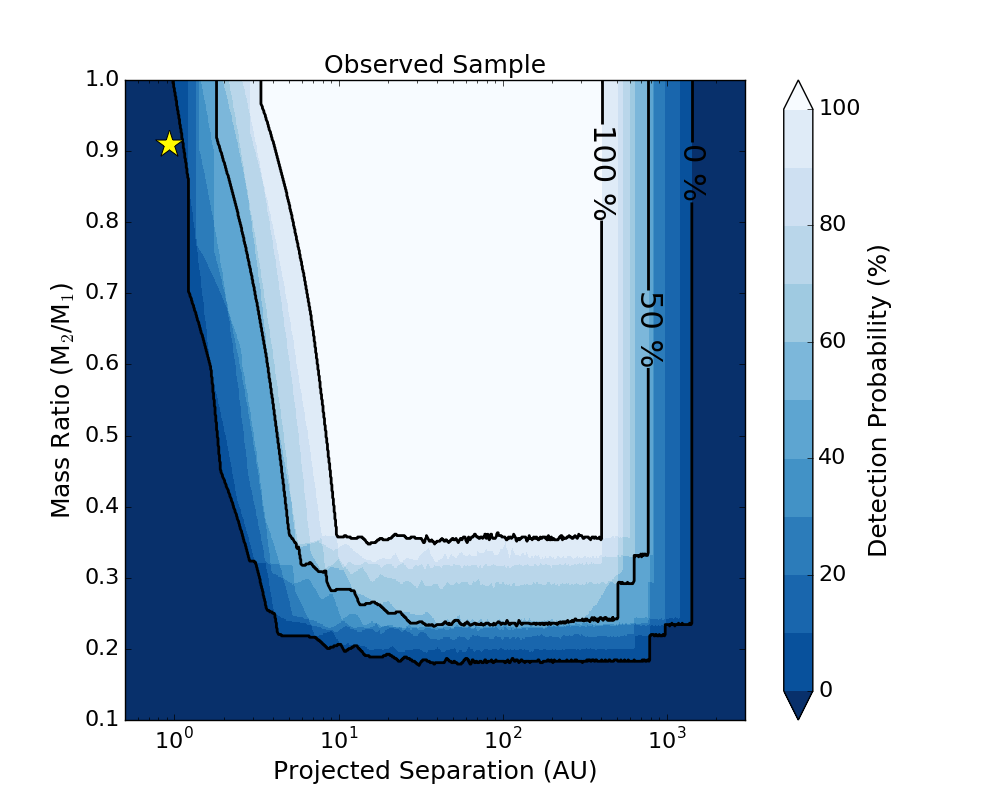}
    \caption{Detection probability map for our sample using the mass ratio sensitivity limits for our 12 targets. Black contours denote the 0\%, 50\% and 100\% completeness regions at the 5-$\sigma$ level. The yellow star shows the position of the unresolved W0146$+$4234 binary discovered in \citet{Dupuy2015}, located in the 0\% detection region.
    }
    \label{f:detection_map_SNAP}
\end{figure}

\section{Additional mid and late-T samples}
\label{additional_samples}

\begin{small}
\begin{table*}
\caption{Additional sample of $\geqslant$T8 brown dwarfs.}
\begin{tabular}{ p{2.65cm} c c c c c c c c c c c }
\hline\hline
Object ID & RA      & Dec.    & SpT & Distance & $J$   & $H$   & Ref. & log($L_{\mathrm{bol}}/L_\odot$) & Mass \\
   & (J2000) & (J2000) & (NIR) & (pc)     & (mag) & (mag) &      & & (M$_{\mathrm{Jup}}$) \\
\hline
\multicolumn{12}{l}{From \citet{Gelino2011}} \\
\hline
WISE J0458$+$6434A$^{\mathrm{a,b}}$ & 04:58:53.90 & $+$64:34:51.9 & T8.5  & $10.5\pm1.4$	& $17.50\pm0.09$  & $17.81\pm0.13$ & (1) & $-5.93\pm0.23$ & $31\pm7$ \\
WISE J0750$+$2725$^{\mathrm{a}}$  & 07:50:03.78 & $+$27:25:44.8 & T9.0    & $15.8\pm2.0$ & $18.69\pm0.04$  & $19.00\pm0.06$ & (2) & $-5.87\pm0.22$ & $34\pm8$ \\
WISE J1322$-$2340                 & 13:22:33.66 & $-$23:40:17.1 & T8.0    & $10.4\pm2.0$ & $17.21\pm0.10$  & $17.01\pm0.14$ & (2) & $-5.81\pm0.21$ & $35\pm8$ \\
WISE J1614$+$1739$^{\mathrm{a}}$  & 16:14:41.45 & $+$17:39:36.7 & T9.0    & $10.0\pm2.0$	& $19.08\pm0.06$  & $18.47\pm0.22$ & (2) & $-6.43\pm0.25$ & $20\pm6$ \\
WISE J1617$+$1807$^{\mathrm{a}}$  & 16:17:05.75	& $+$18:07:14.3	& T8.0    & $15.4\pm2.0$	& $17.66\pm0.08$  & $18.23\pm0.08$ & (3) & $-5.78\pm0.16$ & $36\pm7$ \\
WISE J1653$+$4444                 & 16:53:11.05	& $+$44:44:23.9	& T8.0   & $12.1\pm2.0$	& $17.59\pm0.03$  & $17.53\pm0.05$ & (2) & $-5.83\pm0.18$ & $35\pm7$ \\
WISE J1741$+$2553                 & 17:41:24.26	& +25:53:19.7	& T9.0    & $5.7\pm2.0$	& $16.48\pm0.02$  & $16.24\pm0.04$ & (2) & $-5.78\pm0.32$ & $36\pm9$ \\
\hline
\multicolumn{12}{l}{From \citet{Aberasturi2014}} \\
\hline
ULAS J0034$-$0052$^{\mathrm{a}}$  & 00:34:02.76	& $-$00:52:08.0 & T8.5  & $12.6\pm0.6$	& $18.15\pm0.08$  & $18.49\pm0.04$  & (4) & $-6.02\pm0.17$ & $29\pm6$ \\
2MASS J0729$-$3954                & 07:28:59.47	& $-$39:53:46.3 & T8.0 & $6.0\pm1.0$	& $15.92\pm0.08$  & $15.98\pm0.18$  & (5) & $-5.77\pm0.19$ & $36\pm8$ \\
2MASS J0939$-$2448                & 09:39:35.87	& $-$24:48:38.0 & T8.0    & $10.0\pm2.0$	& $15.98\pm0.11$  & $15.80\pm0.15$  & (6) & $-5.36\pm0.22$ & $49\pm9$ \\
ULAS J1238$+$0953$^{\mathrm{a}}$  & 12:38:28.57	& $+$09:53:51.3 & T8.5  & $18.5\pm4.3$	& $18.95\pm0.02$  & $19.20\pm0.02$  & (7) & $-6.03\pm0.27$ & $29\pm8$ \\
\hline \\ [-2.5ex]
\multicolumn{12}{l}{
  \begin{minipage}{0.94\textwidth}
    \textbf{Notes.}\\
    Magnitudes are on the 2MASS filter system except for $^{\mathrm{a}}$ on the MKO-NIR filter system.\\
    $^{\mathrm{b}}$ Primary component only, see text for secondary component and binary properties.\\
    Bolometric luminosities and masses were derived in this work. Masses were estimated adopting uniform age distributions in the range 2$-$8 Gyr. \\
    \textbf{References.}\\
    Distances for targets from \citet{Gelino2011} are the ``adopted'' distances from table 8 in \citet{Kirkpatrick2012}, except for WISE J0458$+$6434A from \citet{Gelino2011}. Distances for targets from \citet{Aberasturi2014} are those listed in Table 1 in that paper.\\
    Spectral types and photometry from:
    (1) \citet{Gelino2011};
    (2) \citet{Kirkpatrick2011};
    (3) \citet{Burgasser2011}
    (4) \citet{Warren2007};
    (5) \citet{Looper2007};
    (6) \citet{Tinney2005};
    (7) \citet{Burningham2008}.
 \end{minipage}}
\label{t:late-Ts}
\end{tabular}
\end{table*}
\end{small}

In addition to a search for planetary-mass companions, the aim of this survey is to place the first statistically robust constraints to date on the binary properties of ultracool $\geqslant$T8 brown dwarfs. To improve our statistics, we include in our analysis (Section~\ref{stats}) published binary surveys that probed similar spectral type objects. We only consider multiplicity studies containing a minimum of two $\geqslant$T8 targets, as a single object would introduce more systematics into our analysis than it would improve the overall statistics. As our program also aims at confirming the existence of a statistically significant population of wide ultracool binaries like those discovered by \citet{Liu2012}, we excluded surveys that did not search for companions on separations larger than at least a few tens of AU. We do not consider serendipitous discoveries or publications not presenting a full observed sample, as one-off discoveries would strongly bias our results. From the above selection criteria, we retained the binary surveys by \citet{Gelino2011} and \citet{Aberasturi2014}, from which we define an ``extended'' $\geqslant$T8 sample of 23 targets (including our observed program) and a ``comparison'' T5$-$T7.5 sample of 24 objects. Both additional subsets are presented below.

\subsection{Extended sample of $\geqslant$T8 brown dwarfs}
\label{extended_sample}

\begin{figure*}
\addtocounter{figure}{-1}
    \centering
    \begin{subfigure}{0.49\textwidth}
        \includegraphics[width=\textwidth]{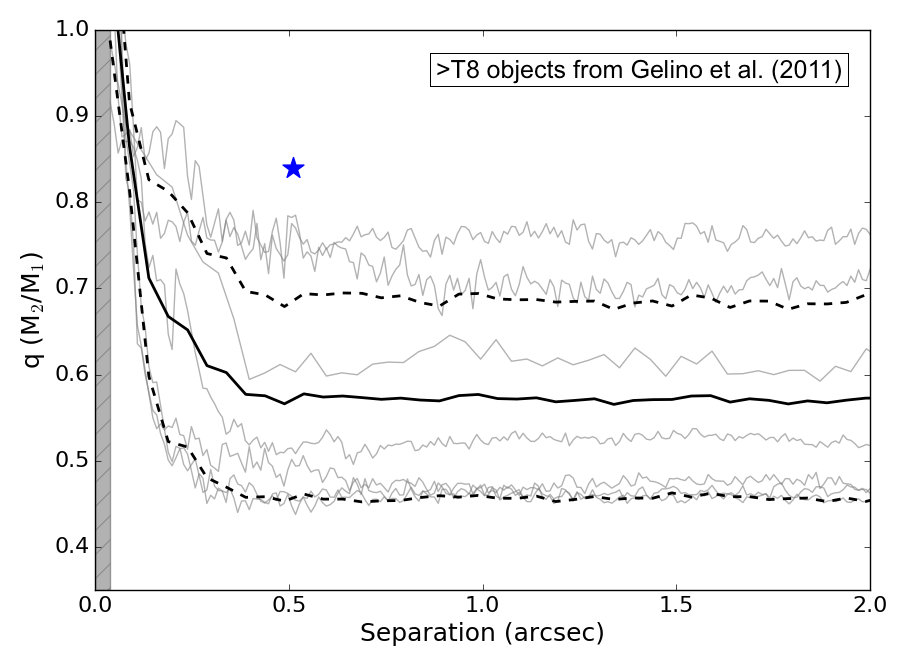}
    \end{subfigure}
    \begin{subfigure}{0.49\textwidth}
        \includegraphics[width=\textwidth]{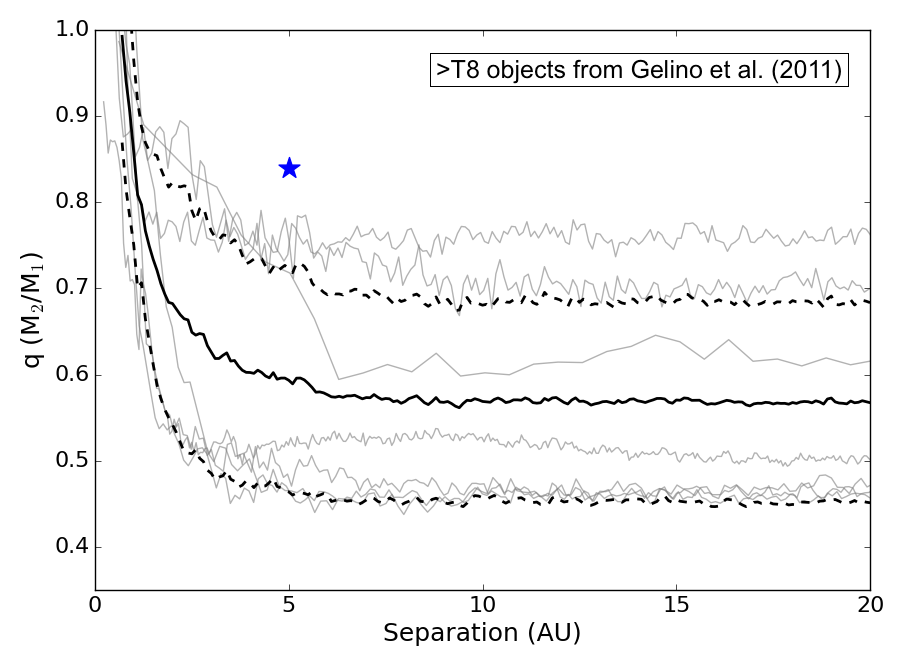}
    \end{subfigure}
    \begin{subfigure}{0.49\textwidth}
        \includegraphics[width=\textwidth]{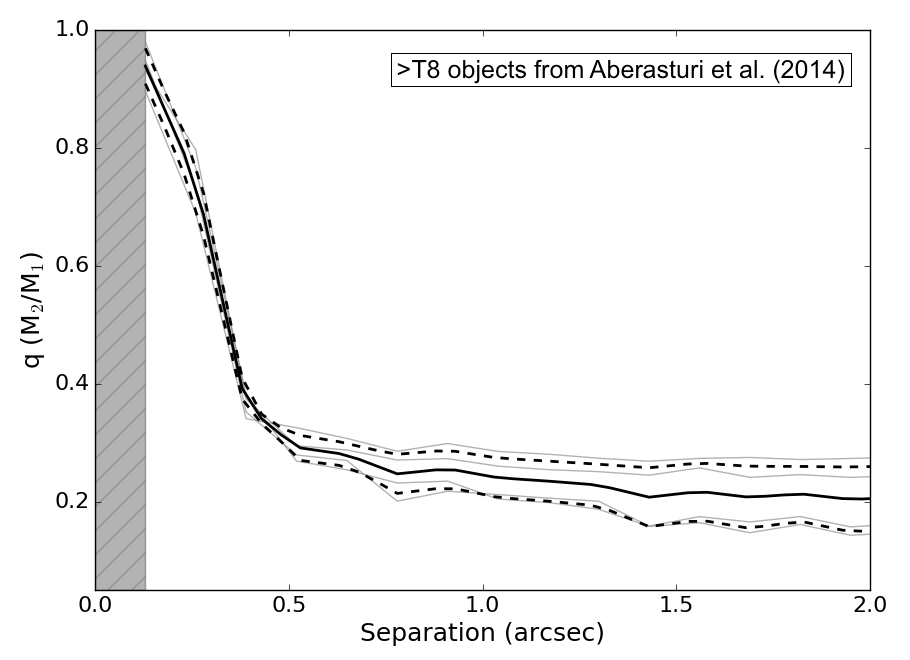}
    \end{subfigure}
    \begin{subfigure}{0.49\textwidth}
        \includegraphics[width=\textwidth]{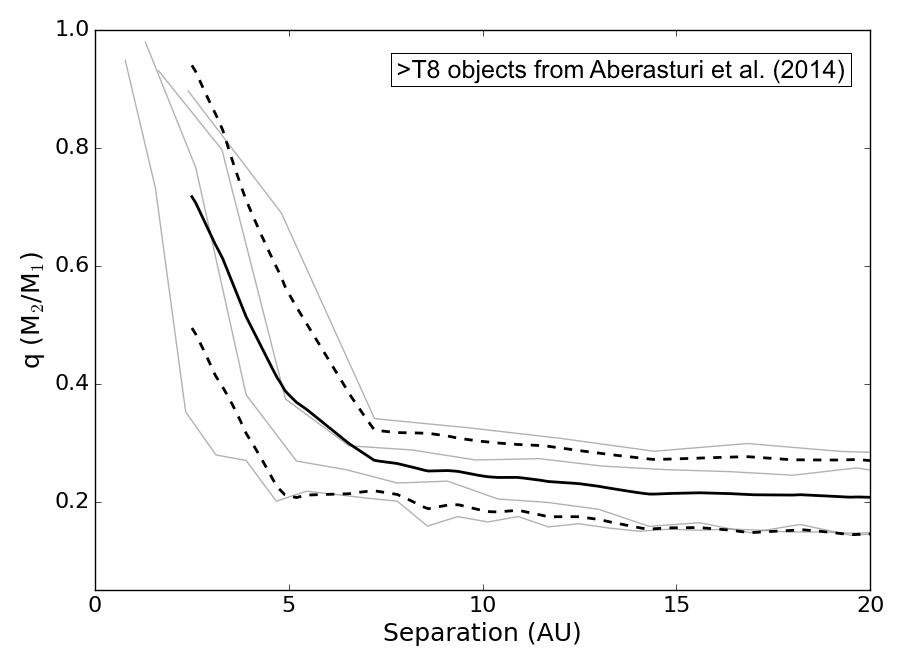}
    \end{subfigure}
    \caption{Mass ratio sensitivity limits for the additional late-T sample. \textbf{Top:} minimum mass ratios detectable around the 7 targets from \citet{Gelino2011} at the 5-$\sigma$ level in the Keck/NIRC2 $H$-band images (grey lines), using the masses calculated in this work and listed in Table~\ref{t:late-Ts}. The shaded region on the left panel represents 4 pixels on the Keck/NIRC2 narrow camera, the radius at which the contrast curves were started. The black solid line shows the mean sensitivity level for the subset and the 1-$\sigma$ standard deviation around the mean (dotted lines). In the right panel, the mean and standard deviation were only calculated at physical separations resolved for all targets. The blue star indicates the position of the secondary companion W0458$+$6434B. The binary companion was masked before computing the contrast curve around the primary W0458$+$6434A. \textbf{Bottom:} same as top panels for the 4 targets from \citet{Aberasturi2014} in the HST/WFC3 F127M images. The shaded region on the left panel represents the WFC3/IR plate scale.}
    \label{f:limits_extended_sample}
\end{figure*}

\begin{figure*}
\addtocounter{figure}{-1}
    \centering   
    \begin{subfigure}{0.45\textwidth}
        \includegraphics[trim={0 0 1cm 1cm}, width=\textwidth]{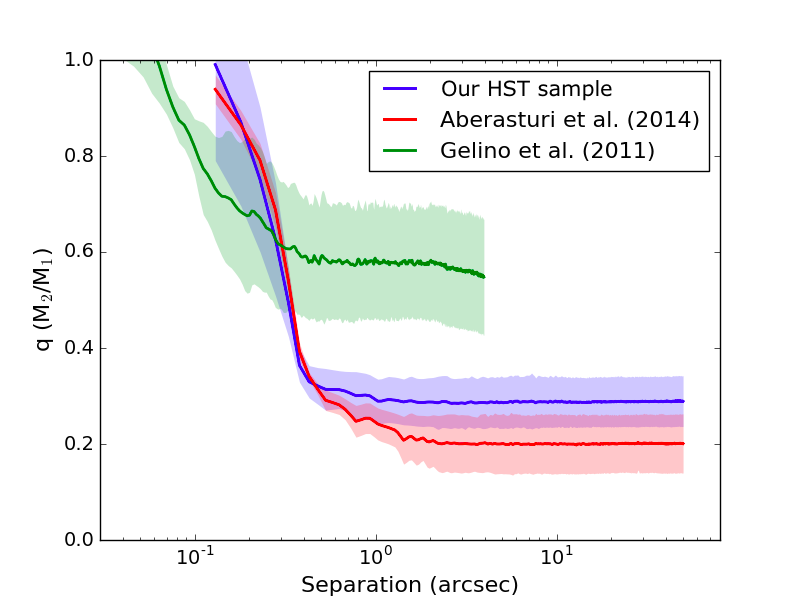}
    \end{subfigure}
    \begin{subfigure}{0.45\textwidth}
        \includegraphics[trim={0 0 1cm 1cm}, width=\textwidth]{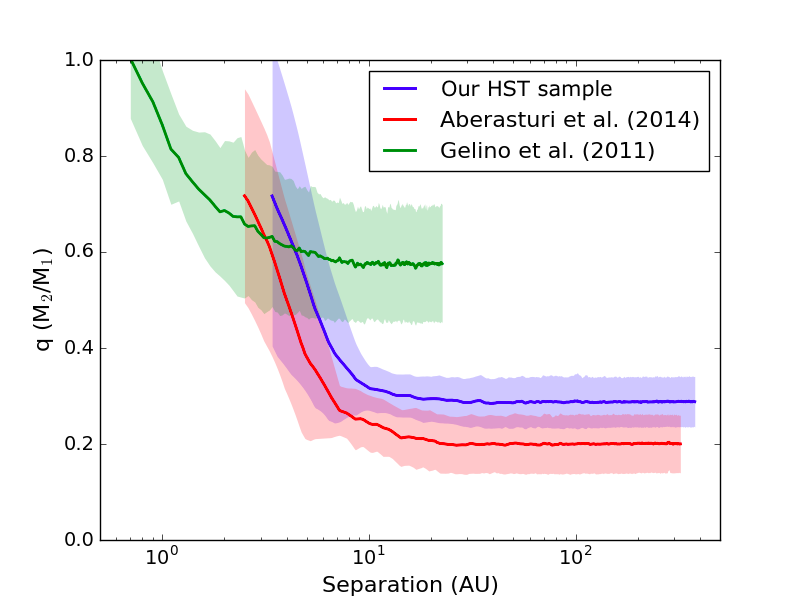}
    \end{subfigure}
    \vspace{2mm}
    \caption{Mean mass ratio sensitivities at the 5-$\sigma$ level for our observed sample and the two subsets from the additional late-T sample, showing the different regions of the parameter space probed by each subset. The shaded areas correspond to the 1-$\sigma$ standard deviation around the mean (solid line). On the right panel, limits of each subset are only shown for physical separations resolved around all targets in any subset. While the HST observations (blue and red curves) probe wide separations and are sensitive to low mass ratios, the Keck images (green) have a smaller inner working angle and may resolve smaller separations.}
    \label{f:limits_late-Ts_all}
\end{figure*}

\begin{figure}
    \centering
    \includegraphics[trim={0.5cm 1cm 1cm 1cm}, width=0.5\textwidth]{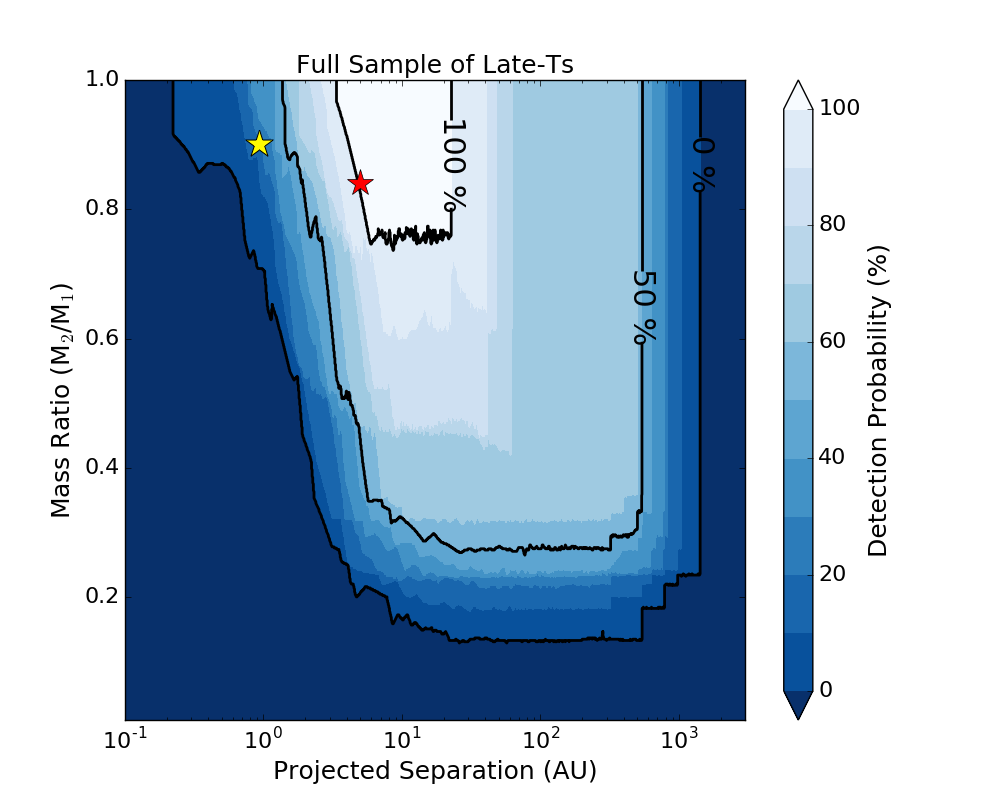}
    \caption{Same as Figure~\ref{f:detection_map_SNAP} for the extended sample combining our observed survey with the $\geqslant$T8 brown dwarfs selected from \citet{Gelino2011} and \citet{Aberasturi2014}. Contours denote the 0\%, 50\% and 100\% completeness regions. The red star shows the position of the T8.5$+$T9 binary discovered in \citet{Gelino2011}, W0458$+$6434AB, using the component masses calculated in this work. The yellow star corresponds to the known W0146$+$4234 binary, unresolved in our observations.}
    \label{f:detection_map_late-Ts}
\end{figure}

To extend our T8 and later sample size, we consider all objects with spectral types $\geqslant$T8 from the studies in \citet{Gelino2011} and \citet{Aberasturi2014}, doubling the overall size of our sample. All additional targets have similar estimated field ages ($\sim$few Gyr) and distances ($<$30 pc) to our core sample. The subset from \citet{Gelino2011} consists of 7 objects and includes a T8.5$+$T9.5 binary system discovered as part of that multiplicity search. The survey conducted by \citet{Aberasturi2014} includes 4 brown dwarfs with spectral types of T8 or later, none of which was found to be a resolved binary. Using the method described in Section~\ref{mass_estimates}, we estimated the mass of each target from its published $J$-band photometry. Photometric information, distances and derived properties for the 11 additional sources are presented in Table~\ref{t:late-Ts}.

\citet{Gelino2011} found that W0458$+$6434 is a $510\pm20$ mas binary with a $\Delta$mag of $\sim$1 in both $J$ and $H$. At the spectrophotometric distance of $10.5\pm1.4$ pc adopted in \citet{Gelino2011} for the system, the measured angular separation corresponds to a physical projected separation of $5.0\pm0.4$ AU. With spectral types of T8.5 and T9.5 \citep{Burgasser2012} for the primary and secondary components, respectively, W0458$+$6434AB is one of the latest-type brown dwarf binaries discovered to date. \citet{Gelino2011} estimated component masses of 15 M$_\mathrm{Jup}$ and 10 M$_\mathrm{Jup}$ at an adopted age of 1 Gyr. From the $J$-band photometry reported for the individual components and applying the method used throughout this work for mass estimation, we derived masses of $31\pm7$ M$_\mathrm{Jup}$ and $26\pm7$ M$_\mathrm{Jup}$ for the two binary components assuming a uniform age distribution in the range $5\pm3$ Gyr. These results are in good agreement with the values reported in \citet{Gelino2011} for the slightly older ages adopted here. Our derived component masses yield a mass ratio of $q = 0.84\pm0.05$ for the system.

Targets from \citet{Aberasturi2014} were observed with WFC3/IR on HST in the F110W, F127M and F164N filters. The F127M bandpass covers the 1.27 $\mu$m peak observed in late-T brown dwarfs and is thus more sensitive to search for faint companions (see \citealp{Aberasturi2014}). We therefore used the HST/WFC3 F127M images from this program and the photometry reported in that paper to derive the detection limits of that subset, after performing the same data reduction as for our core sample. Total exposure times of 1197.7 s were obtained for the F127M observations, providing slightly deeper images than our observed program.

The multiplicity search carried by \citet{Gelino2011} was conducted with the infrared camera NIRC2 together with the Laser Guide Star Adaptive Optics system (LGS AO; \citealp{Wizinowich2004}) on the 10 m Keck II telescope. Images were taken in the $H$ filter and observations are described in the survey paper. The narrow camera (plate scale of 0\farcs009942 pixel$^{-1}$) was used for all observations, with the exception of one target (W0750$+$2725) which was observed with the wide camera (0\farcs039686 pixel$^{-1}$). The Keck/NIRC2 $H$-band images were reduced using custom Python scripts. For each image, we subtracted a mean dark frame generated from all other dithered positions to remove the sky background. We then applied a bad pixel mask and divided by a flat-field image before stacking all dithered frames.

\begin{small}
\begin{table*}
\caption{Comparison sample of T5$-$T7.5 brown dwarfs.}
\begin{tabular}{ p{2.65cm} c c l c c c c c c c c }
\hline\hline
Object ID & RA      & Dec.    & SpT & Distance & $J$   & $H$   & Ref.  & log($L_{\mathrm{bol}}/L_\odot$) & Mass \\
   & (J2000) & (J2000) & (NIR) & (pc)     & (mag) & (mag) &       &   & (M$_{\mathrm{Jup}}$) \\
\hline
\multicolumn{12}{l}{From \citet{Gelino2011}} \\
\hline
WISE J1627$+$3255       & 16:27:25.64   & $+$32:55:24.1 & T6.0 &  $15.4\pm2.0$ & $16.48\pm0.04$  & $16.40\pm0.05$  & (1) &  $-5.37\pm0.12$ & $48\pm8$ \\
WISE J1841$+$7000A$^{\mathrm{a,b}}$ & 18:41:24.74 & $+$70:00:38.0 & T5.0  & $40.2\pm4.9$ & $17.24\pm0.10$  & $17.73\pm0.10$  & (2) & $-4.91\pm0.12$ & $61\pm8$ \\
\hline
\multicolumn{12}{l}{From \citet{Aberasturi2014}} \\
\hline
HD3651B$^{\mathrm{a}}$  & 00:39:18.61	& $+$21:15:12.7 & T7.5  & $11.0\pm0.1$	& $16.16\pm0.03$  & $16.68\pm0.04$  & (3) & $-5.56\pm0.08$ & $43\pm7$ \\
2MASS J0050$-$3322      & 00:50:19.92	& $-$33:22:41.4 & T7.0    & $8.0\pm1.0$	& $15.93\pm0.07$  & $15.84\pm0.19$  & (4) & $-5.67\pm0.12$ & $39\pm7$ \\
SDSS J0325$+$0425       & 03:25:53.11	& $+$04:25:40.0 & T5.5  & $19.0\pm2.0$	& $16.25\pm0.14$  & $> 16.08$       & (5) & $-5.08\pm0.11$ & $57\pm8$ \\
2MASS J0407$+$1514      & 04:07:08.94	& $+$15:14:55.4 & T5.0    & $17.0\pm2.0$	& $16.06\pm0.09$  & $16.02\pm0.21$  & (6) & $-5.08\pm0.11$ & $57\pm8$ \\
2MASS J0510$-$4208      & 05:10:35.32	& $-$42:08:08.2 & T5.0    & $18.0\pm2.0$	& $16.22\pm0.09$  & $16.24\pm0.16$  & (6) & $-5.09\pm0.11$ & $57\pm8$ \\
2MASS J0727$+$1710      & 07:27:19.07	& $+$17:09:52.2 & T7.0    & $9.1\pm0.2$	& $15.60\pm0.06$  & $15.76\pm0.17$  & (7) & $-5.42\pm0.06$ & $47\pm7$ \\
2MASS J0741$+$2351      & 07:41:48.96	& $+$23:51:25.9 & T5.0    & $18.0\pm2.0$	& $16.15\pm0.10$  & $15.84\pm0.18$  & (6) & $-5.07\pm0.11$ & $57\pm8$ \\
2MASS J1007$-$4555      & 10:07:32.99	& $-$45:55:13.3 & T5.0    & $15.0\pm2.0$	& $15.65\pm0.07$  & $15.68\pm0.12$  & (5) & $-5.03\pm0.13$ & $58\pm8$ \\
2MASS J1114$-$2618      & 11:14:48.90	& $-$26:18:27.2 & T7.5  & $10.0\pm2.0$	& $15.86\pm0.08$  & $15.73\pm0.12$  & (8) & $-5.40\pm0.20$ & $47\pm9$ \\
2MASS J1231$+$0847      & 12:31:46.74	& $+$08:47:22.3 & T5.5  & $12.0\pm1.0$	& $15.57\pm0.07$  & $15.31\pm0.11$  & (6) & $-5.21\pm0.08$ & $53\pm8$ \\
SDSS J1346$-$0031       & 13:46:46.04	& $-$00:31:51.3 & T6.5  & $14.6\pm0.5$	& $16.00\pm0.10$  & $15.46\pm0.12$  & (5) & $-5.21\pm0.06$ & $53\pm7$ \\
SDSS J1504$+$1027       & 15:04:11.74	& $+$10:27:18.8 & T7.0    & $15.9\pm2.5$	& $17.03\pm0.23$  & $> 16.90$       & (5) & $-5.52\pm0.18$ & $44\pm8$ \\
SDSS J1628$+$2308       & 16:28:38.99	& $+$23:08:18.4 & T7.0    & $14.0\pm4.0$	& $16.45\pm0.10$  & $16.11\pm0.15$  & (4) & $-5.43\pm0.29$ & $46\pm9$ \\
2MASS J1754$-$1649      & 17:54:54.56	& $+$16:49:18.1 & T5.0    & $14.3\pm1.3$	& $15.81\pm0.07$  & $15.65\pm0.13$  & (9) & $-5.13\pm0.09$ & $56\pm8$ \\
SDSS J1758$+$4633       & 17:58:05.49	& $+$46:33:17.1 & T6.5  & $12.0\pm2.0$	& $16.15\pm0.08$  & $16.25\pm0.21$  & (4) & $-5.45\pm0.16$ & $46\pm8$ \\
2MASS J1828$-$4849      & 18:28:36.01	& $-$48:49:02.6 & T5.5  & $11.0\pm1.0$	& $15.18\pm0.06$  & $14.91\pm0.07$  & (8) & $-5.13\pm0.09$ & $56\pm8$ \\
2MASS J1901$+$4718      & 19:01:05.89	& $+$47:18:09.9 & T5.0    & $15.0\pm2.0$	& $15.86\pm0.07$  & $15.47\pm0.09$  & (5) & $-5.51\pm0.13$ & $44\pm8$ \\
SDSS J2124$+$0100       & 21:24:14.02	& $+$01:00:02.7 & T5.0    & $18.0\pm2.0$	& $16.03\pm0.07$  & $16.18\pm0.20$  & (5) & $-5.02\pm0.11$ & $58\pm8$ \\
2MASS J2154$+$5942      & 21:54:32.98	& $+$59:42:14.4 & T5.0    & $10.0\pm1.0$	& $15.66\pm0.07$  & $15.76\pm0.17$  & (5) & $-5.38\pm0.10$ & $48\pm8$ \\
2MASS J2237$+$7228      & 22:37:20.47	& $+$72:28:35.3 & T6.0    & $13.0\pm2.0$	& $15.76\pm0.07$  & $15.94\pm0.21$  & (10) & $-5.23\pm0.14$ & $53\pm8$ \\
2MASS J2331$-$4718      & 23:31:23.84	& $-$47:18:28.2 & T5.0    & $13.0\pm2.0$	& $15.66\pm0.07$  & $15.51\pm0.15$  & (6) & $-5.16\pm0.14$ & $55\pm8$ \\
2MASS J2359$-$7335      & 23:59:41.09	& $-$73:35:04.9 & T6.5  & $12.3\pm1.9$	& $16.17\pm0.04$  & $16.06\pm0.07$  & (1) & $-5.43\pm0.14$ & $47\pm8$ \\
\hline \\ [-2.5ex]
\multicolumn{12}{l}{
  \begin{minipage}{0.93\textwidth}
    \textbf{Notes.}\\
    Magnitudes are on the 2MASS filter system except for $^{\mathrm{a}}$ on the MKO-NIR filter system.\\
    $^{\mathrm{b}}$ Primary component of only, see text for secondary component and binary properties.\\
    Bolometric luminosities and masses were derived in this work. Masses were estimated adopting uniform age distributions in the range 2$-$8 Gyr. \\
    \textbf{References.}\\
    Distances for targets from \citet{Gelino2011} are the ``adopted'' distances from table 8 in \citet{Kirkpatrick2012}, except for WISE J1841$+$7000A from \citet{Gelino2011}. Distances for targets from \citet{Aberasturi2014} are those listed in Table 1 in that paper.\\
    Spectral types and photometry from:
    (1) \citet{Kirkpatrick2011};
    (2) \citet{Gelino2011};
    (3) \citet{Luhman2007};
    (4) \citet{Dupuy2015};
    (5) \citet{Looper2007};
    (6) \citet{Faherty2009};
    (7) \citet{Vrba2004};
    (8) \citet{Tinney2005};
    (9) \citet{Faherty2012};
    (10) \citet{Mace2013}.
 \end{minipage}}
\label{t:mid-Ts}
\end{tabular}
\end{table*}
\end{small}

The sensitivity limits for all additional targets in the extended late-T sample were obtained following the method applied to our core sample, described in Section~\ref{limits}. We used the measured WFC3/F127M photometry from \citet{Aberasturi2014} and 2MASS or MKO $H$-band photometry for the targets from \citet{Gelino2011} together with the corresponding AMES/COND models \citep{Allard2001} to compute mass detection limits for each object. Sensitivity curves in the Keck images were started at a radius of 4 pixels because of the high noise level inside that radius, except for the target observed with the wide camera, for which we used an initial radius of 1 pixel. The obtained detection limits for each subset are presented in Figure~\ref{f:limits_extended_sample} in terms of mass ratio as a function of angular and physical projected separation.
The regions of the parameter space probed in the two subsets are considerably different as shown in Figure~\ref{f:limits_late-Ts_all}. While the HST observations are deep ($q$ $\sim$ 0.2$-$0.3 at separations $>$10 AU) and have a wide field of view (123\arcsec$\times$136\arcsec), the 0\farcs13 pixel$^\mathrm{-1}$ plate scale of the WFC3/IR instrument only allows us to probe separations down to 0.8$-$2.4 AU at the distances of the targets. In comparison, the NIRC2 images have a resolution of 0\farcs01 pixel$^\mathrm{-1}$ (0\farcs04 pixel$^\mathrm{-1}$ for the wide camera). The 4-pixel radius at which the contrast curves were started (1 pixel for images acquired with the wide camera) corresponds to projected separations of 0.23$-$0.63 AU. With a 10\arcsec $\times$ 10\arcsec field of view and mass ratio limits of $q\sim0.6$, observations from this subset are not sensitive to wide ($>$ 20$-$60 AU) or low-mass ($q < 0.5$) companions.

The average detection probability map for the combined sample of 23 objects (observed and extended samples) was derived from the sensitivity limits of all individual targets, following the approach described in Section~\ref{prob_map}. The resulting map is shown in Figure~\ref{f:detection_map_late-Ts}. As a result of the different facilities and instruments used, the combined survey is only complete down to $q\sim0.75$ and between $\sim$5$-$25 AU, the region of the parameter space where all surveys overlap. The binary discovered in \citet{Gelino2011}, W0458+6434AB (red star), is located inside the 100\% completeness region of the combined survey, meaning that we are sensitive to systems with the physical properties of this system around all targets. The 50\% completeness contour shows that systems with separations in the range $\sim$3$-$500 AU and mass ratios $>$0.3 are detectable around half of the targets in the final sample. The unresolved binary from \citet{Dupuy2015} is located in the 20$-$30\% detection probability region (yellow star).

\subsection{Comparison sample of T5$-$T7.5 brown dwarfs}
\label{comparison_sample}

\begin{figure*}
\addtocounter{figure}{-1}
    \centering   
    \begin{subfigure}{0.48\textwidth}
        \includegraphics[width=\textwidth]{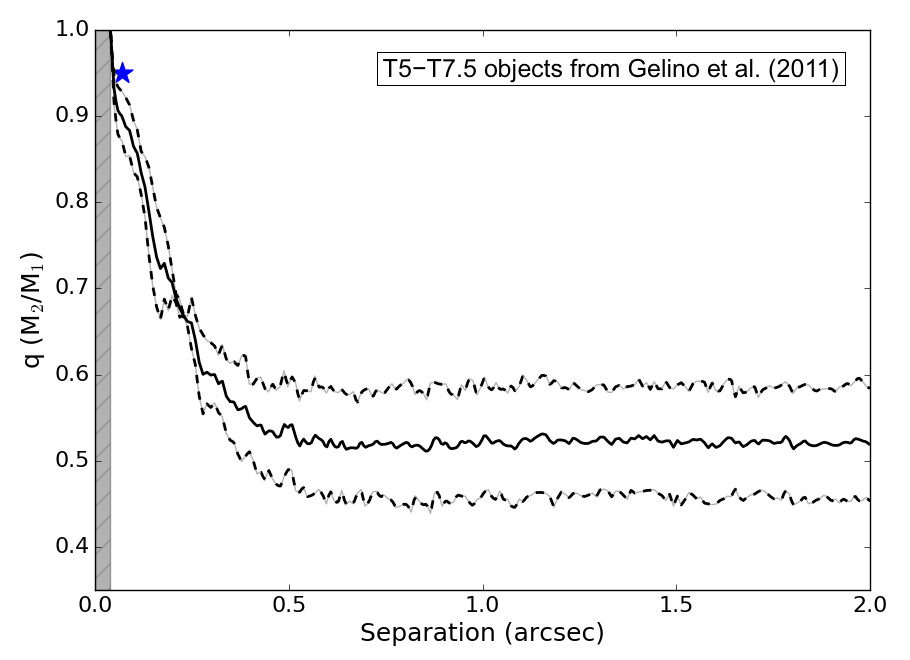}
    \end{subfigure}
    \begin{subfigure}{0.48\textwidth}
        \includegraphics[width=\textwidth]{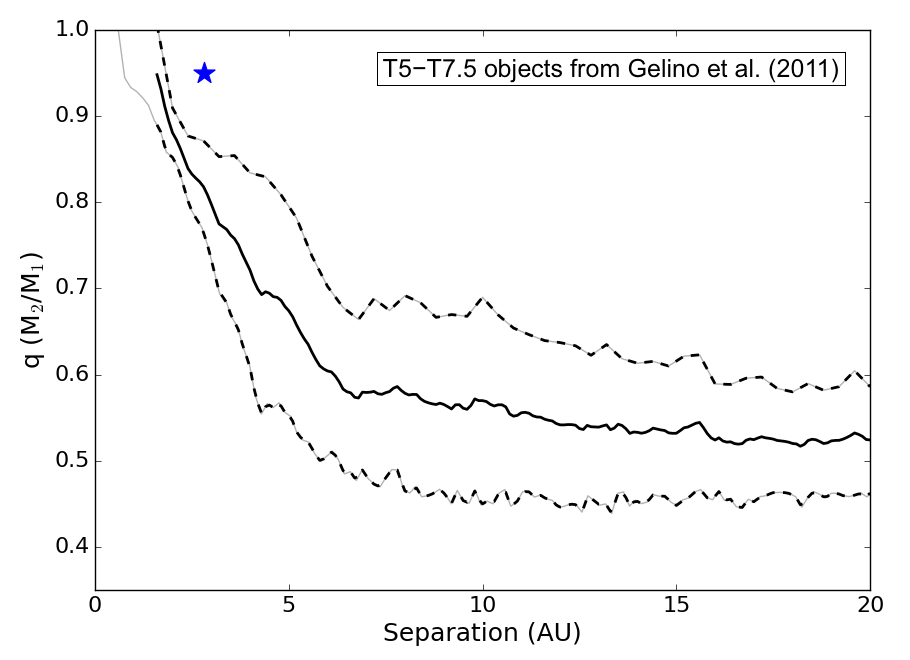}
    \end{subfigure}
    \begin{subfigure}{0.48\textwidth}
        \includegraphics[width=\textwidth]{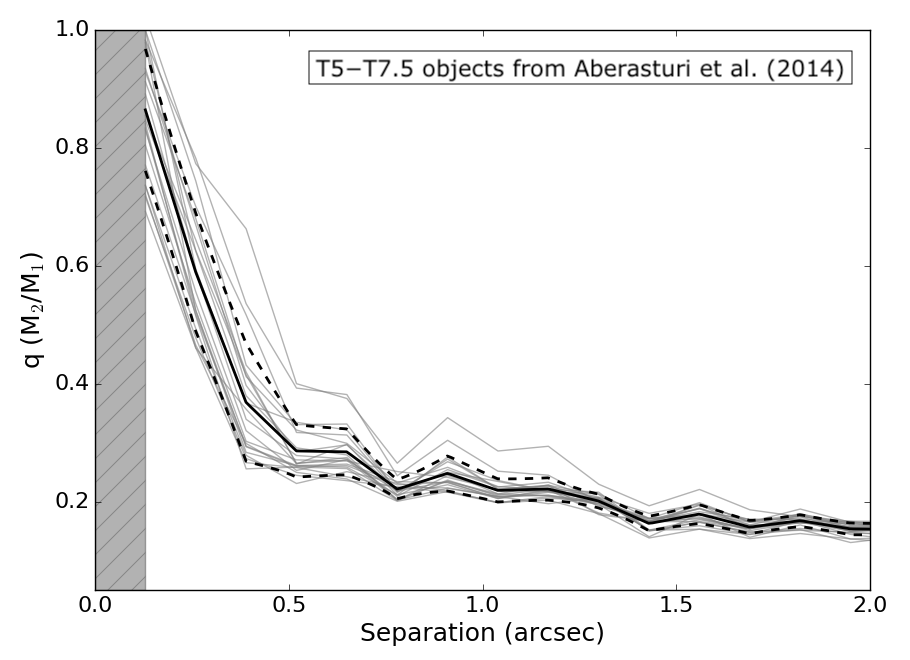}
    \end{subfigure}
    \begin{subfigure}{0.48\textwidth}
        \includegraphics[width=\textwidth]{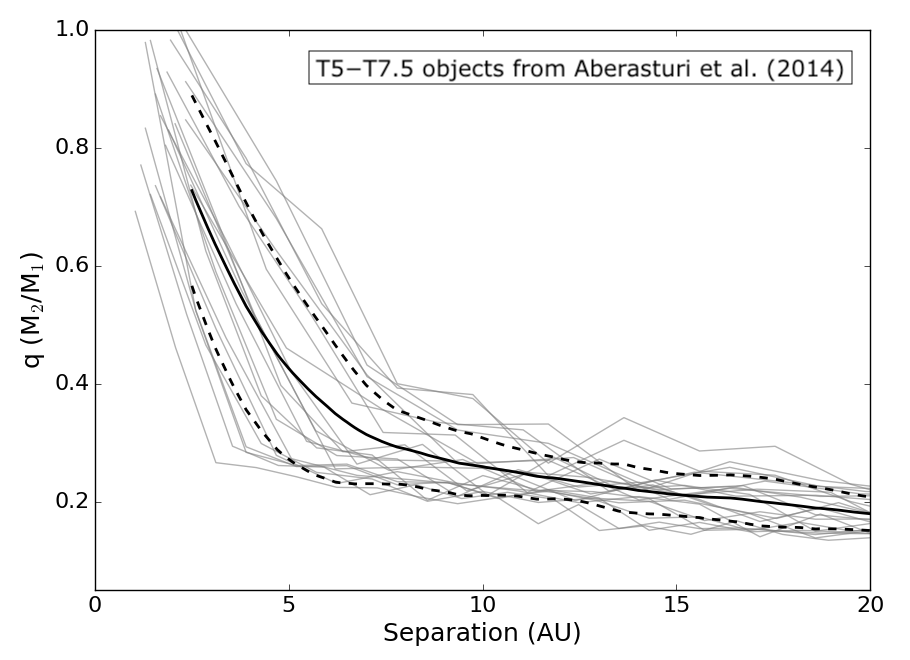}
    \end{subfigure}
    \caption{Same as Figure~\ref{f:limits_extended_sample} for the mid-T sample. \textbf{Top:} minimum mass ratios detectable around the 2 targets from \citet{Gelino2011} at the 5-$\sigma$ level in the Keck/NIRC2 $H$-band images. The blue star indicates the position of the secondary companion W1841$+$7000B. The binary companion was masked before computing the contrast curve around the primary W1841$+$7000A. \textbf{Bottom:} same as top panels for the 22 targets from \citet{Aberasturi2014} in the HST/WFC3 F127M images.}
    \label{f:limits_comparison_sample}
\end{figure*}

\begin{figure}
    \centering
    \includegraphics[trim={0.5cm 1.2cm 1cm 1cm}, width=0.5\textwidth]{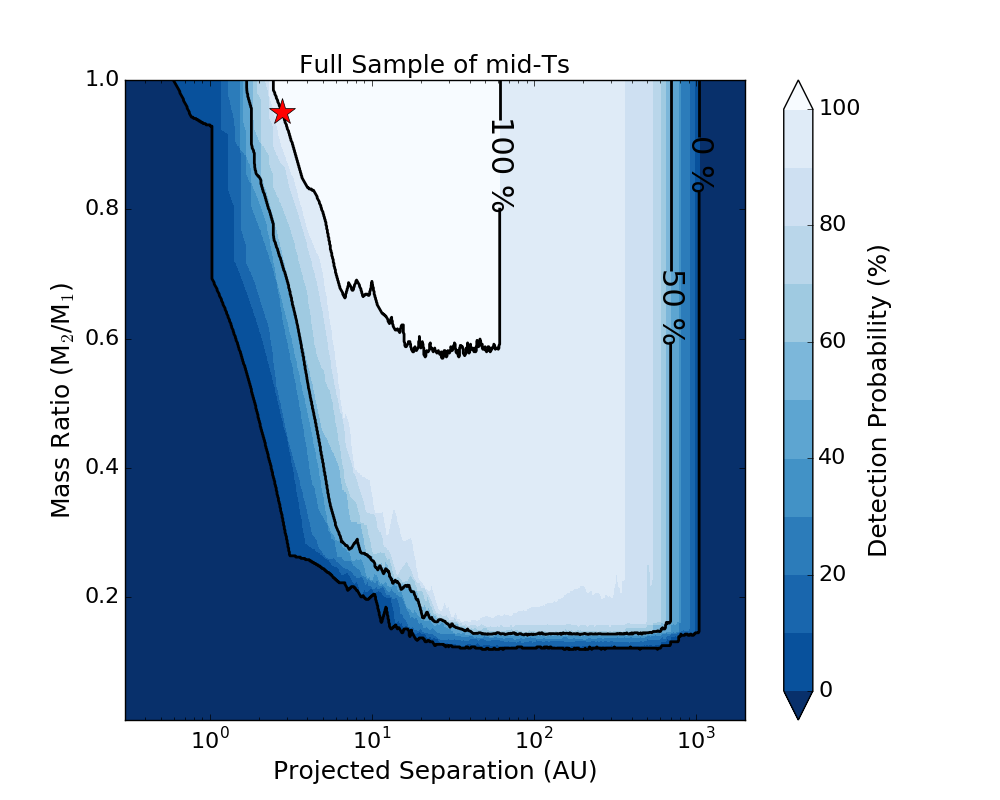}
    \caption{Same as Figure~\ref{f:detection_map_SNAP} for the sample of T5$-$T7.5 sample selected from \citet{Gelino2011} and \citet{Aberasturi2014}. Contours denote the 0\%, 50\% and 100\% completeness regions. The red star shows the position of T5$+$T5 binary discovered in \citet{Gelino2011}, W1841$+$7000AB.}
    \label{f:detection_map_mid-Ts}
\end{figure}

In order to compare our results to the binary properties of earlier-type objects, we also compiled a sample of 24 mid-T (T5$-$T7.5) brown dwarfs from \citet{Gelino2011} (2 objects) and \citet{Aberasturi2014} (22 objects). The two subsets considered come from the same surveys as the additional late-T sample presented in Section~\ref{extended_sample}, allowing for a direct comparison of the obtained results. The comparison mid-T sample is presented in Table~\ref{t:mid-Ts}. Luminosities and masses were derived following the approach described in Section~\ref{mass_estimates}, adopting uniform age distributions in the range 2$-$8 Gyr. With earlier spectral types, and thus higher effective temperatures at the same adopted ages, targets from this subset have larger estimated masses ($\sim$40$-$60 M$_\mathrm{Jup}$) than the late-T objects from our observed program and the extended sample ($<$40 M$_\mathrm{Jup}$). \citet{Gelino2011} identified W1841$+$7000 as a T5$+$T5 near equal-mass binary with a 2.8 AU projected separation ($70\pm14$ mas). We estimated component masses of $61\pm8$ M$_\mathrm{Jup}$ and $58\pm8$ M$_\mathrm{Jup}$ adopting an age of 2$-$8 Gyr and using the photometric measurements for individual components reported in \citet{Gelino2011}, implying a mass ratio of $q = 0.95\pm0.05$. \citet{Aberasturi2014} found no binary system among the 22 objects selected for this subset.

Like for the additional late-T sample, we used the HST WFC3/IR F127M images from the program in \citet{Aberasturi2014} and Keck/NIRC2 $H$-band data for the two targets from \citet{Gelino2011}, applying the same data reduction as in Section~\ref{extended_sample}. Detection limits for all targets were derived in the same way as for the additional late-T objects and are shown in Figure~\ref{f:limits_comparison_sample}. For similar achieved mass limits in each subset from the two programs considered, we obtained slightly better mass ratio sensitivities overall than for the late-T targets as a result of the higher primary masses of the mid-T targets. The sensitivity limits were finally combined to create a detection probability map (see Section~\ref{prob_map}) for the full sample of mid-T brown dwarfs, presented in Figure~\ref{f:detection_map_mid-Ts}. Again, the use of multiple instruments probing different regions of the parameter space resulted in a rather restricted 100\% completeness region. The observed binary, W1841$+$7000AB, was found to be at an 83\% detection probability level.

\section{Measured binary properties}
\label{stats}

Despite a null detection, results from our survey may still be used to place statistical limits on the binary properties for $\geqslant$T8 field brown dwarfs. The absence of resolved binaries in our observed program is consistent with the current census for the binary properties of the latest-type brown dwarfs. Substellar binary rate is believed to decrease with spectral type within the field population \citep{Allen2007, Kraus2012} and multiple systems are hence expected to be rare among the latest-type objects. Multiplicity surveys in the field indicate a clear tendency towards near equal-flux systems. As most field binaries are found to have high mass ratios ($q>0.8$; \citealp{Burgasser2006}), we expect our observed program to be sensitive to most binaries beyond $\sim$5 AU, since we are complete down to $q\sim0.75$ at 5 AU and $q\sim0.4$ from 8 AU. Given our survey sensitivity, it is unlikely that we missed a significant number of such systems. However, the observed peak of the separation distribution for field brown dwarfs ($\sim$4 AU; \citealp{Allen2007, Burgasser2007}) is close to our resolution limits. We are thus only sensitive to somewhat wide binaries, thought to be uncommon at such late spectral types. While a population of very low mass ratio systems lying below our detection limits seems unlikely, a fraction of close binaries could still remain undetected due to survey incompleteness. This observational bias must be carefully taken into account when investigating the binary properties of our probed targets.

With no new companion detected as part of our study, we cannot place any new constraints on the mass ratio or separation distributions of the latest-type brown dwarf binary systems based on our observed sample. We are however able to constrain the binary frequency of old objects with estimated masses $<$40 M$_\mathrm{Jup}$ for the separations and mass ratios probed in this study. We used a Markov Chain Monte Carlo (MCMC) approach to estimate the binary fraction most compatible with the observed data, assuming a range of possible distributions of companion populations. The MCMC sampling tool accounts for the survey detection limits and the presumed shapes of companion population distributions, therefore correcting for observational biases. The statistical tool is described in Section~\ref{MCMC} and results from its application to our core program and the extended and comparison samples are presented in Section~\ref{results}.

\subsection{Bayesian statistical analysis: MCMC tool}
\label{MCMC}

We developed an MCMC sampling tool designed for Bayesian parameter estimation. The tool was built using the \textit{emcee} \citep{Foreman-Mackey2013} Python implementation of the affine-invariant ensemble sampler for MCMC proposed by \citet{Goodman2010}. The core of this method is based on Bayesian parameter estimation. Bayes' theorem states:
\begin{equation}
\label{eq:Bayes}
    P(\theta \mid D) \propto P(D \mid \theta)\: P(\theta) ,
\end{equation}
where $\theta$ represents the model and $D$ the data. $P(\theta)$, the prior distribution, is the initial probability density of the model. $P(D \mid \theta)$, the likelihood function, gives the probability of the data given the model. $P(\theta \mid D)$, the posterior distribution, is the probability of the model given the data. For any model $\theta$, we are able to calculate the likelihood function, that is, the probability that the data $D$ would have been measured given the hypothesised model. Using Bayes' theorem (Eq.~\ref{eq:Bayes}) we may then compute the probability of a hypothesised model being true given the observed data, that is, the posterior distribution.

The MCMC sampling method iteratively generates sequences of samples for each parameter describing the model, calculating the likelihood function for each set of parameters so as to approximate the desired posterior distribution. At each step, the algorithm randomly attempts to move the walkers in the parameter space. Moving to a point in a higher probability density region of the posterior distribution is always accepted. Attempting to move to a less probable point is accepted or rejected based on the current and trial positions. As a result, while the sampler occasionally visits low probability density regions, it tends to remain in higher probability density parts of the parameter space, returning final output samples representative of the sought posterior distributions for each model parameter. The affine-invariant ensemble diverges from the usual ``random walk'' Metropolis-Hasting algorithm (\citealp{Metropolis1953, Hastings1970}) by using the current positions of all the other walkers in the ensemble to move a given walker. The intuition behind this is that other walkers have already sampled an important part of the parameter space and provide valuable information about the underlying distributions. This significantly improves performances by reducing the time required for the algorithm to identify and explore the most relevant regions of the parameter space, making the affine-invariant MCMC very competitive.

Based on previous substellar multiplicity surveys (e.g. \citealp{Close2003,Burgasser2006,Burgasser2007,Allen2007}), companion populations were assumed to follow a lognormal distribution in projected separation $\rho$ and a power law distribution in mass ratio $q \equiv M_\mathrm{2}/M_\mathrm{1}$. The MCMC sampler explores four model parameters describing the companion populations:
\begin{description}
    \item $\rho_0$, the peak of the lognormal distribution in projected separation.
    \item $\sigma$, the standard deviation of the normal distribution in $\log(\rho)$.
    \item $\gamma$, the index of the power law distribution in the mass ratio $q$.
    \item $f$, the binary frequency of a given separation range.
\end{description}
The lognormal distribution in projected separation $\rho$ is given by:
\begin{equation} \label{eq:lognorm}
    P(\rho \mid \mu, \sigma) = \frac{1}{\sqrt{2 \pi}\: \sigma\: \rho}\: e^{-\left(\log_{10}(\rho)\: -\: \mu\right)^2 \:\mathbin{/}\: 2 \sigma^2} ,
\end{equation}
\noindent where $\mu$ is the mean of the underlying normal distribution in $\log(\rho)$. The mean $\mu$ is a function of $\rho_0$ and $\sigma$ and is found by solving for the root of $\partial P(\rho) / \partial \rho$ at $\rho = \rho_0$ at any given step. Eq.~\ref{eq:lognorm} may be truncated to be restricted to a defined range of separations.

The mass ratio distribution ranges from 0 to 1 and is described by the equation:
\begin{equation} \label{eq:powerlaw}
    P(q \mid \gamma) = (\gamma + 1)\: q^\gamma .
\end{equation}

Based on our lack of knowledge of any of these parameters for the very low-mass objects studied in this work, prior distributions were chosen to be flat distributions, set to unity over a chosen range and to zero elsewhere. We defined priors in the ranges 0.3$-$10 AU for $\rho_0$, 0.03$-$1 for $\sigma$, 1$-$12 for $\gamma$ and 0$-$1 for $f$. We assumed no prior knowledge about $f$ so as to explore the full range of possible values. Ranges for the other three model parameters were chosen so as to span a wide enough space to likely cover the expected peak value of each parameter based on previous studies (e.g. \citealp{Reid2006,Burgasser2007}), while limiting the region of parameter space to be explored.
As null or low number of detections does not allow us to constrain the shapes of the separation and mass ratio distributions, wider ranges for prior probabilities in $\rho_0$, $\sigma$ and $\gamma$ only result in a broader output distribution in $f$ due to the many more possible companion populations tested by the sample. We therefore restrained prior distributions to what we consider plausible regions based on past studies.
Walkers were started in a tight 4-dimensional ball, centred around a chosen point expected to be close to the maximum probability point for each parameter. This approach reduces the risk of walkers getting stuck in low-probability regions of the parameter space. The walkers quickly expanded out to explore and fill the relevant parts of parameter space. The initial positions of the walkers were drawn from Gaussian distributions centred around $\rho_0$ = 3 AU, $\sigma$ = 0.5, $\lambda$ = 4 and $f$ = 0.1, with standard deviations of 0.1 AU, 0.01, 0.1 and 0.01, respectively.

Let $N_*$ be the number of objects in the observed sample and $d$ the number of binaries detected in that sample. For each set of parameters generated by the MCMC tool, a synthetic population of $n = 10^5$ companions is drawn from the lognormal and power law distributions in projected separation and mass ratio, respectively. Each simulated companion (with separation $\rho$ and mass ratio $q$) is then injected into the detection probability map for the survey (see Section~\ref{prob_map}) to get the probability $p_i$ that such a companion would have been retrieved in the observations. Assuming a binary rate $f$ for the sample studied, the total number of companions expected to be detected, $k$, for the achieved detection limits is given by:
\begin{equation} \label{eq:Ndetections}
    k = \sum_{i=1}^{n}(p_i) \times f \times \frac{N_*}{n} .
\end{equation}
The obtained value for $k$ may then be compared to the number of binaries $d$ detected in the observed data in order to estimate the likelihood of the data for a given a set of model parameters. We used Poisson statistics to define the likelihood function $\mathcal{L}$:
\begin{equation} \label{eq:poisson}
    \mathcal{L}(d \mid k) = \frac{k^{d}\: e^{-k}}{d!} ,
\end{equation}
where $k$, the mean expected number of detections for the model parameters considered (given by Eq.~\ref{eq:Ndetections}) is the mean of the Poisson probability mass function. Eq.~\ref{eq:poisson} thus gives the probability of detecting $d$ companions given that an average of $k$ binaries are expected to be detected if the binary population in the observed sample is described by parameters $\rho_0$, $\sigma$, $\gamma$ and $f$.

For a survey with a null detection, the code explores all four population parameters throughout the ranges of allowed values but only really allows for the investigation of the binary frequency $f$. In that case, the returned posterior probability distribution for the binary fraction may be used to determine an upper limit for $f$ that is most compatible with the observed data, marginalised over the other three model parameters.

\begin{figure}
    \addtocounter{figure}{-1}
    \centering   
    \begin{subfigure}{0.45\textwidth}
        \includegraphics[width=\textwidth]{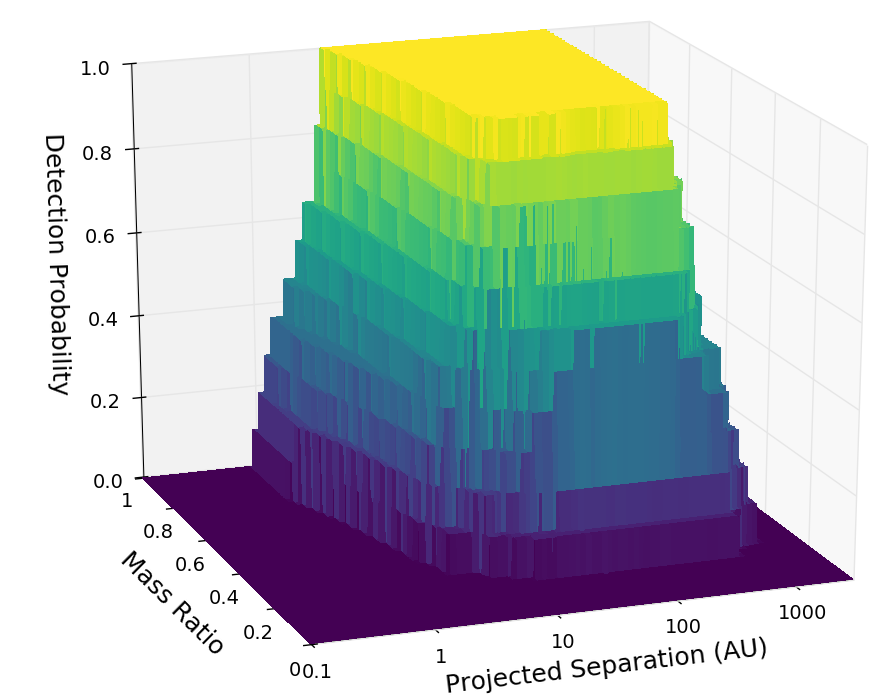}
    \end{subfigure}
    \vspace{2mm}
    \begin{subfigure}{0.45\textwidth}
        \includegraphics[width=\textwidth]{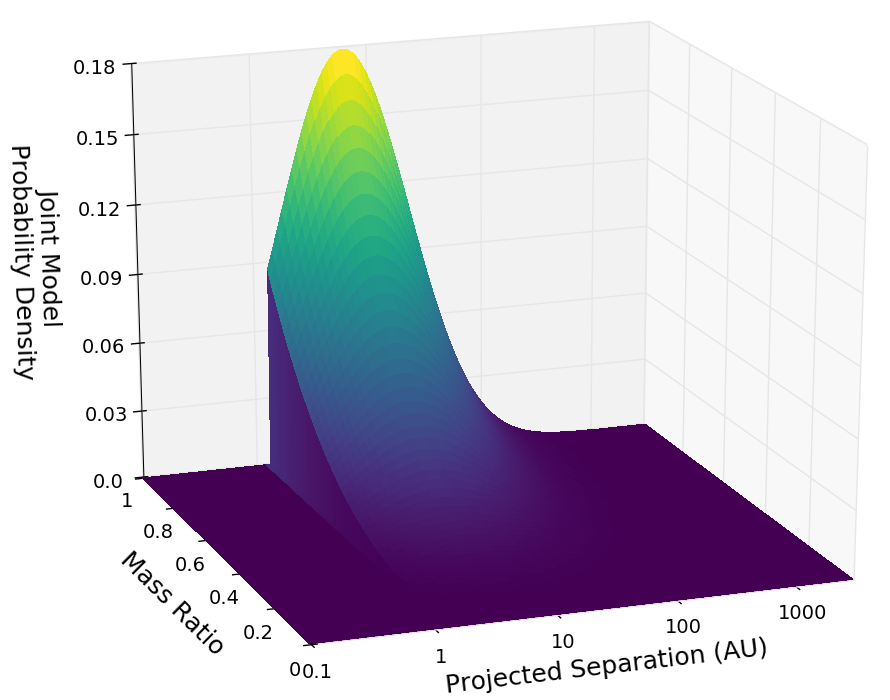}
    \end{subfigure}
    \vspace{2mm}
    \begin{subfigure}{0.45\textwidth}
        \includegraphics[width=\textwidth]{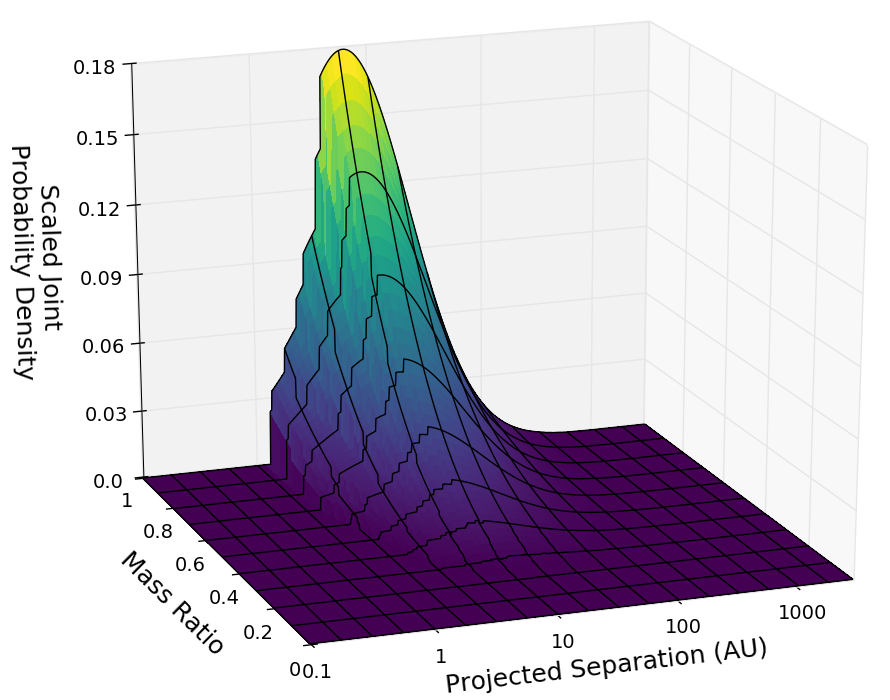}
    \end{subfigure}
    \vspace{1mm}
  \caption{Calculation of the probabilities of observing specific binary systems for given model parameters and achieved sensitivity limits. \textbf{Top:} Detection probability for our observed sample (same as Figure~\ref{f:detection_map_SNAP}). \textbf{Middle:} Joint model probability density of companion distribution assuming populations described by $\rho_0$ = 4 AU, $\sigma$ = 0.5 and $\gamma$ = 3. The separation distribution was truncated at 1.5 AU and 1000 AU. \textbf{Bottom:} Same as middle panel scaled by the detection probability from the top panel. This provides the companion distribution expected to be observed for the achieved limits and model parameters considered. The probability of observing a system with separation $\rho$ and mass ratio $q$ is given by the volume under the final density function for the region delimited by the black lines which encompasses the observed $\rho$ and $q$ values.}
    \label{f:distributions}
\end{figure}

In cases where one or more companions are present in the sample studied, the separations and mass ratios of the observed companions may be taken into account in the MCMC code to further constrain the remaining three parameters. This is done by estimating the probabilities of the detected companions being drawn from the model distributions considered at any step. Sensitivity limits must be taken into account when computing these probabilities in order to truly compare the model to the data and account for observational biases. As our detection limits vary throughout the parameter space, the distributions of companions expected to be observed differ from the model distributions. In order to compare the model to our data, we must transform the parameter space of the model to an observed one. The top panel in Figure~\ref{f:distributions} shows the detection probability at every point in the separation-mass ratio space for our observed survey. The middle panel shows the joint model companion distribution assuming model parameters $\rho_0$ = 4 AU, $\sigma$ = 0.5 and $\gamma$ = 3, truncated at 1.5 AU. The bottom panel shows the same 2-dimensional density function mapped onto the observed parameter space, that is, multiplied by the detection probability in every point. This provides an expected observed distribution of companions for the model parameters considered, given the achieved detection limits.

The $\log(\rho)$ space is then divided into bins of 0.25 and the $q$ domain into bins of 0.1, as shown by the black lines in the bottom panel of Figure~\ref{f:distributions}. The probability of observing any given companion is found by estimating the probability of $\rho$ and $q$ falling in the region enclosing the observed parameters, given by the volume under the scaled density function in that region.
The code computes this probability for the projected separation and mass ratio of every companion detected in the observations. The likelihood $\mathcal{L}$ from Eq.~\ref{eq:poisson} is then multiplied by each of the returned probabilities. The product of all individual probabilities is larger when the observed distributions of companions are well approximated by the scaled model distributions. As a result, this allows the algorithm to favour model distributions from which the observed data were more likely to be drawn, while accounting for detection limits and preventing a bias towards better sampled regions of the parameter space.
The likelihood obtained at each step of the ensemble therefore provides us with the probability of seeing the observed data if the companion population is described by the model parameters considered at that step. The MCMC then uses Bayes' Theorem (Eq.~\ref{eq:Bayes}) and the provided prior distributions to compute the probability of a set of model parameters given the observed data set. The output posterior distributions generated by the sampler finally return the probability density function for each parameter that is most compatible with the observed data.

\subsection{Results}
\label{results}

\subsubsection{Observed sample}
\label{results_SNAP}

The MCMC sampling tool described in Section~\ref{MCMC} was applied to our observed HST sample to investigate the brown dwarf binary rate of our survey. At projected separations of 2 AU, we are sensitive to near equal-mass binaries around $\sim$80\% of our observed sample, to $q>0.8$ companions around half of our targets, and down to $q\sim0.6$ for $\sim$40\% for our sample. Given the known preference for high mass ratios in field substellar binaries, we consider 2 AU a suitable lower limit on the separation range reliably accessible to the observations. As a result, the lognormal distribution in projected separation was truncated so as to only explore the separation range 2$-$1000 AU. The code was run with $2\times10^3$ walkers taking $5\times10^3$ steps each. We found that $\sim$50 steps were sufficient for the sampler to expand from the initial positions to a reasonable sampling of the parameter space and to get settled around the maximum density regions. We thus discarded the initial 50 steps of the ``burn-in'' phase and considered the rest of the samples as representative of the posterior densities. A mean acceptance rate (fraction of steps accepted for each walker) of 0.38 was reached after a few hundred steps. \citet{Foreman-Mackey2013} suggest as a rule of thumb that the acceptance fraction should be between 0.2 and 0.5 and we trust the obtained value to be an acceptable sign of convergence. However, larger samples were required in order to obtain smooth output distributions and be able define confidence intervals for the posterior probability functions. The final number of walkers and steps chosen was found to be a good compromise between the need for a high number of iterations and the expensive associated computing times, while providing a stable acceptance rate within the preferred range.

\begin{figure}
    \centering
    \includegraphics[trim={0.5cm 0.2cm 0.5cm 0.5cm}, width=0.47\textwidth]{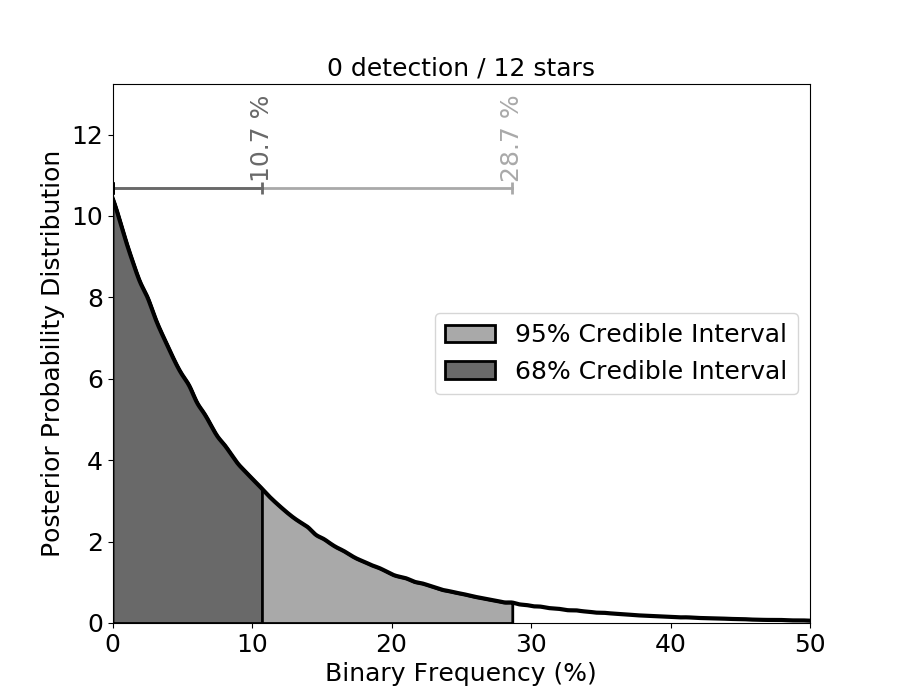}
    \caption{Posterior probability distribution of the binary frequency $f$ for our observed sample of 12 targets obtained with the MCMC sampler described in Section~\ref{MCMC}. The null detection in the program allows us to place an upper limit of $f<10.7$\% (1-$\sigma$) on the binary fraction of our survey at separations $>$2 AU.}
    \label{f:freq_SNAP}
\end{figure}

The null detection of our survey did not allow us to place any new constraints on the separation and mass ratio distributions of $\geqslant$T8 brown dwarf binaries. We were however able to investigate the binary fraction $f$ of our observed sample. Figure~\ref{f:freq_SNAP} shows the posterior probability distribution for the binary frequency $f$ of our survey, given the observations. With no new detected companion in the observed sample, we were only able to place an upper limit on the observed binary rate. We used a highest posterior density approach to determine the boundaries of a Bayesian credible interval for the output posterior distribution. For a given level of credibility $\alpha$, we can define a credible interval bounded by $f_\mathrm{min}$ and $f_\mathrm{max}$ as the shortest interval that contains a fraction $\alpha$ of the probability. This can be thought of as a horizontal line placed over the posterior density intersecting the posterior in $f_\mathrm{min}$ and $f_\mathrm{max}$ such that the region between these two values has a probability $\alpha$. If there is no detection, like in our observed program, the posterior density for $f$ is a one-tail distribution (Figure~\ref{f:freq_SNAP}) and $f_\mathrm{min}$ = 0. The highest density region approach has the useful property that any point within the interval has a higher probability than any other point of the posterior (for a unimodal distribution), thus providing a collection of the most likely values of the parameter.
We consider the $\alpha$ = 68\% and $\alpha$ = 95\% credible intervals, which correspond to 1-$\sigma$ and 2-$\sigma$ Gaussian limits, respectively. Using this approach, we inferred a binary frequency of $f<10.7$\% ($<28.7$\%) at the 1-$\sigma$ (2-$\sigma$) confidence level for our observed sample on separations between 2$-$1000 AU.

\begin{figure}
    \centering
    \includegraphics[trim={0.5cm 0.2cm 0.5cm 0.2cm},  width=0.47\textwidth]{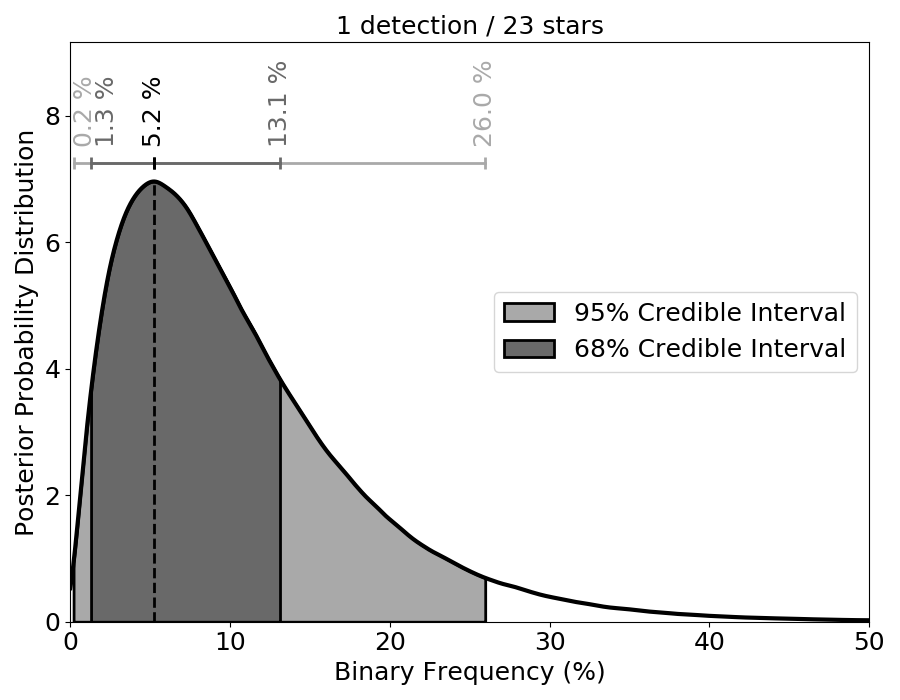}
    \caption{Posterior probability distribution of the binary frequency of T8$-$Y0 brown dwarfs on the separation range 1.5$-$1000 AU for the extended sample of 23 objects, combining our observed HST program and the $\geqslant$T8 sources from the surveys in \citet{Gelino2011} and \citet{Aberasturi2014}. The detection of one companion in the additional subset provides well-defined 1-$\sigma$ and 2-$\sigma$ confidence intervals around a most likely value for the observed binary fraction, $f = 5.2\%$.}
    \label{f:freq_late-Ts}
\end{figure}

\begin{figure}
    \centering
    \includegraphics[trim={0.5cm 0.2cm 0.5cm 0.5cm}, width=0.47\textwidth]{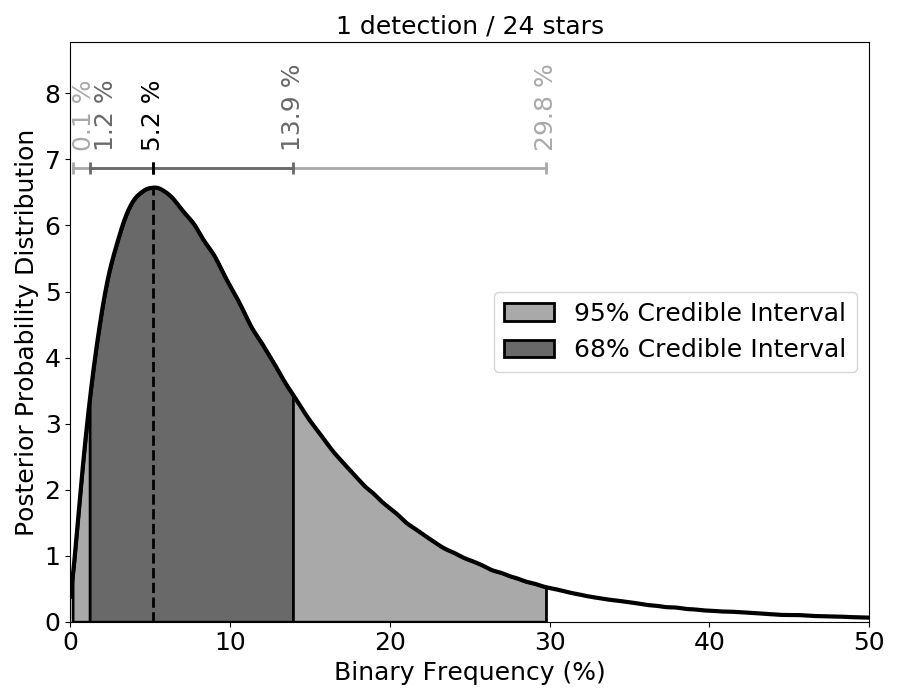}
    \caption{Same as Figure~\ref{f:freq_late-Ts} for the sample of 24 T5$-$T7.5 brown dwarfs compiled from the surveys in \citet{Gelino2011} and \citet{Aberasturi2014}. The obtained distribution is very similar to that obtained for the sample of late-T objects for the same separation range (1.5$-$1000 AU).}
    \label{f:freq_mid-Ts}
\end{figure}

\subsubsection{Extended sample}
\label{results_late-Ts}

To improve our statistics, we performed the same analysis on an extended sample, combining our observed sample with the additional subset of $\geqslant$T8 objects presented in Section~\ref{extended_sample}. We used the MCMC sampling tool described in Section~\ref{MCMC} to run the same statistical analysis on the extended sample of 23 objects as that applied to our observed HST sample. The detection probability map for the combined survey of late-T brown dwarfs (shown in Figure~\ref{f:detection_map_late-Ts}) was used as an input for the code. We defined the same prior distributions and initial walker positions as for our observed sample in Section \ref{results_SNAP}. With the slightly improved average inner working angle of the additional subset, we are sensitive to smaller separations and thus explored the separation range 1.5$-$1000 AU for the extended sample. The binary separation measured in \citet{Gelino2011} for W0458$+$6434AB together with our derived mass ratio for the system were used as additional inputs when computing the likelihood for each set of model parameters. The output posterior probability distribution for the binary rate $f$ is shown in Figure~\ref{f:freq_late-Ts}. While a single detection was still insufficient to reliably constrain the companion distributions in separation and mass ratio, the presence of one binary in the additional subset allowed us to place new limits on the measured binary fraction. The sampler returned a smooth distribution peaking at 5.2\%, the most likely value for $f$ given the observed data. Confidence intervals were inferred from the output distribution following the approach described in Section~\ref{results_SNAP}, yielding a binary frequency of $f_\mathrm{T8-Y0} = 5.2 ^{+7.9}_{-3.9}$ ($^{+20.8}_{-5.0}$)\% at the 1-$\sigma$ (2-$\sigma$) level for separations $>$1.5 AU.

\subsubsection{Comparison sample}
\label{results_mid-Ts}

The MCMC sampling tool detailed in Section~\ref{MCMC} was then run on the mid-T sample presented in Section~\ref{comparison_sample} in order to compare the results obtained for late-Ts to the mid-T binary population. We used the same input parameters (number of walkers and steps, prior distributions, initial walker positions) as those used for our observed and extended late-T samples to constrain the binary rate over separations of 1.5$-$1000 AU. The detection probability map shown in Figure~\ref{f:detection_map_mid-Ts} and the properties the binary system W1841$+$7000AB were used as inputs to compute the posterior probabilities of the parameters describing the underlying companion population distributions.
As for our $\geqslant$T8 sample, a single detection was not sufficient to confidently constrain the separation and mass ratio distributions. The output posterior distribution for the binary frequency $f$ is shown in Figure~\ref{f:freq_mid-Ts}. We inferred a binary fraction of $f_\mathrm{T5-T7.5} = 5.2 ^{+8.7}_{-4.0}$\% at the 1-$\sigma$ level ($5.2 ^{+24.6}_{-5.1}$\% at the 2-$\sigma$ level) for T5$-$T7.5 field brown dwarfs.
The results obtained for the $>1.5$ AU binary rate of mid-Ts are comparable to those derived for $\geqslant$T8 brown dwarfs in Section~\ref{results_late-Ts} for the same separation range.

\subsubsection{Combined mid and late-T samples}
\label{results_all-Ts}

\begin{figure}
    \centering
    \includegraphics[trim={0.5cm 1cm 1cm 1.1cm}, width=0.5\textwidth]{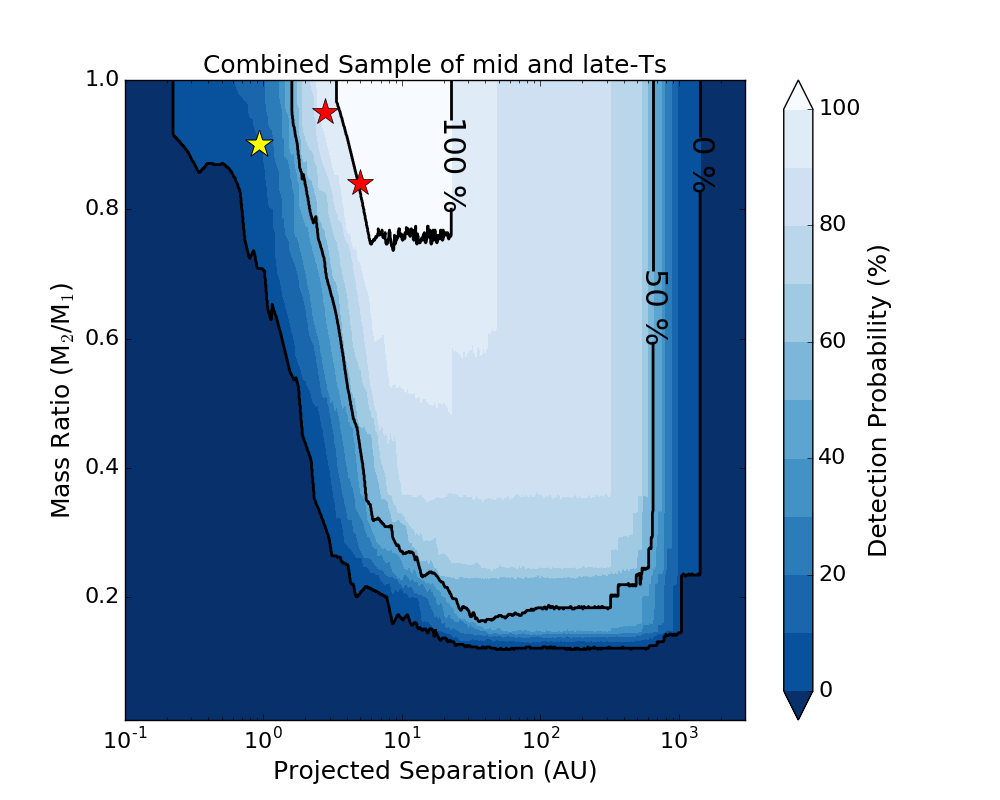}
    \caption{Detection probability map for the combined samples of mid and late-T brown dwarfs, compiled from our observed program and the surveys from \citet{Gelino2011} and \citet{Aberasturi2014}. Contours denote the 0\%, 50\% and 100\% completeness regions. The red stars show the positions of the two binaries discovered by \citet{Gelino2011}, using the component masses calculated in this work. The yellow star corresponds to the known unresolved W0146$+$4234AB system, located in the 0$-$10\% detection probability region of the full sample.}
    \label{f:detection_map_all-Ts}
\end{figure}

\begin{figure}
    \centering
    \includegraphics[trim={0.5cm 0.2cm 0.5cm 0.7cm}, width=0.47\textwidth]{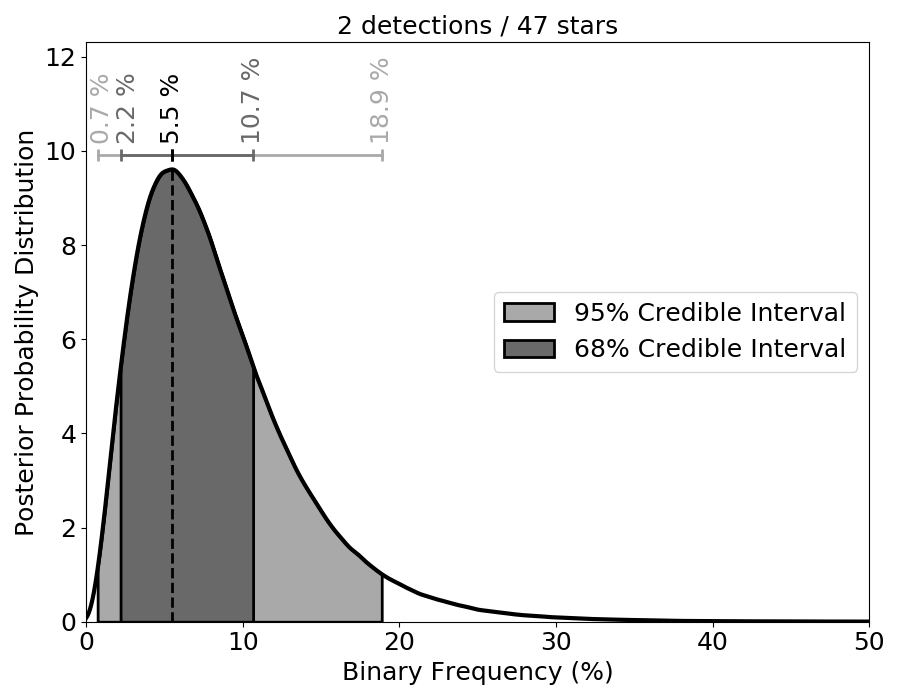}
    \caption{Posterior probability distribution of the $>$1.5 AU binary frequency of $\geqslant$T5 brown dwarfs for the combined sample of 47 objects compiled from our program and the surveys in \citet{Gelino2011} and \citet{Aberasturi2014}.}
    \label{f:freq_all-Ts}
\end{figure}

\begin{figure*}
    \centering
    \includegraphics[width=0.74\textwidth]{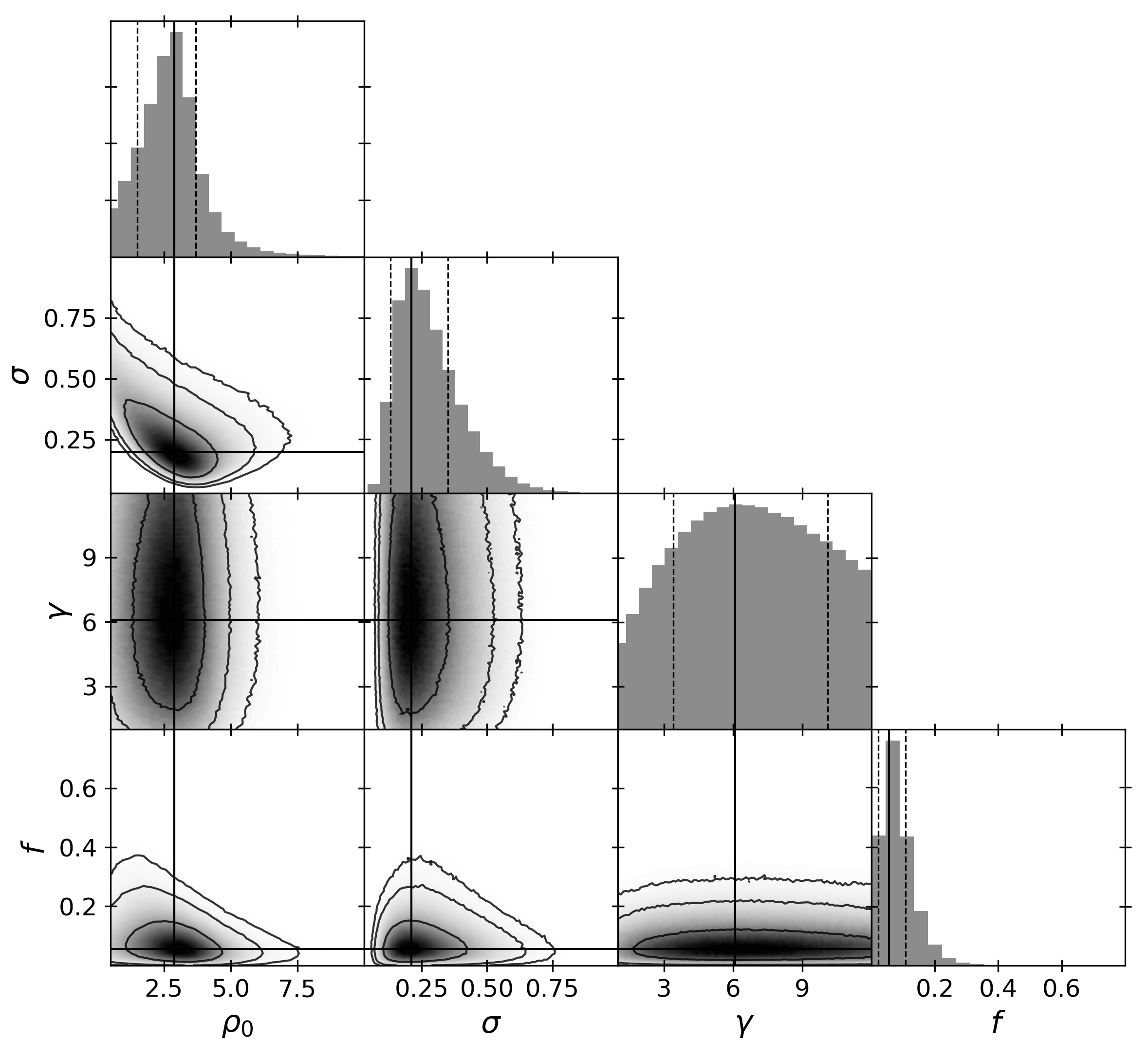}
    \caption{Marginalised posterior probability distributions of all binary parameters from our MCMC analysis (diagonal) and correlation among all pairs of parameters (triangle plot). Normalised histograms at the ends of rows are marginalised over all other parameters. In histograms, solid lines show the best-fit values and dashed lines show the 68\% (1-$\sigma$) credible intervals calculated using a highest density region approach. The black contour lines in the correlation plots correspond to regions containing 68\%, 95\% and 99\% of the posterior.}
    \label{f:corner_plot}
\end{figure*}

As the binary fractions of our compiled samples of T5$-$T7.5 and $\geqslant$T8 brown dwarfs are in excellent agreement, we may combine the two samples so as to more tightly constrain the binary frequency of the $\geqslant$T5 substellar population. The detection probability map for the full sample of 47 objects is shown in Figure~\ref{f:detection_map_all-Ts}, together with the respective positions of the two binaries from \citet{Gelino2011} (red stars) and the unresolved binary W0146$+$4234 (yellow star). Our MCMC tool was used to perform the same statistical analysis on the combined sample as that applied to the individual subsets in previous sections. With a larger sample size and a total of two binary systems, we were able to strengthen the constraints placed separately on mid and late-Ts. The posterior probability distribution for the binary fraction of $\geqslant$T5 brown dwarfs is presented in Figure~\ref{f:freq_all-Ts}. We inferred a binary frequency of $f_\mathrm{T5-Y0} = 5.5 ^{+5.2}_{-3.3} (^{+13.4}_{-4.8})$ \% at the 1-$\sigma$ (2-$\sigma$) credibility level for the full sample of $\geqslant$T5 objects on separations 1.5$-$1000 AU. The peak of the output distribution for $f$ was found to be close to the values obtained for the individual mid and late-T subsets and well within their respective 1-$\sigma$ credible intervals. The increased sample size and the presence of two binary companions in the final sample provided additional constraints to the MCMC tool, resulting in a sharper output posterior distribution for $f$ and narrower credible intervals than those obtained for the separate samples.

The presence of two companions in the combined sample also allowed us to constrain the parameters describing the separation and mass ratio distributions. The full output from our MCMC analysis is presented in Figure~\ref{f:corner_plot}. The best-fit values for the binary parameters of T5$-$Y0 brown dwarfs on separations in the range 1.5$-$1000 AU are: $f_\mathrm{T5-Y0} = 5.5^{+5.2}_{-3.3}$\%, $\rho_0 = 2.9^{+0.8}_{-1.4}$ AU, $\sigma = 0.21^{+0.14}_{-0.08}$ and $\gamma = 6.1^{+4.0}_{-2.7}$, where the errors correspond to 68\% confidence intervals, estimated using the highest density region approach described previously. The power law index $\gamma$ is the only parameter that was not strongly constrained by the MCMC sampler. While the remaining parameters converge to a sharply-defined peak in the posterior distributions, a wider range of possible values was found by the MCMC tool for the power law index, as a result of our attempt to fit a power law through only two data points.

All other MCMC parameters show some covariance in Figure~\ref{f:corner_plot}. In particular, we note that the marginalised posterior distributions for $\rho_0$ and $\sigma$ are asymmetric, with $\rho_0$ having a larger probability density in the lower tail (at values smaller than the best-fit peak value), while $\sigma$ shows a broader upper tail. This reflects the observational incompleteness at very tight separations and shows that the code successfully attempts to fit a companion population with a significant fraction of unresolved systems.
In terms of the correlation between the two parameters, we observe that, as expected, a peak at smaller separations requires a broader width in the lognormal distribution in order to reproduce the observed data. With a best-fit value for $\rho_0$ at 2.9 AU, our sample is nevertheless found to be more compatible with a peak inside our resolution limits than with a population of preponderantly unresolved systems lying just below the probed separation range. We are sensitive to equal-mass binaries around half of our sample from 1.5 AU and down to $\sim$0.5$-$1 AU for a third of the targets. With resolved binaries at 2.8 AU and 5 AU, respectively, and one unresolved system at 0.93 AU, a peak in separation around $\rho_0$ = 0.5$-$2 AU should statistically have resulted in the detection of more tightly-bound binaries, which we do not detect despite the achieved sensitivity limits. While this does not exclude the possibility of secondary peak at even smaller separations (e.g. $\sim$0.2 AU), it strongly suggests that we are not seeing the edge of a distribution peaking just outside our completeness region and the resolution limits of direct imaging surveys (typically 1$-$3 AU), as it has been speculated in the literature (e.g. \citealp{Burgasser2007}).

\subsubsection{False negative analysis and overall binary fraction}
\label{false_neg}

\citet{Dupuy2015} found W0146$+$4234 to be a tight, near-equal mass binary with a projected angular separation of $87.5 \pm 2.1$ mas, corresponding to a projected physical separation of $0.93^{+0.12}_{-0.16}$ AU. The target was part of the original sample for our HST observations but is not resolved in the WFC3 data due to the large plate scale of the IR channel (130 mas pixel$^{-1}$). As a result, we treated this object as an unresolved single source in our analysis, like all other targets with no detected companion. We may however use the fact that W0146$+$4234 is a known binary as a further validation of our results. We can indeed use the posterior distributions from our MCMC analysis to estimate the number of binaries expected to be uncovered or missed within a fixed separation range and check that it is consistent with our observations and additional knowledge of the sample.

We started by estimating the number of companions we expected to detect or miss in our HST program in the 2$-$1000 AU separation range reliably probed by our observations. This was done by selecting $10^5$ random walkers and steps from our final MCMC run in Section \ref{results_all-Ts} (Figure~\ref{f:corner_plot}) to draw sets of values for $f$, $\rho_0$, $\sigma$ and $\gamma$. This allows for the correlation between the different parameters to be taken into account, which is not possible by drawing values from the marginalised posterior distributions. Following the method implemented in our MCMC tool, we then generated a synthetic population of $10^5$ companions based on the drawn $\rho_0$, $\sigma$ and $\gamma$ values. The simulated companions were injected into the detection probability map for our observed sample (see Figure~\ref{f:detection_map_SNAP}) to find the probability $p_i$ that each simulated companion would have been retrieved in our data. The probability that a companion remains undetected is then given by 1-$p_i$.
As the binary fraction $f$ from the MCMC output was computed from separations of 1.5 AU, this parameter had to be corrected to an equivalent 2$-$1000 AU binary fraction. The required scaling factor was found by calculating the ratio of the areas under the separation distribution (defined by the drawn $\rho_0$ and $\sigma$ values) over the two separation ranges considered (2$-$1000 and 1.5$-$1000 AU in this case).
We finally used Eq.~\ref{eq:Ndetections} with the adjusted binary fraction and the obtained $p_i$ (1-$p_i$) values to estimate the number of companions we expected to detect (miss) in our HST program from separations of 2 AU.

\begin{small}
\begin{table*}
\caption{Comparison between predictions from our MCMC output and the observed binary population over various separation ranges. $N$ is the size of the sample considered. The number of companions predicted by the MCMC analysis corresponds to the most likely value and the range given in square brackets represents the 68\% confidence interval, using a highest posterior density approach. Cases for which only an upper limit is provided correspond to the 68\% highest density region of a one-tail distribution peaking at or near 0.}
\begin{tabular}{l c c c c c c c}
\hline\hline
Sample & $N$ & Separation range & \multicolumn{3}{c}{Predictions from MCMC analysis} & \multicolumn{2}{c}{Observations} \\
\cmidrule(lr){4-6} \cmidrule(lr){7-8}
& &  (AU) & Total expected & Detectable & Undetectable & Detected & Undetected \\
\hline
\multirow{2}{*}{Observed program} & \multirow{2}{*}{12} & 0.1$-$2  & 0.3 [0.0$-$0.6]  & $<$ 0.1  & 0.3 [0.0$-$0.5]   & 0 & $\geqslant$ 1 \\
  & & 2$-$1000 & 0.6 [0.2$-$1.2]  & 0.5 [0.2$-$1.0] & $<$ 0.2  & 0 & ... \\[0.2cm]
\multirow{2}{*}{T5$-$Y0 sample} & \multirow{2}{*}{47} & 0.1$-$1.5  & 0.4 [0.0$-$0.8]  & $<$ 0.1   & 0.3 [0.0$-$0.7]  & 0 & $\geqslant$ 1 \\
  & & 1.5$-$1000 & 2.4 [1.2$-$5.0]  & 2.2 [1.2$-$4.5] & 0.2 [0.0$-$0.5] & 2 & ... \\
\hline
\label{t:n_detections}
\end{tabular}
\end{table*}
\end{small}

The obtained results are presented in Table~\ref{t:n_detections}. We found that for the obtained parameter distributions, a total of $\sim$0.6 objects out of our 12 targets should be a $>$2 AU binary. Taking into account our detection limits, we expected to detect an average of 0.5 systems, and found that we are missing out on less than 0.2 companions. This is consistent with our null detection, suggesting that 0$-$1 binaries were to be uncovered in our program, and confirms the excellent completeness of our survey on these separations.
We then extrapolated our posterior distributions to smaller separations and carried the same analysis for the separation range 0.1$-$2 AU, scaling again the drawn binary frequency accordingly. We found that we would have expected up to 0.1 companions to be retrieved at those tight separations for our 12 targets, and that $\sim$0.3 binaries may remain unresolved (see Table~\ref{t:n_detections}). This is in good agreement with our null detection at small separations (see Section \ref{PSF sub}) and with the presence of the known 0.93 AU binary W0146$+$4234AB. As we know that W0146$+$4234 is a binary system and our results do not predict more than 0.3$\pm$0.3 companions (68\% confidence) in total around our 12 objects within 2 AU, these results suggest that the remaining targets in our sample are unlikely to be binaries.

We performed the same analysis on the combined sample of mid and late-Ts and list our results in Table~\ref{t:n_detections}. For the full separation range 0.1$-$1000 AU, the obtained binary parameters predict a total of $\sim$2.8 binaries among the 47 objects, consistent with the 3 known binaries in the full sample. We found that on the probed separation range (1.5$-$1000 AU), around 2.4 companions were expected to be retrieved given our sensitivity limits, and around 0.2 likely remain undetected. This is in excellent agreement with the 2 resolved binaries from \citet{Gelino2011} and suggests that we unlikely missed more than 0$-$1 binaries on these separations. Similarly, at separations $<$1.5 AU, less than 0.1 companions were expected to be retrieved, while about $\sim$0.3 undetected binaries may still lie in the data (Table~\ref{t:n_detections}). This is again consistent with the lack of detection on these separations and with the presence of the unresolved W0146$+$4234 binary system at 0.93 AU.

Finally, from this analysis we found a binary rate ranging from 0$-$4\% (68\% confidence level) on the separation range 0.1$-$1.5 AU for the full sample of 47 targets, with a mean around $\sim$2\%. Assuming that our derived separation and mass ratio distributions hold at such small separations, this brings the overall (0.1$-$1000 AU) binary fraction of T5$-$Y0 brown dwarfs to an estimated $f_\mathrm{tot} = 8\pm6$\% (1-$\sigma$ level). It is important to emphasise that these values rely entirely on the assumption that the resolved population may be extrapolated onto the unseen part of the parameter space. Looking at separations $>$10 AU, we found the wide binary fraction for late-T and Y brown dwarfs to be below $\sim$1\%.

\section{Discussion}
\label{discussion}

\subsection{The binary frequency of ultracool brown dwarfs}

Given the broad pixel scale of the WFC3/IR channel and the emerging evidence for preferred tight orbits for late-type binaries \citep{Burgasser2003,Burgasser2006}, we expected $\sim$0$-$2 binaries to be uncovered around the 12 targets probed in our survey. The absence of new discoveries is consistent with the current census for the binary properties of the latest-type brown dwarfs. With the inclusion of additional subsets from \citet{Gelino2011} and \citet{Aberasturi2014}, we were able to constrain the binary fraction of $\geqslant$T8 brown dwarf binaries with separations $>$1.5 AU to $f_\mathrm{T8-Y0} = 5.2^{+7.9}_{-3.9}\%$ at the 1-$\sigma$ level, placing the first statistically robust constraints to date on the binary frequency of the very coolest (T$_\mathrm{eff}$ $<$ 800 K), lowest-mass ($<$ 40 M$_\mathrm{Jup}$) known brown dwarfs.

The sample of T5$-$T7.5 brown dwarfs gathered in Section~\ref{comparison_sample} has a comparable size to the extended $\geqslant$T8 subset and our statistical analysis uncovered a binary frequency of $f_\mathrm{T5-T7.5} = 5.2 ^{+8.7}_{-4.0}$\% on separations $>$1.5 AU at the 1-$\sigma$ level for this sample. While the results obtained for the mid and late-T samples are statistically consistent, these constraints are based on small number statistics and larger sample sizes are required to confirm and further constrain the substellar multiplicity fraction at such late spectral types. Combining the two samples into a larger $\geqslant$T5 sample, we were able to more tightly constrain the binary rate of T5$-$Y0 ultracool brown dwarfs at separations of 1.5$-$1000 AU to $f_\mathrm{T5-Y0} = 5.5 ^{+5.2}_{-3.3}$\% at the 1-$\sigma$ level. Using the outputs of our MCMC analysis, we extrapolated the inferred population distributions down to 0.1 AU to derive an overall binary fraction of $f_\mathrm{tot} = 8\pm6$\% (1-$\sigma$) for T5 to Y0 brown dwarfs.

Our results are consistent with those obtained by \citet{Aberasturi2014} for T5$-$T8.5 objects. For similar population distributions to those used in this work (power law in mass ratio and lognormal in separation) that study set an upper limit on the total binary rate of 17\% at the 95\% confidence level. This is in good agreement with the 2-$\sigma$ upper limit ($\sim$20\%) derived here for the overall binary rate of $\geqslant$T5 brown dwarfs at the same confidence level. \citet{Opitz2016} searched for close-in, near-equal mass companions to five Y brown dwarfs (including one of our science targets, W0713$-$2917) and found no evidence of binarity down to separations of $\sim$0.5$-$1.9 AU. The lack of uncovered binary system in that study is consistent with the binary statistics established here for very late-type T and Y ultracool brown dwarfs.

\subsection{Decreasing binary fraction with spectral type}

There is evidence that stellar binary pairs with later-type primaries decline in number and have closer separations and more equal mass ratios (\citealp{Duquennoy1991, Fischer1992, Delfosse2004, Kouwenhoven2007, Raghavan2010}). High-resolution imaging surveys in the substellar regime found substantially lower binary rates than in the stellar population and this decrease of binary fraction with primary mass is also observed to persist throughout the brown dwarf regime. Recent studies probing M field stars (\citealp{Fischer1992,Bergfors2010}) concluded that M-dwarfs constitute a smooth intermediate stage in binary properties between higher-mass stars and brown dwarfs and even suggest a trend of decreasing binary fraction with stellar mass within just the M-star spectral range \citep{Janson2012}. In the substellar regime, \citet{Reid2006} investigated the binary properties of 52 M8$-$L7.5 ultracool field dwarfs, probing separations down to 1.5 AU, and found an observed binary fraction of $12^{+7}_{-3}$\%, for an overall, bias-corrected L-dwarf binary frequency of $24^{+6}_{-2}$\%, assuming a lognormal distribution in separation and a power law in mass ratio. These results are consistent with prior surveys for late-M and L brown dwarfs in the field ($15\pm7$\% for M8.0$-$L0.5, \citealp{Close2003}; $15\pm5$\% for M8$-$L8, \citealp{Gizis2003}) for separations $>$1$-$3 AU.

Figure~\ref{f:BF_SpT} shows the binary fraction of solar-type stars, low-mass stars and brown dwarfs as a function of spectral type in the Galactic field. While some values in Figure~\ref{f:BF_SpT} only represent an observed binary frequency defined over a certain range of separations or companion masses (open symbols), and may be missing a significant fraction of binaries, the filled symbols are considered to be ``overall'' binary fractions and clearly highlight the trend of a declining binary rate with spectral type. The separation and mass ratio ranges over which each data point in Figure~\ref{f:BF_SpT} was estimated are plotted in Figure~\ref{f:comp_distributions}. Our results (red) strongly support the idea that the continuously decreasing binary fraction with decreasing primary mass persists down to the very latest spectral types.

\begin{figure}
    \centering
    \includegraphics[width=0.45\textwidth]{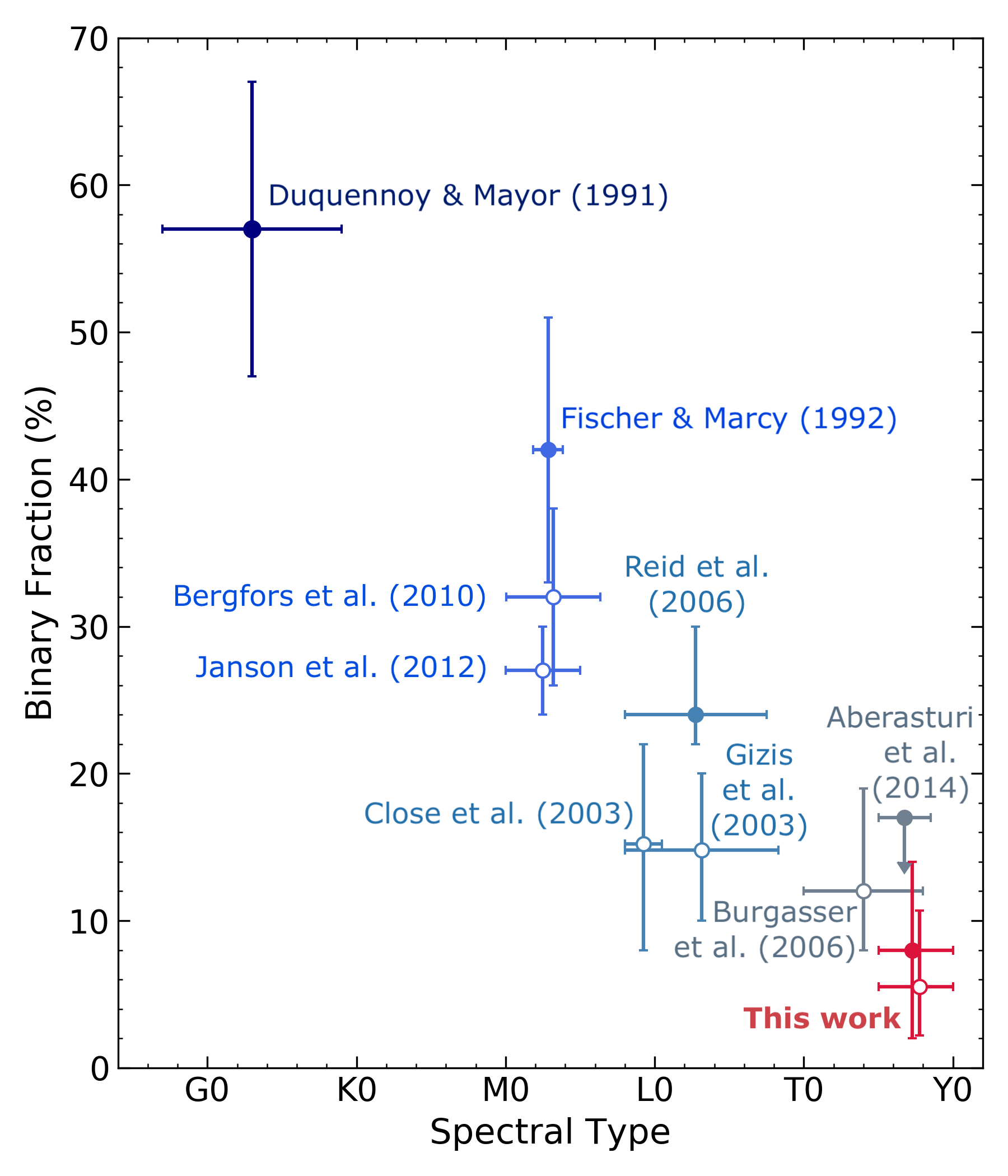}
    \caption{Stellar and substellar binary fractions as a function of the spectral type in the Galactic field showing a clear decline in binary frequency with decreasing primary mass. Filled symbols correspond to total binary fractions, estimated over complete ranges of separations and mass ratios. The $>$1.5 AU (open symbol) and overall (filled symbol) binary frequencies determined in this work for the T5$-$Y0 population are shown in red. The data point from \citet{Aberasturi2014} corresponds to the upper limit estimated in that paper for a lognormal distribution in separation and a power law in mass ratio, similar to the distributions considered here. Error bars on binary fractions correspond to 1-$\sigma$ uncertainties, except for the value from \citet{Aberasturi2014} which corresponds to a 2-$\sigma$ confidence level (see text). The ranges of separations and mass ratios considered in each survey are plotted in Figure~\ref{f:comp_distributions}. Some points were slightly shifted to make the figure clearer.}
    \label{f:BF_SpT}
\end{figure}

\begin{figure*}
    \addtocounter{figure}{-1}
    \centering   
    \begin{subfigure}{0.47\textwidth}
    \includegraphics[width=\textwidth]{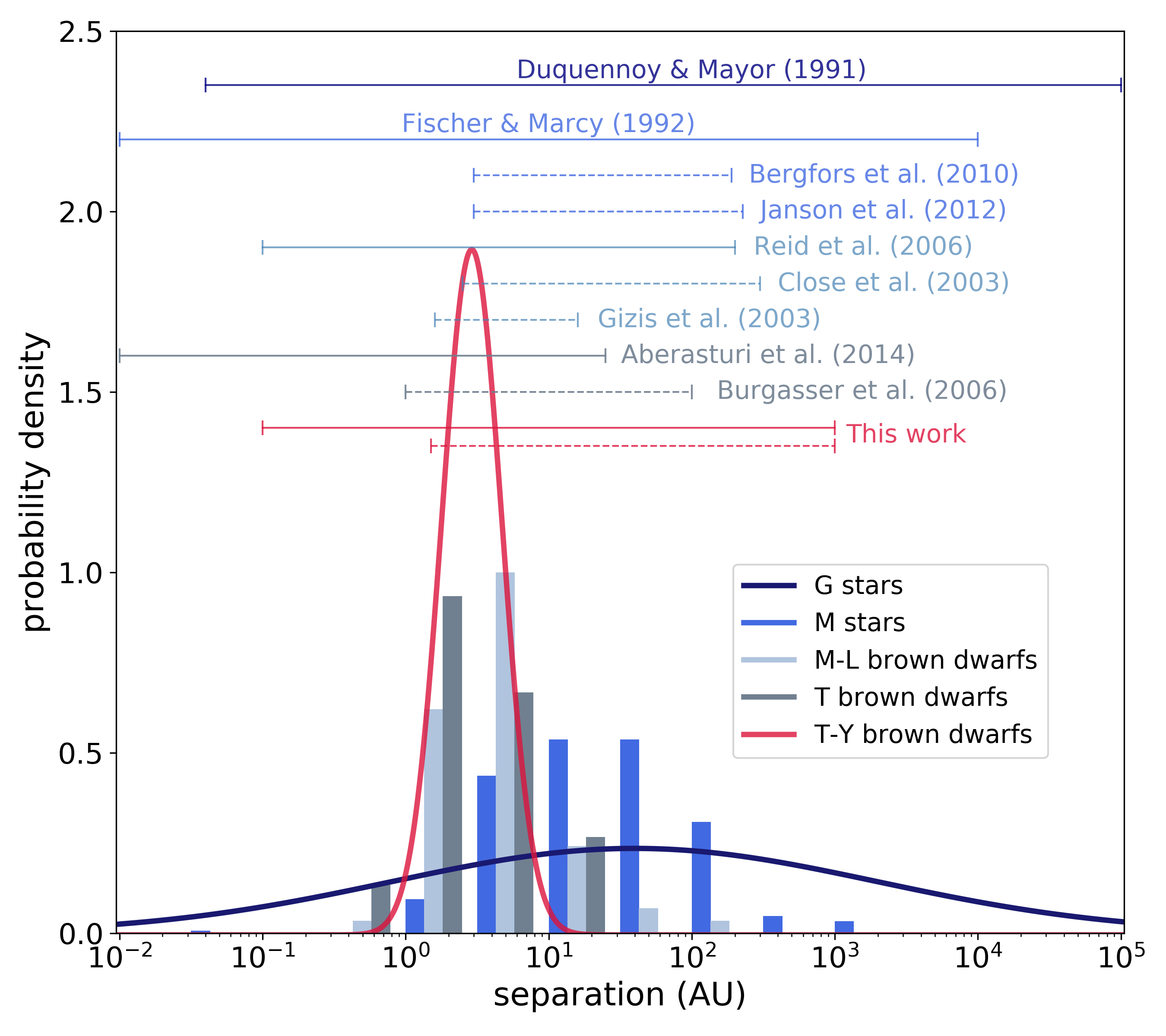}
    \end{subfigure}
    \begin{subfigure}{0.47\textwidth}
    \includegraphics[width=\textwidth]{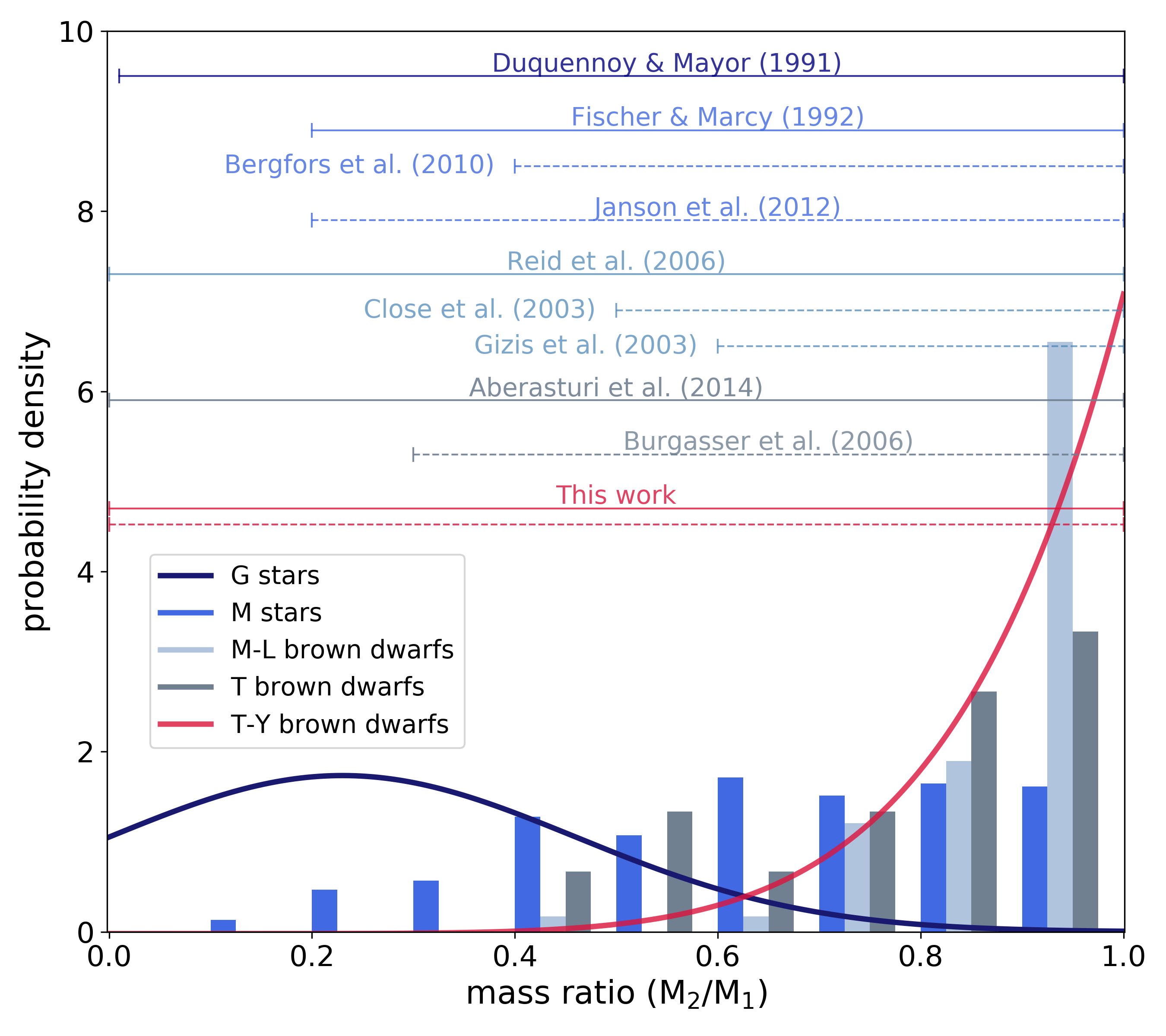}
    \end{subfigure}
    \caption{Separation (left) and mass ratio (right) distributions for companions to stellar and substellar objects, showing the clear shift towards smaller orbital separations and higher mass ratios around lower-mass primaries. The data used to compile the figure are described in the text. The distributions derived in this work for late-T and Y brown dwarfs are shown in red. The horizontal lines show the ranges considered for each of the binary frequencies plotted in Figure~\ref{f:BF_SpT}, where the solid lines represent surveys that estimated overall binary fractions.}
    \label{f:comp_distributions}
\end{figure*}

At the lower end of the substellar mass range, \citet{Burgasser2006} searched for companions to 22 T0$-$T8 nearby ultracool dwarfs and estimated a bias-corrected binary fraction of $12^{+7}_{-4}$\% for the observed sample. Of the five binaries detected in that program, we note that two systems were subsequently determined via spectral decomposition to have L-type primary components. As a result, we argue that the strictly-defined T0$-$T8 binary fraction of that sample (see Figure~\ref{f:BF_SpT}) should be lower than the derived value. In addition, only one out of 13 $\geqslant$T5 targets was identified as a binary, leaving a total of two T0$-$T4.5 binaries out of seven objects. Although these are small number statistics and are not meaningful without a thorough statistical analysis, this also points towards the idea of a lower binary fraction at later spectral types within the T spectral sequence. \citet{Aberasturi2014} determined total binary fractions of $<$16$-$25\% (95\% confidence level) for T5$-$T8.5 brown dwarfs, with an upper limit of 17\% assuming similar population shapes to those used in this work (lognormal in separation and power law in mass ratio). The uncertainties in the binary fractions from the studies mentioned above were generally defined as Poisson errors, corresponding to 1-$\sigma$ Gaussian intervals. We must therefore keep in mind that the data point from \citet{Aberasturi2014} in Figure~\ref{f:BF_SpT} is not directly comparable to the other values plotted here and that the 1-$\sigma$ upper limit for that study resides at a lower value.

In this paper, we established a binary frequency of $5.5 ^{+5.2}_{-3.3}$\% for the T5$-$Y0 population on the separation range 1.5$-$1000 AU. Based on the output of our statistical analysis, we inferred a corresponding overall binary frequency of $8\pm6$\% for $\geqslant$T5 ultracool dwarfs, placing the furthermost points along the spectral type sequence in Figure~\ref{f:BF_SpT}. Our results appear to be consistent with the idea of a decreasing substellar binary frequency with spectral type in the Galactic field. The trend seen in the stellar population is thus believed to continue across and throughout the brown dwarf mass regime and to persist all the way down to the lowest-mass and coolest late-T and Y brown dwarfs.

\subsection{Companion separation and mass ratio distributions}
\label{companion_distributions}

With only one binary in the extended sample of T8$-$Y0 objects, we were not able to place any new constraints on the separation or mass ratio distributions of companions to $\geqslant$T8 field objects. The only detection (from \citealp{Gelino2011}) is inside the 100\% completeness region of the combined survey (see Figure~\ref{f:detection_map_late-Ts}). As a result, we cannot distinguish between a truly low binary fraction and population distributions peaking outside the dynamic and resolution ranges covered by the survey. The same is true of the only mid-T binary present in the T5$-$T7.5 sample, located at a 83\% detection probably level for that sample. Combining the two subsets, on the other hand, allowed us to improve the constraints obtained in the individual analyses. From our MCMC run performed for the full sample of 47 objects, we inferred a peak in separation at $\rho_0 = 2.9^{+0.8}_{-1.4}$ AU with a logarithmic width of $\sigma = 0.21^{+0.14}_{-0.08}$, and a power law index of $\gamma = 6.1^{+4.0}_{-2.7}$ for the mass ratio distribution. As discussed in Section~\ref{false_neg}, these results are perfectly consistent with the presence of the known tight binary, W0146$+$4234AB, unresolved in our data, and based on our detection limits, suggest that we are unlikely to be missing more than one other binary companion in total for the 47 targets.

The derived parameters for the underlying population distributions are in good agreement with results obtained in previous studies. Numerous surveys have observed changes in the shapes of companion distributions with primary spectral type, with a separation distribution peaking at closer separations and mass ratios shifting towards unity for lower-mass objects. Figure~\ref{f:comp_distributions} shows the separation and mass ratio distributions of companions to Sun-like stars, low-mass stars and brown dwarfs, clearly showing the shift in companion distributions with primary mass. The data for the stars come from the distributions derived in \citet{Duquennoy1991} for G-stars and the binaries in \citet{Fischer1992}, \citet{Bergfors2010} and \citet{Janson2012} for M-stars. For late-M and L dwarfs, we considered all field systems from the Very Low Mass Binaries Archive\footnote{\url{http://www.vlmbinaries.org}} with primary masses $<$0.1 M$_\odot$. The T-dwarf histogram used data from table 8 in \citet{Huelamo2015} for 15 confirmed T-binaries (see references therein). The T$-$Y distributions plotted in red correspond to the distributions derived in this work for T5$-$Y0 brown dwarfs. The horizontal lines show the ranges of separations and mass ratios over which the binary fractions in Figure~\ref{f:BF_SpT} were estimated. Studies that estimated an ``overall'' binary fraction are represented with a solid line (corresponding to the filled symbols in Figure~\ref{f:BF_SpT}). As shown in Figure~\ref{f:comp_distributions}, the physical and dynamical ranges considered in these surveys cover a sufficiently large part of the separation-mass ratio space to provide what we consider a complete view of the underlying binary population, assuming that the plotted distributions are representative of the true companion populations.

We note that the data for G and M stars in Figure~\ref{f:comp_distributions} contain a mix of semi-major axes and projected separations. \citet{Dupuy2011} computed conversion factors ($a/\rho$) between projected separations ($\rho$) and semi-major axes ($a$) for Solar-type stars and very low-mass binaries for various cases of discovery biases. The majority of the stellar binaries considered here have separations larger than the inner working angle of the discovery observations. We can thus assume that we are likely in the ``no discovery bias'' case for these binaries, leading to corresponding conversion factors ranging from $\sim$0.8$-$2. Nevertheless, we argue that this is not a major concern since shifting the separation distributions in Figure~\ref{f:comp_distributions} by such factors would not significantly affect the overall shapes of the distributions in logarithmic space and the results discussed above would still hold. The separation values for the histograms of brown dwarfs, on the other hand, correspond to projected separations only and are directly comparable to our results as we worked in observed projected separation space in the analysis carried in this paper.

\citet{Duquennoy1991} found that companions to Sun-like stars show a broad peak centred around $\sim$30 AU and exhibit a continuous increase towards small secondary masses down to the hydrogen-burning limit (Figure~\ref{f:comp_distributions}). \citet{Fischer1992} observed a comparably broad distribution around M-stars, with a peak around 3$-$30 AU, at slightly closer separations than G-stars. That study revealed a roughly flat companion mass function (down to the substellar boundary), similar to the field mass function at low masses.
In the brown dwarf regime, \citet{Reid2006} derived a lognormal distribution in semi-major axis with a peak at $\sim$6.3 AU and a standard deviation of 0.3 for late-M and L objects. This corresponds to a peak in projected separation ($\rho_0$) around $\sim$5$-$8 AU, using the conversion factors from \citet{Dupuy2011} for visual very low-mass binaries. The inferred width of the logarithmic Gaussian represents a much narrower distribution than for stellar primaries, reinforcing the idea of a tighter binary population in the substellar regime. \citet{Reid2006} report a strong preference for high mass ratios among M and L substellar binaries, with a best-fit of $3.6\pm1$ for the power law index. Studies by \citet{Allen2007} and \citet{Burgasser2007} obtained similar results for M-L-T objects and confirmed the tighter separations and higher mass ratios of brown dwarf binary systems relative to the stellar field population.
In this paper, we constrained the companion projected separations of T5$-$Y0 objects to follow a lognormal distribution peaking around 2.9 AU, with a logarithmic width of $\sim$0.21. The obtained distribution is in very good agreement with the observed projected separations of the known T binaries shown in Figure~\ref{f:comp_distributions}, which peak around $\sim$3 AU. Our results predict a slightly narrower distribution than the T-dwarf histogram in Figure~\ref{f:comp_distributions} due to a couple of wider ($>$10 AU) systems present in those data. The power law index derived here for the mass ratio distribution of companions to $\geqslant$T5 objects is also consistent with the brown dwarf data collected for Figure~\ref{f:comp_distributions}. Our results thus support the idea of a steady shift towards tighter orbital configurations and more equal mass ratios with decreasing primary mass.

\subsection{Effects of observational biases and incompleteness}

We comment on the fact that the survey is biased by the limited sensitivity at the lowest separations and flux ratios. Our results are based on an analysis of the resolved binary population of brown dwarfs, which we extrapolated to the unseen part of the observed parameter-space. The presence of a significant number of undetected binaries could therefore result in considerably different companion populations.

For the low primary masses of our observed targets the achieved mass ratio limits correspond to secondary masses of $\sim$5$-$10 M$_\mathrm{Jup}$ for adopted ages of 5 Gyr. Despite an apparent strong preference for near equal-mass configurations \citep{Allen2007,Burgasser2007}, we cannot exclude the possibility of an undetected population of very low mass companions, even if a bimodal mass ratio distribution seems unlikely in the current context of formation models.
Similarly, the observed peak of the separation distribution for resolved field binaries ($\sim$4 AU; \citealp{Allen2007, Burgasser2007}) is close to the resolving limit of our survey, like for most direct imaging programs. It is possible that this observational feature is a direct consequence of the imaging resolution limit rather than a real peak and a significant fraction of very tight binaries could still remain undetected (see \citealp{Burgasser2007}).

For instance, the close T9$+$Y0 binary W0146$+$4234AB discovered by \citet{Dupuy2015} was part of our observed HST program but the tight $87.5\pm2.1$ mas angular separation of the system did not allow us to resolve the two components in the images. Our analysis was thus limited and biased by our resolution limits at small separations. While our results proved to be consistent with the presence of the unresolved W0146$+$4234 binary system in our data, they do not predict many more missing binaries and are not compatible with a significant number of very short-orbit binaries (down to $\sim$0.5 AU), if such a population exists. We believe that given the completeness level of the survey down to separations of $\sim$1 AU, the MCMC sampler would have converged towards such a population had it been compatible with our observed data. Despite allowing $\rho_0$ to take values down to 0.3 AU in our MCMC runs, we found a best-fit value of $2.9^{+0.8}_{-1.4}$ AU at the 68\% confidence level, with a sharply-defined peak and no sign of bimodal distribution over the probed range. While these results do not rule out the possibility of a substantial, secondary population of tighter binaries below $\sim$0.5 AU, they do suggest that the data from our compiled sample of 47 objects is not highly compatible with a separation distribution peaking just outside the resolving limit of imaging programs. We discuss the unresolved binary fraction further in Section~\ref{tight_binaries}.

\subsection{The frequency of unresolved binaries}
\label{tight_binaries}

\citet{Burgasser2007} suggested that substellar binaries with separations $<$3 AU may be as frequent or possibly even more prevalent than currently known resolved systems. The true peak of the separation distribution for brown dwarf binary pairs could thus lie below the current resolving limit of imaging surveys. A number of alternative detection methods are available to probe the shortest-orbit binaries, currently unreachable with high angular resolution imaging. Radial velocity measurements, monitoring of astrometric variability and the spectral binary technique are all sensitive to very close separation systems and may provide a valuable and robust insight into the unresolved binary fraction.

\citet{Bardalez2015} found evidence that spectral binary searches are starting to uncover a significant population of tight binaries. Out of a sample of 33 spectral binary candidates, the authors report 3 resolved binaries and 5 known binary systems that remain unresolved, suggesting a high ratio of unresolved-to-resolved binaries among this sample.
An accurate measurement of the unresolved binary fraction must take into account the occurrence of successful spectral binary candidacy for the brown dwarf population. For example, \citet{Bardalez2014} investigated spectra of 738 objects from the SpeX Prism Library, from which only 35 were retained as spectral binary candidates based on visual or spectral index selection. From these, 14 were found to likely be binaries after spectral fitting. If all final candidates are confirmed, this would lead to a spectral binary rate of only $\sim$2\% ($14/738$), although this method is mainly conceived to retrieve M/L$+$T binaries. Likewise, \citet{Kellogg2017} identified 30 out of 420 M, L and T dwarfs as candidate spectral binaries based on spectral index criteria and via spectral fitting. This places an upper limit of $\sim$7\% for the binary fraction of this sample. However, as that study is a spectroscopic survey of brown dwarfs with unusual colours, the sample is thus biased towards spectral binaries. The authors also note the possibility of contaminants from highly variable T-dwarfs that may resemble spectral binaries (e.g. 2MASS J21392676+0220226 was originally classified as a L8.5$+$T3.5 binary \citep{burgasser2007b,Burgasser2010} but was later identified as a high-amplitude variable by \citet{Radigan2012}). Further studies of unbiased samples are thus required to confirm the occurrence of unresolved spectral binaries relative to the resolved binary population.

\citet{Sahlmann2014} searched for astrometric signatures of giant planets around M8$-$L2 dwarfs. The authors derived a binary fraction of $\sim$10\% within $\sim$1 AU and placed an upper limit of 9\% on the occurrence of planets with masses $>$5 M$_\mathrm{Jup}$ in the separation range 0.01$-$0.8 AU. Given the $\sim$15\% resolved binary frequency of late-M to early-L brown dwarfs \citep{Close2003}, the results from this work could reflect a population of very tight binaries comparable to that of resolved systems.
\citet{Basri2006} conducted a radial velocity survey targeting mid-M to late-L field dwarfs. This study revealed a binary fraction of $\sim$11\% at separations 0$-$6 AU and after accounting for the overlap in separation range between their survey and the direct imaging programs, the authors conclude that their results are consistent with an estimated overall binary fraction of $\sim$20\%. For a resolved binary rate of $\sim$15\% \citep{Close2003, Gizis2003} from separations of $\sim$2$-$3 AU, this suggests an unresolved binary fraction of $\sim$5\% for M and L dwarfs.
\citet{Blake2010} searched for substellar and giant planetary companions to field brown dwarfs and found a rate of $2.5^{+8.6}_{-1.6}$\% for $<$1 AU binaries among late-M and L objects. In young regions, \citet{Joergens2008} found that the binary fraction of low-mass stars and brown dwarfs within 1 AU is $<$10 \% and that radial velocity programs do not reveal an excess of companions at closer separations. The authors concluded that direct imaging surveys do not miss a significant fraction of brown dwarf binaries and that the observed decrease in binary frequency with stellar mass is also confirmed at separations $<$3 AU.

These results are consistent with those obtained by direct imaging programs in \citet{Allen2007} and \citet{Burgasser2007} for M-L-T dwarfs that estimated binary fractions of 3$-$4\% and 2$-$3\% within 1 AU, respectively, by extrapolating results from the resolved binary population. Extending the outputs of our Bayesian analysis to the unresolved separation range, we inferred in this paper a binary fraction at separations $<$1.5 AU of $\sim$2$\pm$2\% for the full T5$-$Y0 sample in good agreement with the values cited above.
Overall, these results point towards an unresolved binary fraction of about $\sim$20$-$60\% that of resolved systems, although large discrepancies remain between various surveys and methods. We may therefore regard the observed peak around $\sim$3$-$4 AU for substellar binaries as a real feature and conclude that the declining binary frequency with spectral type is not a result of the shrinking separation distribution and observational incompleteness. Further studies with a reliable sensitivity at these small orbital periods are however required to confirm these results, which are strongly limited by the small statistics of the unresolved substellar binary population currently available.

\subsection{The dearth of wide binaries in the field}

We note that no wide ($>$10 AU separation) binary was uncovered in our observed program or around the targets probed by \citet{Gelino2011} and \citet{Aberasturi2014}. From the posterior distributions of our MCMC analysis on the full T5$-$Y0 sample we estimated that $<$1\% of mid-T to Y brown dwarfs are found in binary systems with orbital separation $>$10 AU. This is in good agreement with empirical estimates in the literature (e.g. \citealp{Allen2007,Burgasser2007}), that agree on a wide binary fraction of $\sim$1\% at separation $>$15$-$20 AU for the M-L-T dwarf field population. While direct imaging provides weak constraints on closely-separated binaries (see Section~\ref{tight_binaries}), imaging surveys typically have very good completeness levels at separations $>$10 AU out to hundreds of AU and the lack of wide substellar field binaries is a very robust result.

While probing separations $<$10 AU is challenging for young objects due to the large distances to young regions and moving groups, wider separations are accessible for the young substellar population, which may be compared to the observed wide binary population of the field. A number of wide systems with primary masses comparable to our sample have been detected in young ($<$15 Myr) regions at separations of 15$-$800 AU (Taurus, \citealp{Luhman2009,Todorov2010}; Ophiuchus, \citealp{Close2007}; Chamaeleon, \citealp{Luhman2004}; TW Hydra, \citealp{Chauvin2005}) and \citet{Biller2011} confirmed the existence of a statistically significant population of very low mass ($<$0.1 M$_\odot$), wide separation ($>$10-100 AU) binaries in $>$2 Myr star-forming regions (Upper Sco, Taurus, and Chamaeleon). This presents a conundrum, as this wide population is not found around older ($\sim$few Gyr) field brown dwarfs. Young substellar binary systems show a much broader range of separations than their older, field analogues, with separations spanning 3$-$4 orders of magnitude and 25\% of known companions having orbital separations larger than 20 AU \citep{Burgasser2006}. \citet{Bouy2006} and \citet{Close2007} claim wide binary fractions of at least 5\% in young regions, significantly larger than for field objects. Young binaries also show a flatter mass ratio distribution than observed in the Galactic field, with a statistically significant shortfall in $q>0.8$ systems \citep{Burgasser2006}.

While probing star-forming regions facilitates the search for very low-mass objects, which are significantly more luminous at young ages, the majority of these systems have secondary masses of $\sim$5$-$25 M$_\mathrm{Jup}$ that fall within our achieved detection limits ($\sim$2$-$5 M$_\mathrm{Jup}$ at 1 Gyr, $\sim$5$-$10 M$_\mathrm{Jup}$ at 5 Gyr and $\sim$8$-$15 M$_\mathrm{Jup}$ at 10 Gyr). We would therefore most certainly have detected such companions, had they been present around our probed targets. The lack of wide systems in the field agrees with predictions from formation scenarios that only allow very tight systems to survive to field ages \citep{Reipurth2001,Padoan2004,Goodwin2007}. If dynamical evolution processes are responsible for the depletion of such wide, low-mass binaries in the field population, systems such as those discovered by \citet{Liu2012} may simply be uncommon.

The possibility that most field brown dwarfs were born under different conditions than objects from known young star-forming regions, where wide binaries are prevented from forming \citep{Close2007}, must also be considered, as this may hinder a direct comparison between field and young binary populations.
Regions of similar ages but different densities must be probed to determine the effect of natal environment on the subsequent binary rate. \citet{Biller2011} and \citet{Todorov2014} investigated binarity as a function of environment, probing brown dwarfs in Taurus, Chamaeleon and the denser Upper Scorpius association \citep{Preibisch2008} and found comparable binary fractions in Upper Scorpius and the more diffuse clusters. Similarly, \citet{King2012} investigated the multiplicity of low-mass ($>$0.1 M$_\odot$) stars across the same clusters and found no obvious trend over a factor of nearly 20 in density, suggesting that, within the density range encompassed by these regions, natal environment does not significantly affect the formation of low-mass binaries. However, \citet{Lada2003} argued that the field population is mainly dominated by stars that originated from even richer clusters and these results may therefore not be relevant in a direct comparison to field objects.

\section{Summary}
\label{conclusions}

We searched for low-mass companions to 12 nearby brown dwarfs with spectral types of T8 or later using WFC3/IR observations from the Hubble Space Telescope. Our observed sample is one of the largest subsets of very late-type (T$_\mathrm{eff}$ $<$ 800 K) and exclusively low-mass ($<$40 M$_\mathrm{Jup}$) brown dwarfs studied as part of a multiplicity search. We found no evidence for wide binary companions in our survey despite reaching sensitivity limits of 5$-$10 M$_\mathrm{Jup}$ or $q$ $\sim$ 0.2$-$0.4 for ages of 5 Gyr at separations $>$0\farcs5 (3.5$-$10 AU). PSF subtraction did not reveal the presence of tighter binaries, down to separations of $\sim$0\farcs1 (0.7$-$2.5 AU).

From our newly developed statistical tool based on an MCMC sampling method, we inferred an upper limit on the binary frequency of our observed sample of $<$10.7\% (1-$\sigma$) at separations $>$2 AU. Our statistical analysis allows us to marginalise over a range of possible companion population distributions, poorly constrained at the bottom of the substellar mass regime, while taking into account the survey's detection limits to correct for observational biases and incompleteness. Combining our observed program with prior studies, we derived a binary fraction of $f_\mathrm{T8-Y0} = 5.2^{+7.9}_{-3.9}$\% (1-$\sigma$) for the $\geqslant$T8 substellar population on the separation range 1.5$-$1000 AU, placing the first statistically robust constraints to date on the binary fraction of T8$-$Y0 ultracool brown dwarfs. We obtained comparable results for earlier-type T5$-$T7.5 objects ($f_\mathrm{T5-T7.5} = 5.2^{+8.7}_{-4.0}$\%) and further constrained the binary frequency of T5$-$Y0 objects to $f_\mathrm{T5-Y0} = 5.5^{+5.2}_{-3.3}$\% (1-$\sigma$) at separations $>$1.5 AU.

We derived best-fit values of $\rho_0 = 2.9^{+0.8}_{-1.4}$ AU and $\sigma = 0.21^{+0.14}_{-0.08}$ for the peak and logarithmic width of the lognormal distribution in projected separation, and found a power law index of $\gamma = 6.1^{+4.0}_{-2.7}$ for the mass ratio distribution. These outputs support the idea of tighter and higher mass ratio binary systems for lower-mass primaries. From these results, we were able to estimate the overall (0.1$-$1000 AU) binary frequency of T5$-$Y0 brown dwarfs to $f_\mathrm{tot} = 8\pm6$\%, with a 2$\pm$2\% binary rate within 1.5 AU and less than $\sim$1\% beyond 10 AU. Our results are consistent with previous studies and suggest that the decline in binary fraction with decreasing primary mass seen in the field stellar population continues across the substellar mass regime, down to the very coolest and lowest-mass known brown dwarfs.

\section*{Acknowledgements}

We would like to thank our referee, Adam Burgasser, for his very insightful comments and suggestions. This survey is based on observations made with the NASA/ESA Hubble Space Telescope, obtained from the Data Archive at the Space Telescope Science Institute. These observations are associated with program 12873. Some of the data presented in this paper were obtained from the Mikulski Archive for Space Telescopes (MAST). STScI is operated by the Association of Universities for Research in Astronomy, Inc., under NASA contract NAS5-26555. This research has made use of the Keck Observatory Archive (KOA), which is operated by the W. M. Keck Observatory and the NASA Exoplanet Science Institute (NExScI), under contract with the National Aeronautics and Space Administration. This research has benefited from the SpeX Prism Spectral Libraries, maintained by Adam Burgasser at \url{http://pono.ucsd.edu/~adam/browndwarfs/spexprism} and has made use of the Very-Low-Mass Binaries Archive housed at \url{http://www.vlmbinaries.org} and maintained by Nick Siegler, Chris Gelino, and Adam Burgasser.

\bibliographystyle{mnras}

\begin{thebibliography}{}
\makeatletter
\relax
\def\mn@urlcharsother{\let\do\@makeother \do\$\do\&\do\#\do\^\do\_\do\%\do\~}
\def\mn@doi{\begingroup\mn@urlcharsother \@ifnextchar [ {\mn@doi@}
  {\mn@doi@[]}}
\def\mn@doi@[#1]#2{\def\@tempa{#1}\ifx\@tempa\@empty \href
  {http://dx.doi.org/#2} {doi:#2}\else \href {http://dx.doi.org/#2} {#1}\fi
  \endgroup}
\def\mn@eprint#1#2{\mn@eprint@#1:#2::\@nil}
\def\mn@eprint@arXiv#1{\href {http://arxiv.org/abs/#1} {{\tt arXiv:#1}}}
\def\mn@eprint@dblp#1{\href {http://dblp.uni-trier.de/rec/bibtex/#1.xml}
  {dblp:#1}}
\def\mn@eprint@#1:#2:#3:#4\@nil{\def\@tempa {#1}\def\@tempb {#2}\def\@tempc
  {#3}\ifx \@tempc \@empty \let \@tempc \@tempb \let \@tempb \@tempa \fi \ifx
  \@tempb \@empty \def\@tempb {arXiv}\fi \@ifundefined
  {mn@eprint@\@tempb}{\@tempb:\@tempc}{\expandafter \expandafter \csname
  mn@eprint@\@tempb\endcsname \expandafter{\@tempc}}}

\bibitem[\protect\citeauthoryear{{Aberasturi}, {Burgasser}, {Mora}, {Solano},
  {Mart{\'{\i}}n}, {Reid}  \& {Looper}}{{Aberasturi}
  et~al.}{2014}]{Aberasturi2014}
{Aberasturi} M.,  {Burgasser} A.~J.,  {Mora} A.,  {Solano} E.,  {Mart{\'{\i}}n}
  E.~L.,  {Reid} I.~N.,   {Looper} D.,  2014, \mn@doi [\aj]
  {10.1088/0004-6256/148/6/129}, \href
  {http://adsabs.harvard.edu/abs/2014AJ....148..129A} {148, 129}

\bibitem[\protect\citeauthoryear{{Allard}, {Hauschildt}, {Alexander}, {Tamanai}
   \& {Schweitzer}}{{Allard} et~al.}{2001}]{Allard2001}
{Allard} F.,  {Hauschildt} P.~H.,  {Alexander} D.~R.,  {Tamanai} A.,
  {Schweitzer} A.,  2001, \mn@doi [\apj] {10.1086/321547}, \href
  {http://adsabs.harvard.edu/abs/2001ApJ...556..357A} {556, 357}

\bibitem[\protect\citeauthoryear{{Allen}}{{Allen}}{2007}]{Allen2007}
{Allen} P.~R.,  2007, \mn@doi [\apj] {10.1086/521207}, \href
  {http://adsabs.harvard.edu/abs/2007ApJ...668..492A} {668, 492}

\bibitem[\protect\citeauthoryear{{Allers} \& {Liu}}{{Allers} \&
  {Liu}}{2010}]{Allers2010}
{Allers} K.~N.,  {Liu} M.~C.,  2010, in American Astronomical Society Meeting
  Abstracts \#215. p.~335

\bibitem[\protect\citeauthoryear{{Baraffe}, {Chabrier}, {Allard}  \&
  {Hauschildt}}{{Baraffe} et~al.}{2003}]{Baraffe2003}
{Baraffe} I.,  {Chabrier} G.,  {Allard} F.,   {Hauschildt} P.,  2003, in
  {Mart{\'{\i}}n} E.,  ed.,  IAU Symposium Vol. 211, Brown Dwarfs. p.~41

\bibitem[\protect\citeauthoryear{{Bardalez Gagliuffi}, {Gelino}  \&
  {Burgasser}}{{Bardalez Gagliuffi} et~al.}{2015}]{Bardalez2015}
{Bardalez Gagliuffi} D.~C.,  {Gelino} C.~R.,   {Burgasser} A.~J.,  2015,
  \mn@doi [\aj] {10.1088/0004-6256/150/5/163}, \href
  {http://adsabs.harvard.edu/abs/2015AJ....150..163B} {150, 163}

\bibitem[\protect\citeauthoryear{{Basri} \& {Reiners}}{{Basri} \&
  {Reiners}}{2006}]{Basri2006}
{Basri} G.,  {Reiners} A.,  2006, \mn@doi [\aj] {10.1086/505198}, \href
  {http://adsabs.harvard.edu/abs/2006AJ....132..663B} {132, 663}

\bibitem[\protect\citeauthoryear{{Beichman}, {Gelino}, {Kirkpatrick},
  {Cushing}, {Dodson-Robinson}, {Marley}, {Morley}  \& {Wright}}{{Beichman}
  et~al.}{2014}]{Beichman2014}
{Beichman} C.,  {Gelino} C.~R.,  {Kirkpatrick} J.~D.,  {Cushing} M.~C.,
  {Dodson-Robinson} S.,  {Marley} M.~S.,  {Morley} C.~V.,   {Wright} E.~L.,
  2014, \mn@doi [\apj] {10.1088/0004-637X/783/2/68}, \href
  {http://adsabs.harvard.edu/abs/2014ApJ...783...68B} {783, 68}

\bibitem[\protect\citeauthoryear{{Bergfors} et~al.,}{{Bergfors}
  et~al.}{2010}]{Bergfors2010}
{Bergfors} C.,  et~al., 2010, \mn@doi [\aap] {10.1051/0004-6361/201014114},
  \href {http://adsabs.harvard.edu/abs/2010A%26A...520A..54B} {520, A54}

\bibitem[\protect\citeauthoryear{{Biller}, {Allers}, {Liu}, {Close}  \&
  {Dupuy}}{{Biller} et~al.}{2011}]{Biller2011}
{Biller} B.,  {Allers} K.,  {Liu} M.,  {Close} L.~M.,   {Dupuy} T.,  2011,
  \mn@doi [\apj] {10.1088/0004-637X/730/1/39}, \href
  {http://adsabs.harvard.edu/abs/2011ApJ...730...39B} {730, 39}

\bibitem[\protect\citeauthoryear{Biretta}{Biretta}{2014}]{Biretta2014}
Biretta J.,  2014, Space Telescope WFC Instrument Science Report, 1, 10

\bibitem[\protect\citeauthoryear{{Blake}, {Charbonneau}  \& {White}}{{Blake}
  et~al.}{2010}]{Blake2010}
{Blake} C.~H.,  {Charbonneau} D.,   {White} R.~J.,  2010, \mn@doi [\apj]
  {10.1088/0004-637X/723/1/684}, \href
  {http://adsabs.harvard.edu/abs/2010ApJ...723..684B} {723, 684}

\bibitem[\protect\citeauthoryear{{Bouy}, {Mart{\'{\i}}n}, {Brandner},
  {Zapatero-Osorio}, {B{\'e}jar}, {Schirmer}, {Hu{\'e}lamo}  \& {Ghez}}{{Bouy}
  et~al.}{2006}]{Bouy2006}
{Bouy} H.,  {Mart{\'{\i}}n} E.~L.,  {Brandner} W.,  {Zapatero-Osorio} M.~R.,
  {B{\'e}jar} V.~J.~S.,  {Schirmer} M.,  {Hu{\'e}lamo} N.,   {Ghez} A.~M.,
  2006, \mn@doi [\aap] {10.1051/0004-6361:20054252}, \href
  {http://adsabs.harvard.edu/abs/2006A%26A...451..177B} {451, 177}

\bibitem[\protect\citeauthoryear{{Burgasser}}{{Burgasser}}{2007}]{burgasser2007b}
{Burgasser} A.~J.,  2007, \mn@doi [\aj] {10.1086/520878}, \href
  {http://adsabs.harvard.edu/abs/2007AJ....134.1330B} {134, 1330}

\bibitem[\protect\citeauthoryear{{Burgasser}, {Kirkpatrick}, {Reid}, {Brown},
  {Miskey}  \& {Gizis}}{{Burgasser} et~al.}{2003}]{Burgasser2003}
{Burgasser} A.~J.,  {Kirkpatrick} J.~D.,  {Reid} I.~N.,  {Brown} M.~E.,
  {Miskey} C.~L.,   {Gizis} J.~E.,  2003, \mn@doi [\apj] {10.1086/346263},
  \href {http://adsabs.harvard.edu/abs/2003ApJ...586..512B} {586, 512}

\bibitem[\protect\citeauthoryear{{Burgasser}, {Kirkpatrick}, {Cruz}, {Reid},
  {Leggett}, {Liebert}, {Burrows}  \& {Brown}}{{Burgasser}
  et~al.}{2006}]{Burgasser2006}
{Burgasser} A.~J.,  {Kirkpatrick} J.~D.,  {Cruz} K.~L.,  {Reid} I.~N.,
  {Leggett} S.~K.,  {Liebert} J.,  {Burrows} A.,   {Brown} M.~E.,  2006,
  \mn@doi [\apjs] {10.1086/506327}, \href
  {http://adsabs.harvard.edu/abs/2006ApJS..166..585B} {166, 585}

\bibitem[\protect\citeauthoryear{{Burgasser}, {Reid}, {Siegler}, {Close},
  {Allen}, {Lowrance}  \& {Gizis}}{{Burgasser} et~al.}{2007}]{Burgasser2007}
{Burgasser} A.~J.,  {Reid} I.~N.,  {Siegler} N.,  {Close} L.,  {Allen} P.,
  {Lowrance} P.,   {Gizis} J.,  2007, Protostars and Planets V, \href
  {http://adsabs.harvard.edu/abs/2007prpl.conf..427B} {pp 427--441}

\bibitem[\protect\citeauthoryear{{Burgasser}, {Cruz}, {Cushing}, {Gelino},
  {Looper}, {Faherty}, {Kirkpatrick}  \& {Reid}}{{Burgasser}
  et~al.}{2010}]{Burgasser2010}
{Burgasser} A.~J.,  {Cruz} K.~L.,  {Cushing} M.,  {Gelino} C.~R.,  {Looper}
  D.~L.,  {Faherty} J.~K.,  {Kirkpatrick} J.~D.,   {Reid} I.~N.,  2010, \mn@doi
  [\apj] {10.1088/0004-637X/710/2/1142}, \href
  {http://adsabs.harvard.edu/abs/2010ApJ...710.1142B} {710, 1142}

\bibitem[\protect\citeauthoryear{{Burgasser} et~al.,}{{Burgasser}
  et~al.}{2011}]{Burgasser2011}
{Burgasser} A.~J.,  et~al., 2011, \mn@doi [\apj] {10.1088/0004-637X/735/2/116},
  \href {http://adsabs.harvard.edu/abs/2011ApJ...735..116B} {735, 116}

\bibitem[\protect\citeauthoryear{{Burgasser}, {Gelino}, {Cushing}  \&
  {Kirkpatrick}}{{Burgasser} et~al.}{2012}]{Burgasser2012}
{Burgasser} A.~J.,  {Gelino} C.~R.,  {Cushing} M.~C.,   {Kirkpatrick} J.~D.,
  2012, \mn@doi [\apj] {10.1088/0004-637X/745/1/26}, \href
  {http://adsabs.harvard.edu/abs/2012ApJ...745...26B} {745, 26}

\bibitem[\protect\citeauthoryear{{Burningham} et~al.,}{{Burningham}
  et~al.}{2008}]{Burningham2008}
{Burningham} B.,  et~al., 2008, \mn@doi [\mnras]
  {10.1111/j.1365-2966.2008.13885.x}, \href
  {http://adsabs.harvard.edu/abs/2008MNRAS.391..320B} {391, 320}

\bibitem[\protect\citeauthoryear{{Burningham} et~al.,}{{Burningham}
  et~al.}{2013}]{Burningham2013}
{Burningham} B.,  et~al., 2013, \mn@doi [\mnras] {10.1093/mnras/stt740}, \href
  {http://adsabs.harvard.edu/abs/2013MNRAS.433..457B} {433, 457}

\bibitem[\protect\citeauthoryear{{Caloi}, {Cardini}, {D'Antona}, {Badiali},
  {Emanuele}  \& {Mazzitelli}}{{Caloi} et~al.}{1999}]{Caloi1999}
{Caloi} V.,  {Cardini} D.,  {D'Antona} F.,  {Badiali} M.,  {Emanuele} A.,
  {Mazzitelli} I.,  1999, \aap, \href
  {http://adsabs.harvard.edu/abs/1999A%26A...351..925C} {351, 925}

\bibitem[\protect\citeauthoryear{{Chauvin}, {Lagrange}, {Dumas}, {Zuckerman},
  {Mouillet}, {Song}, {Beuzit}  \& {Lowrance}}{{Chauvin}
  et~al.}{2005}]{Chauvin2005}
{Chauvin} G.,  {Lagrange} A.-M.,  {Dumas} C.,  {Zuckerman} B.,  {Mouillet} D.,
  {Song} I.,  {Beuzit} J.-L.,   {Lowrance} P.,  2005, \mn@doi [\aap]
  {10.1051/0004-6361:200500116}, \href
  {http://adsabs.harvard.edu/abs/2005A%26A...438L..25C} {438, L25}

\bibitem[\protect\citeauthoryear{{Close}, {Siegler}, {Potter}, {Brandner}  \&
  {Liebert}}{{Close} et~al.}{2002}]{Close2002}
{Close} L.~M.,  {Siegler} N.,  {Potter} D.,  {Brandner} W.,   {Liebert} J.,
  2002, \mn@doi [\apjl] {10.1086/339795}, \href
  {http://adsabs.harvard.edu/abs/2002ApJ...567L..53C} {567, L53}

\bibitem[\protect\citeauthoryear{{Close}, {Siegler}, {Freed}  \&
  {Biller}}{{Close} et~al.}{2003}]{Close2003}
{Close} L.~M.,  {Siegler} N.,  {Freed} M.,   {Biller} B.,  2003, \mn@doi [\apj]
  {10.1086/368177}, \href {http://adsabs.harvard.edu/abs/2003ApJ...587..407C}
  {587, 407}

\bibitem[\protect\citeauthoryear{{Close} et~al.,}{{Close}
  et~al.}{2007}]{Close2007}
{Close} L.~M.,  et~al., 2007, \mn@doi [\apj] {10.1086/513417}, \href
  {http://adsabs.harvard.edu/abs/2007ApJ...660.1492C} {660, 1492}

\bibitem[\protect\citeauthoryear{{Cushing} et~al.,}{{Cushing}
  et~al.}{2011}]{Cushing2011}
{Cushing} M.~C.,  et~al., 2011, \mn@doi [\apj] {10.1088/0004-637X/743/1/50},
  \href {http://adsabs.harvard.edu/abs/2011ApJ...743...50C} {743, 50}

\bibitem[\protect\citeauthoryear{{Delfosse} et~al.,}{{Delfosse}
  et~al.}{2004}]{Delfosse2004}
{Delfosse} X.,  et~al., 2004, in {Hilditch} R.~W.,  {Hensberge} H.,
  {Pavlovski} K.,  eds,  Astronomical Society of the Pacific Conference Series
  Vol. 318, Spectroscopically and Spatially Resolving the Components of the
  Close Binary Stars. pp 166--174

\bibitem[\protect\citeauthoryear{{Duch{\^e}ne} \& {Kraus}}{{Duch{\^e}ne} \&
  {Kraus}}{2013}]{Duchene2013}
{Duch{\^e}ne} G.,  {Kraus} A.,  2013, \mn@doi [\araa]
  {10.1146/annurev-astro-081710-102602}, \href
  {http://adsabs.harvard.edu/abs/2013ARA%26A..51..269D} {51, 269}

\bibitem[\protect\citeauthoryear{{Duch{\^e}ne}, {Bontemps}, {Bouvier},
  {Andr{\'e}}, {Djupvik}  \& {Ghez}}{{Duch{\^e}ne} et~al.}{2007}]{Duchene2007}
{Duch{\^e}ne} G.,  {Bontemps} S.,  {Bouvier} J.,  {Andr{\'e}} P.,  {Djupvik}
  A.~A.,   {Ghez} A.~M.,  2007, \mn@doi [\aap] {10.1051/0004-6361:20077270},
  \href {http://adsabs.harvard.edu/abs/2007A%26A...476..229D} {476, 229}

\bibitem[\protect\citeauthoryear{{Dupuy} \& {Liu}}{{Dupuy} \&
  {Liu}}{2011}]{Dupuy2011}
{Dupuy} T.~J.,  {Liu} M.~C.,  2011, \mn@doi [\apj]
  {10.1088/0004-637X/733/2/122}, \href
  {http://adsabs.harvard.edu/abs/2011ApJ...733..122D} {733, 122}

\bibitem[\protect\citeauthoryear{{Dupuy} \& {Liu}}{{Dupuy} \&
  {Liu}}{2012}]{Dupuy2012}
{Dupuy} T.~J.,  {Liu} M.~C.,  2012, \mn@doi [\apjs]
  {10.1088/0067-0049/201/2/19}, \href
  {http://adsabs.harvard.edu/abs/2012ApJS..201...19D} {201, 19}

\bibitem[\protect\citeauthoryear{{Dupuy}, {Liu}  \& {Leggett}}{{Dupuy}
  et~al.}{2015}]{Dupuy2015}
{Dupuy} T.~J.,  {Liu} M.~C.,   {Leggett} S.~K.,  2015, \mn@doi [\apj]
  {10.1088/0004-637X/803/2/102}, \href
  {http://adsabs.harvard.edu/abs/2015ApJ...803..102D} {803, 102}

\bibitem[\protect\citeauthoryear{{Duquennoy} \& {Mayor}}{{Duquennoy} \&
  {Mayor}}{1991}]{Duquennoy1991}
{Duquennoy} A.,  {Mayor} M.,  1991, \aap, \href
  {http://adsabs.harvard.edu/abs/1991A%26A...248..485D} {248, 485}

\bibitem[\protect\citeauthoryear{{Faherty}, {Burgasser}, {Cruz}, {Shara},
  {Walter}  \& {Gelino}}{{Faherty} et~al.}{2009}]{Faherty2009}
{Faherty} J.~K.,  {Burgasser} A.~J.,  {Cruz} K.~L.,  {Shara} M.~M.,  {Walter}
  F.~M.,   {Gelino} C.~R.,  2009, \mn@doi [\aj] {10.1088/0004-6256/137/1/1},
  \href {http://adsabs.harvard.edu/abs/2009AJ....137....1F} {137, 1}

\bibitem[\protect\citeauthoryear{{Faherty} et~al.,}{{Faherty}
  et~al.}{2012}]{Faherty2012}
{Faherty} J.~K.,  et~al., 2012, \mn@doi [\apj] {10.1088/0004-637X/752/1/56},
  \href {http://adsabs.harvard.edu/abs/2012ApJ...752...56F} {752, 56}

\bibitem[\protect\citeauthoryear{{Filippazzo}, {Rice}, {Faherty}, {Cruz}, {Van
  Gordon}  \& {Looper}}{{Filippazzo} et~al.}{2015}]{Filippazzo2015}
{Filippazzo} J.~C.,  {Rice} E.~L.,  {Faherty} J.,  {Cruz} K.~L.,  {Van Gordon}
  M.~M.,   {Looper} D.~L.,  2015, \mn@doi [\apj] {10.1088/0004-637X/810/2/158},
  \href {http://adsabs.harvard.edu/abs/2015ApJ...810..158F} {810, 158}

\bibitem[\protect\citeauthoryear{{Fischer} \& {Marcy}}{{Fischer} \&
  {Marcy}}{1992}]{Fischer1992}
{Fischer} D.~A.,  {Marcy} G.~W.,  1992, \mn@doi [\apj] {10.1086/171708}, \href
  {http://adsabs.harvard.edu/abs/1992ApJ...396..178F} {396, 178}

\bibitem[\protect\citeauthoryear{{Foreman-Mackey}, {Hogg}, {Lang}  \&
  {Goodman}}{{Foreman-Mackey} et~al.}{2013}]{Foreman-Mackey2013}
{Foreman-Mackey} D.,  {Hogg} D.~W.,  {Lang} D.,   {Goodman} J.,  2013, \mn@doi
  [\pasp] {10.1086/670067}, \href
  {http://adsabs.harvard.edu/abs/2013PASP..125..306F} {125, 306}

\bibitem[\protect\citeauthoryear{{Fruchter} \& {Hook}}{{Fruchter} \&
  {Hook}}{2002}]{Fruchter2002}
{Fruchter} A.~S.,  {Hook} R.~N.,  2002, \mn@doi [\pasp] {10.1086/338393}, \href
  {http://adsabs.harvard.edu/abs/2002PASP..114..144F} {114, 144}

\bibitem[\protect\citeauthoryear{Gagliuffi et~al.,}{Gagliuffi
  et~al.}{2014}]{Bardalez2014}
Gagliuffi D. C.~B.,  et~al., 2014, The Astrophysical Journal, 794, 143

\bibitem[\protect\citeauthoryear{{Garcia}, {Dupuy}, {Allers}, {Liu}  \&
  {Deacon}}{{Garcia} et~al.}{2015}]{Garcia2015}
{Garcia} E.~V.,  {Dupuy} T.~J.,  {Allers} K.~N.,  {Liu} M.~C.,   {Deacon}
  N.~R.,  2015, \mn@doi [\apj] {10.1088/0004-637X/804/1/65}, \href
  {http://adsabs.harvard.edu/abs/2015ApJ...804...65G} {804, 65}

\bibitem[\protect\citeauthoryear{{Gelino} et~al.,}{{Gelino}
  et~al.}{2011}]{Gelino2011}
{Gelino} C.~R.,  et~al., 2011, \mn@doi [\aj] {10.1088/0004-6256/142/2/57},
  \href {http://adsabs.harvard.edu/abs/2011AJ....142...57G} {142, 57}

\bibitem[\protect\citeauthoryear{{Gizis}, {Reid}, {Knapp}, {Liebert},
  {Kirkpatrick}, {Koerner}  \& {Burgasser}}{{Gizis} et~al.}{2003}]{Gizis2003}
{Gizis} J.~E.,  {Reid} I.~N.,  {Knapp} G.~R.,  {Liebert} J.,  {Kirkpatrick}
  J.~D.,  {Koerner} D.~W.,   {Burgasser} A.~J.,  2003, \mn@doi [\aj]
  {10.1086/374991}, \href {http://adsabs.harvard.edu/abs/2003AJ....125.3302G}
  {125, 3302}

\bibitem[\protect\citeauthoryear{Goodman \& Weare}{Goodman \&
  Weare}{2010}]{Goodman2010}
Goodman J.,  Weare J.,  2010, Communications in applied mathematics and
  computational science, 5, 65

\bibitem[\protect\citeauthoryear{{Goodwin} \& {Whitworth}}{{Goodwin} \&
  {Whitworth}}{2007}]{Goodwin2007}
{Goodwin} S.~P.,  {Whitworth} A.,  2007, \mn@doi [\aap]
  {10.1051/0004-6361:20066745}, \href
  {http://adsabs.harvard.edu/abs/2007A%26A...466..943G} {466, 943}

\bibitem[\protect\citeauthoryear{Hastings}{Hastings}{1970}]{Hastings1970}
Hastings W.,  1970, \mn@doi [Biometrika] {10.1093/biomet/57.1.97}, 57, 97

\bibitem[\protect\citeauthoryear{{Hu{\'e}lamo} et~al.,}{{Hu{\'e}lamo}
  et~al.}{2015}]{Huelamo2015}
{Hu{\'e}lamo} N.,  et~al., 2015, \mn@doi [\aap] {10.1051/0004-6361/201525634},
  \href {http://adsabs.harvard.edu/abs/2015A\%26A...578A...1H} {578, A1}

\bibitem[\protect\citeauthoryear{{Janson} et~al.,}{{Janson}
  et~al.}{2012}]{Janson2012}
{Janson} M.,  et~al., 2012, \mn@doi [\apj] {10.1088/0004-637X/754/1/44}, \href
  {http://adsabs.harvard.edu/abs/2012ApJ...754...44J} {754, 44}

\bibitem[\protect\citeauthoryear{{Joergens}}{{Joergens}}{2008}]{Joergens2008}
{Joergens} V.,  2008, \mn@doi [\aap] {10.1051/0004-6361:200810413}, \href
  {http://adsabs.harvard.edu/abs/2008A%26A...492..545J} {492, 545}

\bibitem[\protect\citeauthoryear{{Kellogg}, {Metchev}, {Miles-P{\'a}ez}  \&
  {Tannock}}{{Kellogg} et~al.}{2017}]{Kellogg2017}
{Kellogg} K.,  {Metchev} S.,  {Miles-P{\'a}ez} P.~A.,   {Tannock} M.~E.,  2017,
  \mn@doi [\aj] {10.3847/1538-3881/aa83b0}, \href
  {http://adsabs.harvard.edu/abs/2017AJ....154..112K} {154, 112}

\bibitem[\protect\citeauthoryear{{King}, {Parker}, {Patience}  \&
  {Goodwin}}{{King} et~al.}{2012}]{King2012}
{King} R.~R.,  {Parker} R.~J.,  {Patience} J.,   {Goodwin} S.~P.,  2012,
  \mn@doi [\mnras] {10.1111/j.1365-2966.2012.20437.x}, \href
  {http://adsabs.harvard.edu/abs/2012MNRAS.421.2025K} {421, 2025}

\bibitem[\protect\citeauthoryear{{Kirkpatrick} et~al.,}{{Kirkpatrick}
  et~al.}{2011}]{Kirkpatrick2011}
{Kirkpatrick} J.~D.,  et~al., 2011, \mn@doi [\apj]
  {10.1088/0067-0049/197/2/19}, \href
  {http://adsabs.harvard.edu/abs/2011ApJS..197...19K} {197, 19}

\bibitem[\protect\citeauthoryear{{Kirkpatrick} et~al.,}{{Kirkpatrick}
  et~al.}{2012}]{Kirkpatrick2012}
{Kirkpatrick} J.~D.,  et~al., 2012, \mn@doi [\apj]
  {10.1088/0004-637X/753/2/156}, \href
  {http://adsabs.harvard.edu/abs/2012ApJ...753..156K} {753, 156}

\bibitem[\protect\citeauthoryear{{Kouwenhoven}, {Brown}, {Portegies Zwart}  \&
  {Kaper}}{{Kouwenhoven} et~al.}{2007}]{Kouwenhoven2007}
{Kouwenhoven} M.~B.~N.,  {Brown} A.~G.~A.,  {Portegies Zwart} S.~F.,   {Kaper}
  L.,  2007, \mn@doi [\aap] {10.1051/0004-6361:20077719}, \href
  {http://adsabs.harvard.edu/abs/2007A%26A...474...77K} {474, 77}

\bibitem[\protect\citeauthoryear{{Kraus} \& {Hillenbrand}}{{Kraus} \&
  {Hillenbrand}}{2012}]{Kraus2012}
{Kraus} A.~L.,  {Hillenbrand} L.~A.,  2012, \mn@doi [\apj]
  {10.1088/0004-637X/757/2/141}, \href
  {http://adsabs.harvard.edu/abs/2012ApJ...757..141K} {757, 141}

\bibitem[\protect\citeauthoryear{{Krist}}{{Krist}}{1995}]{Krist1995}
{Krist} J.,  1995, in {Shaw} R.~A.,  {Payne} H.~E.,   {Hayes} J.~J.~E.,  eds,
  Astronomical Society of the Pacific Conference Series Vol. 77, Astronomical
  Data Analysis Software and Systems IV. p.~349

\bibitem[\protect\citeauthoryear{{Lada} \& {Lada}}{{Lada} \&
  {Lada}}{2003}]{Lada2003}
{Lada} C.~J.,  {Lada} E.~A.,  2003, \mn@doi [\araa]
  {10.1146/annurev.astro.41.011802.094844}, \href
  {http://adsabs.harvard.edu/abs/2003ARA%26A..41...57L} {41, 57}

\bibitem[\protect\citeauthoryear{{Liu} et~al.,}{{Liu} et~al.}{2011}]{Liu2011b}
{Liu} M.~C.,  et~al., 2011, \mn@doi [\apj] {10.1088/0004-637X/740/2/108}, \href
  {http://adsabs.harvard.edu/abs/2011ApJ...740..108L} {740, 108}

\bibitem[\protect\citeauthoryear{{Liu}, {Dupuy}, {Bowler}, {Leggett}  \&
  {Best}}{{Liu} et~al.}{2012}]{Liu2012}
{Liu} M.~C.,  {Dupuy} T.~J.,  {Bowler} B.~P.,  {Leggett} S.~K.,   {Best}
  W.~M.~J.,  2012, \mn@doi [\apj] {10.1088/0004-637X/758/1/57}, \href
  {http://adsabs.harvard.edu/abs/2012ApJ...758...57L} {758, 57}

\bibitem[\protect\citeauthoryear{{Looper}, {Kirkpatrick}  \&
  {Burgasser}}{{Looper} et~al.}{2007}]{Looper2007}
{Looper} D.~L.,  {Kirkpatrick} J.~D.,   {Burgasser} A.~J.,  2007, \mn@doi [\aj]
  {10.1086/520645}, \href {http://adsabs.harvard.edu/abs/2007AJ....134.1162L}
  {134, 1162}

\bibitem[\protect\citeauthoryear{{Luhman}}{{Luhman}}{2004}]{Luhman2004}
{Luhman} K.~L.,  2004, \mn@doi [\apj] {10.1086/423666}, \href
  {http://adsabs.harvard.edu/abs/2004ApJ...614..398L} {614, 398}

\bibitem[\protect\citeauthoryear{{Luhman}}{{Luhman}}{2007}]{Luhman2007}
{Luhman} K.~L.,  2007, \mn@doi [\apjs] {10.1086/520114}, \href
  {http://cdsads.u-strasbg.fr/abs/2007ApJS..173..104L} {173, 104}

\bibitem[\protect\citeauthoryear{{Luhman}, {Mamajek}, {Allen}, {Muench}  \&
  {Finkbeiner}}{{Luhman} et~al.}{2009}]{Luhman2009}
{Luhman} K.~L.,  {Mamajek} E.~E.,  {Allen} P.~R.,  {Muench} A.~A.,
  {Finkbeiner} D.~P.,  2009, \mn@doi [\apj] {10.1088/0004-637X/691/2/1265},
  \href {http://adsabs.harvard.edu/abs/2009ApJ...691.1265L} {691, 1265}

\bibitem[\protect\citeauthoryear{{Mace} et~al.,}{{Mace}
  et~al.}{2013}]{Mace2013}
{Mace} G.~N.,  et~al., 2013, \mn@doi [\apjs] {10.1088/0067-0049/205/1/6}, \href
  {http://adsabs.harvard.edu/abs/2013ApJS..205....6M} {205, 6}

\bibitem[\protect\citeauthoryear{{McLean}, {McGovern}, {Burgasser},
  {Kirkpatrick}, {Prato}  \& {Kim}}{{McLean} et~al.}{2003}]{McLean2003}
{McLean} I.~S.,  {McGovern} M.~R.,  {Burgasser} A.~J.,  {Kirkpatrick} J.~D.,
  {Prato} L.,   {Kim} S.~S.,  2003, \mn@doi [\apj] {10.1086/377636}, \href
  {http://adsabs.harvard.edu/abs/2003ApJ...596..561M} {596, 561}

\bibitem[\protect\citeauthoryear{{Metchev}, {Kirkpatrick}, {Berriman}  \&
  {Looper}}{{Metchev} et~al.}{2008}]{Metchev2008}
{Metchev} S.~A.,  {Kirkpatrick} J.~D.,  {Berriman} G.~B.,   {Looper} D.,  2008,
  \mn@doi [\apj] {10.1086/524721}, \href
  {http://adsabs.harvard.edu/abs/2008ApJ...676.1281M} {676, 1281}

\bibitem[\protect\citeauthoryear{{Metropolis}, {Rosenbluth}, {Rosenbluth},
  {Teller}  \& {Teller}}{{Metropolis} et~al.}{1953}]{Metropolis1953}
{Metropolis} N.,  {Rosenbluth} A.~W.,  {Rosenbluth} M.~N.,  {Teller} A.~H.,
  {Teller} E.,  1953, \mn@doi [\jcp] {10.1063/1.1699114}, \href
  {http://adsabs.harvard.edu/abs/1953JChPh..21.1087M} {21, 1087}

\bibitem[\protect\citeauthoryear{{Opitz}, {Tinney}, {Faherty}, {Sweet},
  {Gelino}  \& {Kirkpatrick}}{{Opitz} et~al.}{2016}]{Opitz2016}
{Opitz} D.,  {Tinney} C.~G.,  {Faherty} J.~K.,  {Sweet} S.,  {Gelino} C.~R.,
  {Kirkpatrick} J.~D.,  2016, \mn@doi [\apj] {10.3847/0004-637X/819/1/17},
  \href {http://adsabs.harvard.edu/abs/2016ApJ...819...17O} {819, 17}

\bibitem[\protect\citeauthoryear{{Padoan} \& {Nordlund}}{{Padoan} \&
  {Nordlund}}{2004}]{Padoan2004}
{Padoan} P.,  {Nordlund} {\AA}.,  2004, \mn@doi [\apj] {10.1086/345413}, \href
  {http://adsabs.harvard.edu/abs/2004ApJ...617..559P} {617, 559}

\bibitem[\protect\citeauthoryear{{Peter}, {Feldt}, {Henning}  \&
  {Hormuth}}{{Peter} et~al.}{2012}]{Peter2012}
{Peter} D.,  {Feldt} M.,  {Henning} T.,   {Hormuth} F.,  2012, \mn@doi [\aap]
  {10.1051/0004-6361/201015027}, \href
  {http://adsabs.harvard.edu/abs/2012A%26A...538A..74P} {538, A74}

\bibitem[\protect\citeauthoryear{{Pinfield} et~al.,}{{Pinfield}
  et~al.}{2014}]{Pinfield2014}
{Pinfield} D.~J.,  et~al., 2014, \mn@doi [\mnras] {10.1093/mnras/stu1540},
  \href {http://adsabs.harvard.edu/abs/2014MNRAS.444.1931P} {444, 1931}

\bibitem[\protect\citeauthoryear{{Preibisch} \& {Mamajek}}{{Preibisch} \&
  {Mamajek}}{2008}]{Preibisch2008}
{Preibisch} T.,  {Mamajek} E.,  2008, {The Nearest OB Association:
  Scorpius-Centaurus (Sco OB2)}.
p.~235

\bibitem[\protect\citeauthoryear{{Radigan}, {Jayawardhana}, {Lafreni{\`e}re},
  {Artigau}, {Marley}  \& {Saumon}}{{Radigan} et~al.}{2012}]{Radigan2012}
{Radigan} J.,  {Jayawardhana} R.,  {Lafreni{\`e}re} D.,  {Artigau} {\'E}.,
  {Marley} M.,   {Saumon} D.,  2012, \mn@doi [\apj]
  {10.1088/0004-637X/750/2/105}, \href
  {http://adsabs.harvard.edu/abs/2012ApJ...750..105R} {750, 105}

\bibitem[\protect\citeauthoryear{{Raghavan} et~al.,}{{Raghavan}
  et~al.}{2010}]{Raghavan2010}
{Raghavan} D.,  et~al., 2010, \mn@doi [\apjs] {10.1088/0067-0049/190/1/1},
  \href {http://adsabs.harvard.edu/abs/2010ApJS..190....1R} {190, 1}

\bibitem[\protect\citeauthoryear{{Reid}, {Gizis}, {Kirkpatrick}  \&
  {Koerner}}{{Reid} et~al.}{2001}]{Reid2001}
{Reid} I.~N.,  {Gizis} J.~E.,  {Kirkpatrick} J.~D.,   {Koerner} D.~W.,  2001,
  \mn@doi [\aj] {10.1086/318023}, \href
  {http://adsabs.harvard.edu/abs/2001AJ....121..489R} {121, 489}

\bibitem[\protect\citeauthoryear{{Reid}, {Lewitus}, {Allen}, {Cruz}  \&
  {Burgasser}}{{Reid} et~al.}{2006}]{Reid2006}
{Reid} I.~N.,  {Lewitus} E.,  {Allen} P.~R.,  {Cruz} K.~L.,   {Burgasser}
  A.~J.,  2006, \mn@doi [\aj] {10.1086/505626}, \href
  {http://adsabs.harvard.edu/abs/2006AJ....132..891R} {132, 891}

\bibitem[\protect\citeauthoryear{{Reipurth} \& {Clarke}}{{Reipurth} \&
  {Clarke}}{2001}]{Reipurth2001}
{Reipurth} B.,  {Clarke} C.,  2001, \mn@doi [\aj] {10.1086/321121}, \href
  {http://adsabs.harvard.edu/abs/2001AJ....122..432R} {122, 432}

\bibitem[\protect\citeauthoryear{{Sahlmann}, {Lazorenko}, {S{\'e}gransan},
  {Mart{\'{\i}}n}, {Mayor}, {Queloz}  \& {Udry}}{{Sahlmann}
  et~al.}{2014}]{Sahlmann2014}
{Sahlmann} J.,  {Lazorenko} P.~F.,  {S{\'e}gransan} D.,  {Mart{\'{\i}}n} E.~L.,
   {Mayor} M.,  {Queloz} D.,   {Udry} S.,  2014, \mn@doi [\aap]
  {10.1051/0004-6361/201323208}, \href
  {http://adsabs.harvard.edu/abs/2014A%26A...565A..20S} {565, A20}

\bibitem[\protect\citeauthoryear{{Schneider} et~al.,}{{Schneider}
  et~al.}{2015}]{Schneider2015}
{Schneider} A.~C.,  et~al., 2015, \mn@doi [\apj] {10.1088/0004-637X/804/2/92},
  \href {http://adsabs.harvard.edu/abs/2015ApJ...804...92S} {804, 92}

\bibitem[\protect\citeauthoryear{{Stephens} \& {Leggett}}{{Stephens} \&
  {Leggett}}{2004}]{Stephens2004}
{Stephens} D.~C.,  {Leggett} S.~K.,  2004, \mn@doi [\pasp] {10.1086/381135},
  \href {http://adsabs.harvard.edu/abs/2004PASP..116....9S} {116, 9}

\bibitem[\protect\citeauthoryear{{Tinney}, {Burgasser}, {Kirkpatrick}  \&
  {McElwain}}{{Tinney} et~al.}{2005}]{Tinney2005}
{Tinney} C.~G.,  {Burgasser} A.~J.,  {Kirkpatrick} J.~D.,   {McElwain} M.~W.,
  2005, \mn@doi [\aj] {10.1086/491734}, \href
  {http://adsabs.harvard.edu/abs/2005AJ....130.2326T} {130, 2326}

\bibitem[\protect\citeauthoryear{{Tinney}, {Faherty}, {Kirkpatrick}, {Cushing},
  {Morley}  \& {Wright}}{{Tinney} et~al.}{2014}]{Tinney2014}
{Tinney} C.~G.,  {Faherty} J.~K.,  {Kirkpatrick} J.~D.,  {Cushing} M.,
  {Morley} C.~V.,   {Wright} E.~L.,  2014, \mn@doi [\apj]
  {10.1088/0004-637X/796/1/39}, \href
  {http://adsabs.harvard.edu/abs/2014ApJ...796...39T} {796, 39}

\bibitem[\protect\citeauthoryear{{Todorov}, {Luhman}  \& {McLeod}}{{Todorov}
  et~al.}{2010}]{Todorov2010}
{Todorov} K.,  {Luhman} K.~L.,   {McLeod} K.~K.,  2010, \mn@doi [\apj]
  {10.1088/2041-8205/714/1/L84}, \href
  {http://adsabs.harvard.edu/abs/2010ApJ...714L..84T} {714, L84}

\bibitem[\protect\citeauthoryear{{Todorov}, {Luhman}, {Konopacky}, {McLeod},
  {Apai}, {Ghez}, {Pascucci}  \& {Robberto}}{{Todorov}
  et~al.}{2014}]{Todorov2014}
{Todorov} K.~O.,  {Luhman} K.~L.,  {Konopacky} Q.~M.,  {McLeod} K.~K.,  {Apai}
  D.,  {Ghez} A.~M.,  {Pascucci} I.,   {Robberto} M.,  2014, \mn@doi [\apj]
  {10.1088/0004-637X/788/1/40}, \href
  {http://adsabs.harvard.edu/abs/2014ApJ...788...40T} {788, 40}

\bibitem[\protect\citeauthoryear{{Vrba} et~al.,}{{Vrba}
  et~al.}{2004}]{Vrba2004}
{Vrba} F.~J.,  et~al., 2004, \mn@doi [\aj] {10.1086/383554}, \href
  {http://adsabs.harvard.edu/abs/2004AJ....127.2948V} {127, 2948}

\bibitem[\protect\citeauthoryear{{Warren} et~al.,}{{Warren}
  et~al.}{2007}]{Warren2007}
{Warren} S.~J.,  et~al., 2007, \mn@doi [\mnras]
  {10.1111/j.1365-2966.2007.12348.x}, \href
  {http://adsabs.harvard.edu/abs/2007MNRAS.381.1400W} {381, 1400}

\bibitem[\protect\citeauthoryear{{Wizinowich} et~al.,}{{Wizinowich}
  et~al.}{2004}]{Wizinowich2004}
{Wizinowich} P.~L.,  et~al., 2004, in {Bonaccini Calia} D.,  {Ellerbroek}
  B.~L.,   {Ragazzoni} R.,  eds,  \procspie Vol. 5490, Advancements in Adaptive
  Optics. pp 1--11, \mn@doi{10.1117/12.552489}

\bibitem[\protect\citeauthoryear{{Wright} et~al.,}{{Wright}
  et~al.}{2010}]{Wright2010}
{Wright} E.~L.,  et~al., 2010, \mn@doi [\aj] {10.1088/0004-6256/140/6/1868},
  \href {http://adsabs.harvard.edu/abs/2010AJ....140.1868W} {140, 1868}

\makeatother
\end{thebibliography}
\input{main.bbl}

\label{lastpage}
\end{document}